\documentclass[a4paper,12 pt]{article}

\usepackage[T1]{fontenc}
\usepackage[utf8]{inputenc}
\usepackage[english]{babel}
\usepackage{amsmath}
\usepackage{amssymb}
\usepackage{amsfonts}
\usepackage{eufrak}
\usepackage[babel]{csquotes}
\usepackage[sorting=none]{biblatex}
\usepackage{slashed}
\usepackage{graphicx}
\usepackage{caption}
\usepackage{color}

\newcommand{\be}{\begin{equation}}
\newcommand{\ee}{\end{equation}}
\newcommand{\ba}{\begin{align}}
\newcommand{\ea}{\end{align}}

\date{}
\bibliography{biblio.bib}

\begin{document}
\begin{center}
\textbf{The Gribov Phenomenon in Flat and Curved Spaces}\\
\vspace{1cm}
Marco de Cesare\\
\vspace{1cm}
\textit{Dipartimento di Fisica, Complesso Universitario di
Monte S. Angelo, Via Cintia, 80126 Naples, Italy}
\end{center}
%The gauge principle is one of the most profound concepts in modern Physics. It lies at the foundations of the theories which describe all the known fundamental interactions. Yang-Mills theories are constructed by promoting a global symmetry to a gauge symmetry. The standard model of particle physics is a Yang-Mills theory with gauge group $SU(3)\times SU(2)\times U(1)$.
\cleardoublepage
\textbf{Abstract}

The quantization of Yang-Mills theories relies on the gauge-fixing procedure. However, in the non-Abelian case this procedure leads to the well known Gribov ambiguity. In order to solve the ambiguity a modification of the functional integral formula must be introduced. As a consequence of this, the Green functions get deep modifications in the infrared. We consider in particular the $SU(N)$ case and show that in the pure gauge case the ghost propagator is enhanced, while the gluon propagator is suppressed in this limit, therefore the study of the Gribov ambiguity may shed some light on the mass gap problem and on colour confinement.  
We discuss some recent developments on the subject in the case of a curved background. We argue that the concurrent presence of a spacetime curvature and the Gribov ambiguity may introduce further modifications to the Green functions in the infrared.
%Though in order to provide a full proof of this statement one has to solve some technical problems, which prevent one from finding the explicit form of the propagators in the infrared.
%without whom much more than this thesis would be missing.
\cleardoublepage
%\begin{quote}
%\emph{``There must be some way out of here''\\
%said the joker to the thief}\\
%(Bob Dylan)
%\end{quote}
\tableofcontents
\begin{section}{Introduction}
The first attempt to formulate a unified theory of all the interactions dates back to Plato. In the Timeo he gave an exposition of his theory, according to which each of the four elements is made of tiny symmetric shapes. These are now known as Platonic polyhedra. The elements owe their properties to the geometric properties of the elementary constituents. An explanation of all the natural phenomena, no matter how complicated they could appear, was then proposed, which consisted in reducing them to elementary processes between regular polyhedra.

Though Plato's theory may seem naive from our point of view and besides the lack of a scientific method, it shares some of the characteristic aspects of modern theoretical physics. The idea that matter is made of simple constituents was backed by many in the history of phylosophy even before Plato. In modern times it proved itself to be very fruitful when applied to the physical science, leading to the formulation of the atomic hypothesis, which was at the basis of many important discoveries. Another important aspect in Plato's theory is evidently the role played by symmetry. It is symmetry which tells us which objects are elementary. This is true even in quantum field theory, where elementary particles in Minkowski space are classified as irreducible representations of the Poincarè group\footnote{Though what we regard as an elementary particle actually depends on the energy scale}. %Symmetry also determines the properties of the interaction between two different elements

Leaving aside Plato, we turn to the importance of geometry in modern physics. Before the formulation of general relativity geometry was considered simply as a background in which physical processes take place. It was Einstein's merit to recognize the spacetime metric as a dynamical object, and to identify the gravitational field with its curvature. 
%Einstein-Hilbert's action functional  displays a remarkable symmetry of the theory:
The corner-stone of Einstein's theory is the principle of general covariance, which states that the equations of Physics have the same form in every system of local coordinates on spacetime \emph{i.e.}, they must be written in tensor form. In the Lagrangian formulation this is translated in a special symmetry of Einstein-Hilbert's action functional. The symmetry we are talking about is the invariance under the action of the diffeomorphism group. This is a group of \emph{local} transformations and is infinite dimensional, in contrast to global symmetry groups which describe \emph{e.g.}, rotational invariance, translational invariance, flavour symmetry, etc.

In 1932 Heisenberg postulated that isospin were an approximate symmetry of the strong interaction between nucleons. Because of the tiny difference in mass between the proton and the neutron and the nearly equal composition in neutrons and protons of the lightest stable nuclei, he argued that they behave in the same way with respect to the strong interaction. The nucleon is then described as an isospin doublet, and one could not distinguish two different isospin states if it were not for the electromagnetic interaction. In 1954 Yang and Mills investigated the consequences of assuming a local isospin symmetry. This is in the same spirit of general covariance, as one has in principle the freedom of choosing at each point of spacetime what to call a proton and what a neutron. This assumption has deep consequences, as the promotion of a global symmetry to a local symmetry requires the introduction of a new field with its own dynamics. The new field is a \emph{gauge connection} and defines how to parallel 
transport matter fields from one spacetime point to another. There is a close relationship between the theory formulated by Yang and Mills and the mathematical theory of fibre bundles. Actually this provides a geometric language in which to formulate the physical theory, in much the same way as differential geometry of Riemannian manifolds is the language for general relativity. Gauge transformations play an analogous role to that played by changes of coordinates in general relativity.

%The gauge group of the theory formulated by Yang and Mills in their paper is the isospin group $SU(2)$. The gauge principle can be used to describe the fundamental interactions. The strong interaction is described by a Yang-Mills theory with gauge group $SU(3)$, which expresses the local colour symmetry of the theory. The weak and the electromagnetic interactions are unified in the Glashow-Weinberg-Salam model. This is a gauge theory with gauge group $SU(2)\times U(1)$. The Higgs mechanism is responsible for the mass

The Standard Model of particle physics provides a unified description of three of the fundamental interactions. This is a Yang-Mills theory with gauge group $SU(3)\times SU(2)\times U(1)$. The $SU(3)$ factor accounts for colour symmetry of the strong interaction, while the $SU(2)\times U(1)$ factor pertains to the weak and the electromagnetic interaction. $SU(2)\times U(1)$ is spontaneously broken to $U(1)$ in the Glashow-Weinberg-Salam model by the Higgs mechanism, which gives mass to the vector bosons mediating the weak force. At present all the known fundamental interactions are described in terms of gauge theories. A quantum theory of gravitation does not yet exist, although many different approaches have been proposed.

In 1973 Wilczek, Gross and Politzer discovered that non-Abelian Yang-Mills theories are asymptotically free. This means that the coupling constant of the gauge field to itself and to the fermions becomes smaller and smaller as the energy scale increases\footnote{The validity of this result actually depends on the number of fermions involved}. This is an important property of the strong interaction, implying that perturbation theory becomes very accurate at high energies. This is in contrast to quantum electrodynamics, where perturbation theory breaks down at high energy while it works fine at low energies. The weak interaction is also described by a non-Abelian gauge theory, but the Higgs mechanism breaks the gauge symmetry and the theory becomes well defined perturbatively in the infrared, \emph{i.e.} at low energies.

There are two important properties which a quantum theory of the strong interaction should account for, \emph{i.e.} \emph{colour confinement} and the formation of a \emph{mass gap}. Perturbation theory is of no help in the understanding of these features, because they manifest themselves at low energies, where according to asymptotic freedom the couplings are much greater than one. The meaning of colour confinement is that \emph{coloured} particles cannot be detected as free asymptotic states, only bound states of such particles can be detected. The mass gap is the mass of the lightest particle in the spectrum. The existence of a mass gap implies that no massless particle can form an asymptotic state. The two concepts are distinct but intimately related. No proof exists at present that QCD satisfies these properties.

In 1978 Gribov proposed a possible explanation for the formation of a mass gap. He observed that in non-Abelian theories the gauge fixing procedure is ambiguous. In fact in non-Abelian theories there are gauge equivalent connections which satisfy the same gauge-fixing condition. Fixing the gauge is necessary in order to quantize gauge theories in the Lagrangian approach. This ambiguity can be solved by restricting the domain of functional integration. A way to partially accomplish this is to restrict to the so called Gribov region in the space of gauge connections. When this is done, non perturbative corrections are obtained which deeply modify the behaviour of the theory in the infrared. An effective term appears in the action which accounts for the presence of the boundary of such a region. In particular the gauge field propagator turns out to decrease to zero in the infrared. Refinements of the model allow to calculate glueball masses, and the predictions agree with results obtained by lattice simulations.
 However some technical problems are still unsolved, as no successful way to completely eliminate the redundancies from the functional integral has been found. An alternative way to approach the problem using stochastic quantization was proposed by Zwanziger.

Recently the Gribov ambiguity has been investigated also in curved spacetimes. If one studies quantum field theory with the purpose of obtaining predictions for the outcome of particle physics experiments, the presence of a gravitational field can just be neglected. This is true for the range of energies available at the present day in colliders, and will remain true at least for the next one hundred years. Usually quantum field theories on curved spacetimes are studied with the hope of obtaining some hints at the quantization of gravity. The unification of the four fundamental interactions is supposed to occur at the Planck energy scale. Here the reason for studying non-Abelian gauge theories in curved spacetimes is completely different, and the purpose is to understand if the presence of gravity modifies the confinement picture.

In the first chapter we recall the concept of gauge invariance for Maxwell's equations. The concept of gauge connection is introduced while working on specific examples. It is stressed the role of the electromagnetic field as the entity which allows one to turn a global symmetry into a local one. The dynamics of the electromagnetic field is obtained by the requirement of gauge invariance.

In the second chapter the geometric framework is set up for the study of non-Abelian gauge theories. Concepts already defined in the Abelian case are generalized and analyzed. The language of principal fibre bundles is introduced.

The third chapter is dedicated to some aspects of the quantization of Yang-Mills theories. The Lagrangian formulation of quantum field theory is presented. It is shown how to quantize gauge theories using the method developed by Feynman, De Witt, Faddeev and Popov (FDFP). The role of ghost fields is discussed, and a brief account of BRST (Becchi, Rouet, Stora and Tyutin) symmetry is given.

In the fourth chapter the problem which gives the name to this thesis is introduced. The Gribov ambiguity is studied in flat space. Geometric properties of the Gribov region and of the fundamental modular region are discussed. Gribov's heuristic approach to the problem of quantization in the presence of the ambiguity is presented. This explains the possible role of such ambiguity in the formation of a mass gap.

In the fifth and last chapter the Gribov ambiguity is discussed in the case of a curved background spacetime. Particular attention is paid to the determination of the asymptotic properties of spacetimes not admitting Gribov copies of the naive vacuum. The possibility that spacetime curvature may introduce further modifications to the infrared behaviour of correlation functions is discussed.
\end{section}

\begin{section}{Maxwell's theory}
%%The equations that bear his name describe the dynamics of the electromagnetic field. They are written (in c.g.s e.m. units) in vector form as follows
%%\begin{align}\label{maxwell}
%%\nabla\dot\mathbf{D}=4\pi\rho\\
%%\nabla\dot\mathbf{B}=0\\
%%\nabla\times\mathbf{E}=-frac{1}{c}\frac{\partial \mathbf{B}}{\partial t}\\
%%\nabla\times\mathbf{H}=frac{1}{c}(\frac{\partial \mathbf{D}}{\partial t}+4\pi \mathbf{J})
%%\end{align}
%%Together with these the law of current conservation holds
%%\be
%%\nabla\mathbf{J}+\frac{\partial \mathbf{J}}{\partial t}=0.
%%\ee
%%In order for the problem of determining the evolution of the electromagnetic field, given that of the sources, we should also add the constitution equation, which express the link between the fields in matter with those in the vacuum.
%%Because of the second and third equations the electromagnetic field can be expressed in terms of one vector and one scalar function.
%%\ba
%%\mathbf{B}=\nabla\times\mathbf{A}\\
%%\mathbf{E}=-\nabla\phi-\frac{1}{c}\frac{\partial \mathbf{A}}{\partial t}.
%%\ea
%%$\mathbf{A}$ and $\phi$ are called respectively the vector and the scalar potential. We can immediately see that the $\mathbf{E}$, $\mathbf{B}$ fields are left untouched whenever we perform a transformation on the potentials of the form
%%\ba
%%\mathbf{A}'=\mathbf{A}+\nabla\chi\\
%%\phi '=\phi-\frac{1}{c}\frac{\partial\phi}{\partial t}.
%%\ea
%%We will refer to these transformations as gauge transformations. Thus gauge transformations are 
The first gauge-invariant theory was formulated by Maxwell in order to give a unified description of electric and magnetic phenomena, and light. The equations that bear his name describe the dynamics of the electromagnetic field (in vacuum) in terms of two vector fields $\mathbf{E}$ and $\mathbf{B}$. Because of the remarkable structure of Maxwell equations, we can also give a description of the physics in terms of two non-physical fields called potentials. The relations linking the potentials, which we will refer to as $\phi$ and $\mathbf{A}$, to the fields is the following:
\ba
\mathbf{B}=\nabla\times\mathbf{A},\\
\mathbf{E}=-\nabla\phi-\frac{1}{c}\frac{\partial \mathbf{A}}{\partial t}.
\end{align}
From these equations it is immediate to notice that the following transformations
\ba
\mathbf{A}'=\mathbf{A}+\nabla\chi,\\
\phi '=\phi-\frac{1}{c}\frac{\partial\chi}{\partial t},
\end{align}
leave the physical fields invariant (here $\chi$ denotes a function of position and time). The existence of such transformations is referred to as the \emph{gauge invariance} property of electromagnetism.\\
Gauge invariance is quite different from the symmetries we may already have encountered in point-particle mechanics, such as translation invariance in space and time, rotational invariance and so on. Those symmetries are associated to some particular transformations, always related to the \emph{physical degrees of freedom}, we can perform on the physical system without affecting its properties. A gauge transformation instead does not affect at all the physical degrees of freedom of the system, so the meaning of gauge invariance and the role it plays are quite different and subtler. In fact gauge invariance can be eliminated (completely or in part) by a procedure called gauge-fixing. Consider for example the Coulomb gauge condition\be
\nabla\cdot \mathbf{A}=0.
\ee
In the case in which sources are absent we can also consistently put $\phi=0$. These two conditions completely eliminate the gauge freedom (so long as we are in the whole three-dimensional space). In the Coulomb gauge only the \emph{transverse} degrees of freedom survive, in accordance with the well-known transversality of the electromagnetic radiation.\\
This reasoning shows that gauge invariance actually expresses the presence of redundancies in the description in terms of potentials.\\
\begin{subsection}{A charged particle in an electromagnetic field} 
We begin with the study of the motion of a non-relativistic particle in an electromagnetic field.
The Hamiltonian operator for a particle of mass $m$ and charge $e$ reads as follows
\be
H=\frac{1}{2m}\left(p-\frac{e}{c}A\right)^2+e\phi.
\ee
When we perform a gauge transformation on the potentials the Hamiltonian changes form, but the dynamics of the particle must be unaffected by this change. In other words as a consequence of a gauge transformation, the wave function is multiplied by a phase factor.\\
In order to see this explicitly let us focus our attention on the form of the Hamiltonian, in particular on the term in brackets. Let us consider the action of this operator on the wave function
\be
\left(p-\frac{e}{c}A\right)\psi.
\ee
We'll indicate with an apex the transformed quantities. After a gauge transformation $A$ turns into $A'$ as $\psi$ turns into $\psi'$. So we have to deal with the expression
\be
\left(p-\frac{e}{c}A'\right)\psi'
\ee
and try to figure out what its relation should be with the unprimed expression. We assume that the following equation holds
\be\label{COVARIANT}
\left(\left(p-\frac{e}{c}A\right)\psi\right)'=\left(p-\frac{e}{c}A'\right)\psi',
\ee
where the prime in the LHS means that it should be transformed in the same way as $\psi$ does. Writing $\psi$ as the result of the action of an operator on $\psi$, namely $\psi'=K\psi$, we get
\be
K\left(-i\hbar\nabla-\frac{e}{c}A\right)\psi=\left(-i\hbar\nabla-\frac{e}{c}A\right)K\psi-\frac{e}{c}\nabla\chi K\psi,
\ee
whence it follows
\be
i\hbar\nabla K+\frac{e}{c}\nabla\chi K=0.
\ee
The solution of this equation is
\be
K=\exp{\left( i\frac{e}{c}\chi\right)},
\ee
which is indeed a phase factor, although depending on position and time. A check of consistency of our guess is now in order. We shall prove that, provided $\psi'$ is a solution of
\be
i\hbar\frac{\partial\psi '}{\partial t}=H'\psi '
\ee
$\psi$ is a solution of the unprimed equation
\be
i\hbar\frac{\partial\psi}{\partial t}=H\psi.
\ee
The converse can be proved in a similar way. From our hypotheses it follows that
\be
i\hbar\frac{\partial K}{\partial t}\psi+i\hbar K\frac{\partial\psi}{\partial t}=\frac{1}{2m}\left(p-\frac{e}{c}A'\right)^2\psi ' +\left(e\phi-\frac{e}{c}\frac{\partial\chi}{\partial t}\right)K\psi.
\ee
Using our initial guess and the expression we just found for $K$, this reduces to the expected result.\\

The whole information about the state of the system is contained in the wave function $\psi$. We can multiply it by an arbitrary (albeit constant in space) phase factor and still have the same state. When we take into account the presence of an electromagnetic field, our freedom of phase multiplication increases a lot. We can pick an arbitrary phase at each point of space-time and multiply it by the wave function without affecting the state in any way.\\
Though this incredible freedom has a price. We could not be allowed to change our phases at will without the existence of a gauge-invariant field, having its own dynamics. This is the key point that will lead us to the construction, first made by Yang and Mills \cite{yang1954conservation}, of the more complicated but nevertheless fundamental non-Abelian gauge theories.\\
Before closing this section we want to make it more evident how in our construction we implicitly defined a new concept. Looking back to equation (\ref{COVARIANT}), we see that it can be rewritten as
\be
\left(\nabla-i\frac{e}{c}\mathbf{A}\right)\psi=K^{-1}\left(\nabla-i\frac{e}{c}\mathbf{A}'\right)K\psi.
\ee
From this equation we can obtain once again the transformation law for the potential
\be
\mathbf{A}'=\mathbf{A}-i\frac{c}{e}K^{-1}\nabla K.
\ee
When we write it in this form, it is immediately clear that a gauge transformation depends on the choice of a $U(1)$-valued function on three dimensional space.
We will refer to the ``vector'' $\nabla-i\frac{e}{c}\mathbf{A}$ as the covariant gradient. The component of this vector along the $i$-th direction will be the corresponding covariant derivative and identified by the symbol $D_{i}(\mathbf{A})$. Calculating the commutator of covariant derivatives taken along different directions, we get
\be
[D_{i}(\mathbf{A}),D_{j}(\mathbf{A})]=-i\frac{e\hbar}{c}(\partial_{i}A_{j}-\partial_{j}A_{i}),
\ee
which is, apart from an imaginary factor, the $(i,j)$-component of the electromagnetic field tensor.
\end{subsection}
\begin{subsection}{The Complex Klein-Gordon field}
The Klein-Gordon Lagrangian\footnote{The metric on minkowskian spacetime is taken with signature (1,-1,-1,-1)} for a complex scalar field reads as follows
\be
\mathcal{L}=\partial^{\mu}\phi^{*}\partial_{\mu}\phi-m^2\phi^{*}\phi.
\ee
It is seen to be invariant under \emph{global} $U(1)$ transformations
\begin{align}\label{U(1)}
\phi\rightarrow e^{-i\Lambda}\phi,\\
\phi^{*}\rightarrow e^{i\Lambda}\phi^{*}.
\end{align}
Here $\Lambda$ denotes a constant. As it is well-known, Noether's theorem associates to each continous symmetry of the Lagrangian a conserved quantity. The Noether current associated to this symmetry is the four-vector
\begin{align}
J^{\mu}=&i(\phi^{*}\partial^{\mu}\phi-\phi\partial^{\mu}\phi^{*}),\\
\partial^{\mu}J_{\mu}=&0.
\end{align}
The  Noether charge is the integral over a space-like hypersurface of the time component of the Noether current. It is a real number and it vanishes when $\phi$ is real. That's the reason for taking into account complex fields.\\
We would now like to turn our global symmetry into a \emph{local symmetry}. We can try to accomplish that by making $\Lambda$ a function of position. Though, when calculating the variation of the Lagrangian
\footnote{To calculate the variations we need to take formula (\ref{U(1)}) for an infinitesimal $\Lambda$, \emph{i.e.} $\delta\phi=-i\Lambda\phi$, $\delta\phi^{*}=i\Lambda\phi^{*}$.}
we obtain
\be
\delta\mathcal{L}=\partial_{\mu}\Lambda J^{\mu}.
\ee
So the invariance is lost. But we can try to restore it by adding new terms to the Lagrangian, which compensate for this one \cite{Ryder:QFT}. In first instance we can add the term
\be
\mathcal{L}_{1}=-eJ^{\mu}A_{\mu},
\ee
which is a coupling of the four-current with an external field $A_{\mu}$. Assuming that under a local $U(1)$ transformation the field transforms as
\be\label{GAUGE TRANSFORM Amu}
A_{\mu}\rightarrow A_{\mu}+\frac{1}{e}\partial_{\mu}\Lambda,
\ee
where $e$ is a constant, we get
\be
\delta\mathcal{L}_{1}=-e\delta{J^{\mu}}A_{\mu}-J^{\mu}\partial_{\mu}\Lambda.
\ee
The last term in this expression exactly compensates for the variation of the starting Lagrangian. We need yet another term to compensate for the first one.
Let's calculate the variation in the four-current.
\be
\delta J^{\mu}=2\phi^{*}\phi\partial^{\mu}\Lambda
\ee
This implies
\be
\delta\mathcal{L}+\delta\mathcal{L}_{1}=-2e\phi^{*}\phi\partial^{\mu}\Lambda.
\ee
Thus we  just need to add the term
\be
\mathcal{L}_{2}=e^2\phi^{*}\phi A_{\mu}A^{\mu}.
\ee
The sum of the three contributions is zero
\be
\delta\mathcal{L}+\delta\mathcal{L}_{1}+\delta\mathcal{L}_{2}=0.
\ee
So we found that we can turn a global symmetry into a local one by considering the Lagrangian
\be
\mathcal{L}'=\partial^{\mu}\phi^{*}\partial_{\mu}\phi-m^2\phi^{*}\phi-eJ^{\mu}A_{\mu}+e^2\phi^{*}\phi A_{\mu}A^{\mu}.
\ee
If we introduce the covariant derivative operator $D_{\mu}=\partial_{\mu}+ieA_{\mu}$ we can rewrite it as
\be
\mathcal{L}'=D^{\mu}\phi^{*}D_{\mu}\phi-m^2\phi^{*}\phi.
\ee
The constant $e$ appearing in the formulae plays the role of a coupling constant with the external field $A_{\mu}$, thus allowing its interpretation as the electric charge of the field.
%%The dynamics of the latter is not yet specified, but we can inlude it in the lagrangian by adding the term
%%\be
%%\mathcal{L}_{em}=-\frac{1}{4}F_{\mu\nu}F^{\mu\nu},
%%\ee
%%where $F_{\mu\nu}$ is the electromagnetic field tensor. Thus the full lagrangian $\mathcal{L}'+\mathcal{L}_{em}$ is gauge-invariant
\end{subsection}
\begin{subsection}{The Dirac field}
The Lagrangian for the Dirac field is written as
\be
\mathcal{L}_{D}=\bar{\psi}(i\gamma^{\mu}\partial_{\mu}-m)\psi.
\ee
It exhibits global $U(1)$ invariance in the same way as the complex Klein-Gordon field does. A $U(1)$ transformation acts on the spinor fields as follows
\begin{align}
\psi=&e^{-i\Lambda}\psi,\\
\bar{\psi}=&e^{i\Lambda}\bar{\psi}.
\end{align}
So that under an infinitesimal transformation we have
\be
\delta\mathcal{L}_{D}=\bar{\psi}\gamma^{\mu}\partial_{\mu}\Lambda\psi.
\ee
We add to the Lagrangian the term
\be
\mathcal{L}_{D}^{1}=-e\bar{\psi}\gamma^{\mu}A_{\mu}\psi,
\ee
where the field $A_{\mu}$ transforms according to (\ref{GAUGE TRANSFORM Amu}). The Lagrangian thus obtained is once again consistent with the minimal coupling prescription
\be
\mathcal{L}_{D}'=\bar{\psi}(i\gamma^{\mu}D_{\mu}-m)\psi=\bar{\psi}(i\gamma^{\mu}\partial_{\mu}-m)\psi-\bar{\psi}\gamma^{\mu}A_{\mu}\psi.
\ee
We have so far seen that the way to promote a global symmetry to a local symmetry is independent of the nature of the field to which this symmetry is an attribute, \emph{i.e.} of its being fermionic or bosonic. We also proved that the link between the freedom of multiplying the wave-function by a position dependent phase and the electromagnetic field generalizes also to the full relativistic case. This is accomplished by using the minimal coupling prescription, which makes also evident the important role played by the covariant derivative in the dynamics. Though one important ingredient is missing. The electromagnetic field  was treated as an external non-dynamical field. In order to understand better its nature we should also investigate its dynamics. This will be done in the next section.
\end{subsection}
\begin{subsection}{The Maxwell field}
We will deal with the construction of a Lagrangian for the Maxwell field in vacuum. When sources are present we shall simply add to this one the terms which express the interaction with them. We have already studied these terms in the case of the Klein-Gordon and Dirac field, but the sources should not be necessarily of this kind. For example we can also consider the coupling of the electromagnetic field with a system of charged classical point-particles.

Maxwell equations are concisely written in four-dimensional notation as follows
\begin{align}
\partial_{\mu}F^{\mu\nu}=0,\\
\varepsilon^{\lambda\mu\nu}\partial_{\lambda}F_{\mu\nu}=0.
\end{align}
Those in the first set are properly dynamical, while those in the second set are geometrical constraints. As they are of first order in the field $F_{\mu\nu}$ the Lagrangian should be quadratic in it and contains no derivatives of $F_{\mu\nu}$. It should of course be Lorentz-invariant. The only two quadratic and Lorentz invariant quantities are $F_{\mu\nu}F^{\mu\nu}$ and $\varepsilon^{\lambda k\mu\nu}F_{\lambda k}F_{\mu\nu}$, but the latter should be excluded because it generates a trivial dynamics. So in a suitable system of units of measurement the Lagrangian for the Maxwell field reads as follows
\be
\mathcal{L}_{em}=-\frac{1}{4}F_{\mu\nu}F^{\mu\nu}.
\ee
This Lagrangian is automatically gauge-invariant as we based the whole of our construction on the gauge-invariant field $F_{\mu\nu}$. As it is evident, a mass term  of the type $m^2 A_{\mu}A^{\mu}$ would destroy gauge-invariance. As we will see in the following though, in a three dimensional spacetime it is possible to construct other quantities which are both Lorentz and gauge invariant that can be used to make the electromagnetic field acquire a mass.
%%It is that gauge fields are massless, though their quanta may acquire a mass due to the coupling with other fields.
\end{subsection}
\end{section}

\begin{section}{Gauge fields}\label{capitolo gauge}
In the first chapter we have seen that the simple concept of gauge invariance, arising from the equations of motion of the electromagnetic field equations, is indeed very rich. We have observed that it is associated to a local $U(1)$ symmetry, and that the matter fields are in the fundamental representation of this group. We have also discussed how naturally the concept of a \emph{covariant derivative} emerges and how it is intimately related to the existence of a gauge potential. Starting from this we will construct a class of theories which annoverate Maxwell theory as a special case. Considering gauge groups more complicated than $U(1)$ we can describe the strong and the weak forces.

In this chapter we set the geometric framework for the study of classical gauge fields. The generalization to curved spaces is immediate and we will not spend many words on it, as we will concentrate our efforts in the generalization to different gauge groups and its implications. We will begin our study of gauge theories in the basic case of a trivial principal bundle. This is the standard approach which paves the way for the transition to the quantum theory. Once we set the basis we will present the more abstract geometric formulation in terms of principal fibre bundles. Though the approach presented in the first half of the chapter is not less general. In fact distinct topological sectors arise naturally when considering the asymptotic behaviour of the gauge connection. When the base manifold has the topology of $\mathbb{R}^{n}$, by one-point-compactification these sectors are actually seen to be in one-to-one correspondence with non-trivial principal bundles on the sphere $\mathbb{S}^{n}$.
%We start with an heuristic discussion with the aim of introducing the basic concepts and also to give some insights on the meaning of %the abstract constructions that follow.
\begin{subsection}{Parallel transport, Curvature}
%The concept of parallel transport is introduced in order to relate gauge transformations performed at different points of spacetime.
Let $M$ be the spacetime manifold, which we will assume to be a pseudo-Riemannian manifold whose topology we don't need to specify for the moment, and let us consider a field $\phi$ on $M$ belonging to a certain representation $R$ of the gauge group $G$. In many applications it is useful to take $SU(N)$ as the gauge group. The Lie-Algebra $su(N)$ of such a group is a a real vector space, spanned by anti-Hermitian $N\times N$ matrices which are traceless.

To begin with, we need a generalization of the concept of \emph{covariant derivative} introduced in the previous chapter.
This is quite straightforward to do, and is accomplished by introducing a Lie-algebra-valued field on spacetime \cite{WardWells}. We will call this field the \emph{gauge connection} and  denote it by $A$. To take into account the curvature of spacetime, we simply replace partial derivatives with Levi-Civita derivatives\footnote{This is the only derivative operator which is compatible with the metric structure on $M$. We purposely do not use the expressions covariant derivative and Levi-Civita connection in order to avoid making confusion with the corresponding concepts in gauge theories}. Thus the covariant derivative of the field $\phi$ is given by
\be
D_{\mu}\phi=\nabla_{\mu}\phi+A_{\mu}\phi,
\ee
in the case in which $\phi$ belongs to the fundamental representation, or
\be
D_{\mu}\phi=\nabla_{\mu}\phi+[A_{\mu},\phi],
\ee
if it belongs to the adjoint representation of $G$.
But there is another requirement which should be satisfied, \emph{i.e.} the covariant derivative of the field should transform in the same way. In the case of the fundamental representation the transformation laws are the following
\begin{align}
\phi\rightarrow g^{-1}\phi\\
D_{\mu}\phi\rightarrow g^{-1}D_{\mu}\phi
\end{align}
while in the case of the adjoint representation we have
\begin{align}
\phi\rightarrow g^{-1}\phi g\\
D_{\mu}\phi\rightarrow g^{-1}(D_{\mu}\phi) g.
\end{align}
In both cases we get the following transformation law for the gauge connection
\be
A_{\mu}\rightarrow g^{-1}A_{\mu} g+ g^{-1}\nabla_{\mu}g.
\ee

Using the gauge connection we can introduce a notion of parallel transport in a natural way. We will say that a field $\phi$ is \emph{parallel} if its covariant derivative is identically equal to zero
\be
D_{\mu}\phi=0.
\ee
For a generic $\phi$, given its value at a point $x$, we can define the parallel transported field $\phi^{t}_{A}$ at $x+dx$ by means of the formula
\be\label{transport}
\phi^{t}_{A}(x+dx)=\phi(x)-A_{\mu}dx^{\mu}\phi(x).
\ee
The concept of parallel transport allows us to relate gauge transformations performed at different points of spacetime \cite{ItzykzsonZuber}. Indeed it follows from the transformation law for the connection that
\be\label{compatibilità}
T(x,x+dx;A[g])g(x)=g(x+dx)T(x,x+dx;A)
\ee
Here $T(x,x+dx;A)$ denotes the parallel transport operator associated to the connection $A$, while $A[g]$ is the gauge-transformed connection. 

We now want to find a formula for the parallel transport of a field to a distant point \cite{Ramond:QFT}. If we want to move from a point $x$ to another $y$ along a given path $\Gamma$, we can partition the path with points ${x_{j}}$ and iterate formula (\ref{transport})
\be
\phi^{t}_{A}(y)=\prod_{j}(1-dx^{\mu}A_{\mu})\phi(x).
\ee
The RHS defines the path-ordered integral
\be
Pe^{-\int_{x}^{y}dx^{\mu}A_{\mu}}.
\ee
When $x\equiv y$ so that  $\Gamma$ is a closed curve, the path-ordered integral is an element of the gauge group called the holonomy of $A$ at $x$ around $\Gamma$. From formula (\ref{compatibilità}) we have the following transformation rule for the holonomy
\be
Pe^{-\int_{\Gamma}dx^{\mu}A[g]_{\mu}}=g(x)Pe^{-\int_{\Gamma}dx^{\mu}A_{\mu}}g^{-1}(x).
\ee
The trace of the holonomy defines the Wilson loop
\be
W(\Gamma)=Tr(Pe^{-\int_{\Gamma}dx^{\mu}A_{\mu}}),
\ee
which is gauge invariant and does not depend on the choice of a metric on $M$. Observables of this kind play the lead role in topological field theories \cite{Witten:Jones}.

We need to define another important concept which is the curvature of a connection. This is a two-form given by the commutator of covariant derivatives
\be
F_{\mu\nu}=[D_{\mu},D_{\nu}]=\nabla_{\mu}A_{\nu}-\nabla_{\nu}A_{\mu}+[A_{\mu},A_{\nu}].
\ee
Under a gauge transformation it transforms as an element of the adjoint representation 
\be
F_{\mu\nu}\rightarrow g^{-1}F_{\mu\nu}g.
\ee
It is a simple consequence of the Jacobi identity that it satisfies the Bianchi identity
\be
D_{h}F_{jk}+D_{k}F_{hj}+D_{j}F_{kh}=0.
\ee
It is possible to show that the curvature is zero if and only if the connection is pure-gauge. We can see that the parallel transport of a field along a closed path up to third order terms depends only on the gauge curvature. When the curvature is zero the holonomy is equal to the identity and so the connection is gauge equivalent to the zero connection. The proof of the converse is trivial.\\
Though this property holds only locally. On a manifold with a non-trivial topology closed loops around which the holonomy is not zero may exist even with a connection which is globally flat. An important physical consequence of this is found in the Aharonov–Bohm effect \cite{Nakahara:GTP}\cite{Ryder:QFT}.
\end{subsection}

\begin{subsection}{Dynamics of the gauge field}
As we have seen in the first chapter, gauge fields are usually coupled to matter fields according to the \emph{minimal coupling} prescription\footnote{One can choose to adopt other non-minimal prescriptions, but they lead to non-renormalizable quantum theories.}. This amounts to exchanging ordinary derivatives for covariant derivatives  in the matter Lagrangian.
Still we need to find a Lagrangian which gives the dynamics of the gauge field. In order to accomplish this we can repeat the same arguments used to construct a Lagrangian for the Maxwell field. The Lagrangian should be invariant both under the action of the gauge group and that of the invariance group of the metric on $M$. We also require the equations of motion to be of first order in the field strength\footnote{We remark that in the case of non-Abelian theories this assumption cannot be justified on the basis of a superposition principle, which is valid for the Maxwell theory (see for example \cite{Landau:CaCl}).}. The only terms which satisfy these properties in the case of a four-dimensional spacetime are $Tr(F_{\mu\nu}F^{\mu\nu})$ and $Tr(F_{\mu\nu}(*F)^{\mu\nu})$. Though the latter is seen to be a total divergence and so (for a suitable class of boundary conditions) it does not contribute to the equations of motion. Thus we can write the Yang-Mills action as
\be
\mathcal{S}_{YM}=\frac{1}{4}\int d^4xTr(F_{\mu\nu}F^{\mu\nu}).
\ee
When used in a dynamical context it is customary to refer to the gauge curvature $F_{\mu\nu}$ as the \emph{field strength}. This also helps to make contact with the physical interpretation of this quantity, which is a generalization of the electromagnetic field tensor.

The Cartan-Killing metric is defined by the formula
\be
(A,B)=-Tr(AB).
\ee
It is obviously symmetric and positive-definite on anti-hermitian matrices. We can use it to rewrite the Yang-Mills action in the form given by \cite{WardWells}
\be
\mathcal{S}_{YM}=-\frac{1}{4}\int d^4x(F_{\mu\nu},F^{\mu\nu}).
\ee
Sometimes it is expressed in a slightly different form \cite{Ramond:QFT}. We introduce a basis $\{\tau^{a}\}$ in the Lie-Algebra. We take the vectors $\tau^{a}$ to be normalized according to the formula
\be
(\tau^{a},\tau^{b})=\delta^{ab}.
\ee
Decomposing the field strength in terms of the Lie-Algebra basis vectors\footnote{The basis chosen is orthonormal so we don't have to bother with the distinction between covariant and contravariant colour indices.}, we have
\be
F_{\mu\nu}=F_{\mu\nu}^{a}\tau^{a},
\ee
so that the action is also given by the formula
\be
\mathcal{S}_{YM}=\frac{1}{4}\int d^4xTr((F_{\mu\nu})^{a}(F^{\mu\nu})^{a}).
\ee
The equations of motion are
\be\label{YANG E MILLS}
D_{\mu}F^{\mu\nu}=0.
\ee

We can couple a matter field to the gauge-field using the minimal coupling prescription. The full Lagrangian in the case of the Dirac field reads as follows
\be
\mathcal{L}=\frac{1}{4}Tr(F_{\mu\nu}F^{\mu\nu})+\bar{\psi}\left(\slashed{D}+im\right)\psi.
\ee
And the equations of motion are
\begin{align}
D_{\nu}(F^{\mu\nu})^{a}=\bar{\psi}\gamma^{\mu}\tau^{a}\psi\\
\left(i\slashed{D}-m\right)\psi=0.
\end{align}
$\psi$ is a vector in the fundamental representation of the gauge group, whose components are spinors\footnote{Spinors are in the fundamental representation of the Clifford algebra of the metric $g$ on $M$. The dimension of the algebra depends on the dimensionality of spacetime.}. $\bar{\psi}$ is an adjoint spinor.
%scalar field reads as follows
%\be
%\mathcal{L}=\mathcal{L}_{YM}+\frac{1}{2}D^{\mu}\phi D_{\mu}\phi-\frac{1}{2}m^2\phi^2.
%\ee

%From this we obtain the equations of motion 
%\begin{align}
%D^{\mu}D_{\mu}\phi+m^2\phi^2=0\\
%D^{\mu}F_{\mu\nu}=
%\end{align}
\end{subsection}

\begin{subsection}{Self-dual and anti-self-dual fields}
The Hodge $*$ operator, when acting on two-forms on a four-dimensional spacetime, squares to minus the identity \cite{WardWells}. So its eigenvalues are $\pm i$. Then it is natural to define self-dual-fields
\be\label{SD}
*F_{\mu\nu}=iF_{\mu\nu}
\ee
and anti-self-dual fields
\be\label{ASD}
*F_{\mu\nu}=-iF_{\mu\nu}.
\ee
In both cases the Yang-Mills equations are automatically satisfied as a consequence of the Bianchi identity
\be
D^{\mu}*F_{\mu\nu}=0.
\ee
However these solutions are not so useful in spacetimes with a Lorentzian metric, as the only field verifying either of these properties is the zero field $F_{\mu\nu}=0$. In fact for the gauge group $SU(N)$, the field strength $F_{\mu\nu}$ is anti-hermitian, and so has to be its dual $*F_{\mu\nu}$. However because of the $i$ factor the property of (anti)self-duality can only be satisfied in the trivial case.

When we formulate Yang-Mills theory on a Euclidean background the eigenvalues of the Hodge $*$ operator are $\pm 1$. Then the argument leading to the vanishing of (anti)self-dual fields in the Minkowskian case doesn't apply. Actually such field configurations are very important for the quantum theory as they are absoulte minima of the Euclidean action, known as instantons. It is possible to show that instantons are either self-dual or anti-self-dual connections. Their contribution is important in the study of the low energy regime of confining theories such as QCD, as it was shown in the works of 't Hooft, Polyakov.
In the Euclidean theory definitions (\ref{SD}) and (\ref{ASD}) modify respectively as follows
\be
*F_{\mu\nu}=\pm F_{\mu\nu}.
\ee
In order to prove that self-dual fields are indeed local minima of the Euclidean action, we consider the quantity
\be
-Tr\left[(F_{\mu\nu}-*F_{\mu\nu})(F^{\mu\nu}-*F^{\mu\nu})\right]\geq0,
\ee
whose positive definiteness is evident. From this inequality it follows that
\be\label{LOWER BOUND YM}
-Tr(F_{\mu\nu}F^{\mu\nu})\geq-Tr(F_{\mu\nu}(*F)^{\mu\nu}).
\ee
Then the classical action is at a minimum when the field is self-dual \cite{Ramond:QFT}. An analogous proof may be constructed for anti-self-dual fields, so that also these configurations correspond to minima of the action. Let us focus on the self-dual case. The invariant $Tr(F_{\mu\nu}(*F)^{\mu\nu})$ is a total divergence
\begin{align}
Tr(F_{\mu\nu}(*F)^{\mu\nu})=4\partial_{\mu}W^{\mu}\\
W^{\mu}=Tr\left(A_{\nu}\partial_{\rho}A_{\sigma}+\frac{2}{3}A_{\nu}A_{\rho}A_{\sigma}\right)\varepsilon^{\mu\nu\rho\sigma}
\end{align}
$W_{\mu}$ (actually the three-form to which it is dual in 4-d space) is known in the mathematical literature as the Chern-Simons form of the Chern Character $\mathcal{R}^2=Tr(F_{\mu\nu}(*F)^{\mu\nu})$. Assuming that the spacetime $M$ has the standard topology of $\mathbb{R}^4$, we can integrate inequality (\ref{LOWER BOUND YM}) over a finite region and use integration by parts to obtain
\be
S_{YM}\geq -\oint n_{\mu}W^{\mu},
\ee
where $n_{\mu}$ is the direction normal to the boundary. With the above assumption, if we integrate over a region having the same topology of the whole space, the boundary is diffeomorphic to $S^3$. If we require the field strength to vanish at infinity, then the gauge connection at very large distances from the origin is a pure gauge
\be
A_{\mu}\simeq U^{-1}\partial_{\mu}U.
\ee
Then the lower bound is
\be\label{LOWER BOUND}
-\oint n_{\mu}W^{\mu}=\frac{1}{3}\oint n_{\mu}\varepsilon^{\mu\nu\rho\sigma}Tr\left(U^{-1}\partial_{\nu}UU^{-1}\partial_{\rho}UU^{-1}\partial_{\sigma}U\right).
\ee
In this formula $U$ is a function on $S^{3}$ with values in the gauge group $SU(N)$. Then lower bounds are classified in the same way of mappings from $S^3$ to $SU(N)$, \emph{i.e.} according to the homotopy group $\pi_3(SU(N))$. The value of the lower bound depends only on the asymptotic behaviour of the gauge transformation $U$ and not on details of the connection at finite points in spacetime. It is not possible to deform continuously a map falling in one homotopy class into another one with different asymptotic properties at infinity. Thus the lower bounds we obtained actually pertain to different topological sectors.

Let us evaluate the values of such lower bounds in the case of the gauge group $SU(2)$. The map $U$ in this case is specified by three functions $\phi_1$, $\phi_2$, $\phi_3$, which can be taken for convenience to be the Euler angles. Then the derivatives of $U$ with respect to the Cartesian coordinates on $\mathbb{R}^4$ are written as
\be
\partial_{\mu}U=\sum_{a=1}^{3}\frac{\partial\phi_{a}}{\partial x_{\mu}}\frac{\partial U}{\partial\phi_{a}}.
\ee
It follows that we can rewrite equation (\ref{LOWER BOUND}) in the following way
\be\label{LOWER BOUND PAR} 
-\oint n_{\mu}W^{\mu}=\frac{1}{3}\oint n_{\mu}\varepsilon^{\mu\nu\rho\sigma}\partial_{\nu}\phi^{a}\partial_{\rho}\phi^{b}\partial_{\sigma}\phi^{c} Tr\left(U^{-1}\partial_{a}UU^{-1}\partial_{b}UU^{-1}\partial_{c}U\right).
\ee
Moreover, using the antisymmetry of the tensor $\varepsilon^{\mu\nu\rho\sigma}$
\be
-\oint n_{\mu}W^{\mu}=2\oint n_{\mu}\varepsilon^{\mu\nu\rho\sigma}\partial_{\nu}\phi^{1}\partial_{\rho}\phi^{2}\partial_{\sigma}\phi^{3} Tr\left(U^{-1}\partial_{1}UU^{-1}\partial_{2}UU^{-1}\partial_{3}U\right).
\ee
We express the $SU(2)$ matrix in terms of Euler angles, so that $(\psi_1,\psi_2,\psi_3)=(\psi,\theta,\phi)$.
\be
U(\psi,\theta,\phi)=U_{z}(\phi)U_{x}(\theta)U_{z}(\psi)=e^{i\phi\frac{\sigma_3}{2}}e^{i\theta\frac{\sigma_1}{2}}e^{i\psi\frac{\sigma_3}{2}}
\ee
Then we evaluate the terms of the form $U^{-1}\partial_{a}U$ appearing in the trace in (\ref{LOWER BOUND PAR}).
\begin{align}
U^{-1}\partial_{\theta}U=&\frac{i}{2}(\sigma_{1}\cos\psi-\sigma_2\sin\psi)\\
U^{-1}\partial_{\phi}U=&\frac{i}{2}\left[\sigma_3\cos\theta+\sin\theta(\sigma_2\cos\psi+\sigma_1\sin\psi)\right]\\
U^{-1}\partial_{\psi}U=&i\frac{\sigma_3}{2}
\end{align}
Thus we can simply evaluate the trace
\be
-\frac{i}{8}\left[\cos^2\psi+\sin^2\psi\right]\sin\theta\; Tr(\sigma_1\sigma_2\sigma_3)=\frac{\sin\theta}{4}.
\ee
The term $n_{\mu}\varepsilon^{\mu\nu\rho\sigma}\partial_{\nu}\phi^{a}\partial_{\rho}\phi^{b}\partial_{\sigma}\phi^{c}$ is the Jacobian of the map $\phi$ from $S^3$ to $SU(2)$. Each point in $SU(2)$ is the image under the map $\phi$ of an integer number of points in $S^3$. In order for the map to be continuous, this integer has to be the same for each point. So we can simply shift the integration domain in (\ref{LOWER BOUND PAR}) from $S^3$ to $SU(2)$ and multiply by $n\in\mathbb{Z}$. This is a negative number if the map inverts the orientation of volumes.
The result is \cite{Ramond:QFT}
\be
S_{YM}\geq 2n\int_0^{\pi} d\theta \int_0^{2\pi} d\phi \int_0^{2\pi} d\psi \frac{\sin\theta}{4}=4\pi^2 n.
\ee
The integer number appearing in this formula is called the Pontrjagyn index and it is defined according to the formula
\be
n=-\frac{1}{16\pi^2}\int Tr(F_{\mu\nu}(*F)^{\mu\nu}).
\ee

The original instanton solution for the gauge group $SU(2)$ may be given in quaternionic notation \cite{Atyiah}
\be
A=Im\left\{\frac{xd\bar{x}}{1+|x|^2}\right\}.
\ee
Here $x=x_{1}+ix_{2}+jx_{3}+kx_{4}$ is the quaternionic expression of an $SU(2)$ element and $\bar{x}=x_{1}-ix_{2}-jx_{3}-kx_{4}$ is its conjugate. The rules for quaternionic multiplication are well-known. The quantity $|x|$ is defined by the formula $|x|^2=\bar{x}x=x\bar{x}$. The coefficient $x_{1}$ is the real part of the quaternion $x$, while the rest is referred to as its imaginary part. The field strength associated to the instanton is
\be
F=\frac{dx\wedge d\bar{x}}{(1+|x|^2)^2}.
\ee
It can be verified that this field configuration is self-dual.
\end{subsection}

\begin{subsection}{Non-Abelian charge}
As it is evident from the considerations we made in the previous chapter, the constancy of the electric charge of an elementary particle, say the electron, is strongly interwoven with gauge invariance of the Maxwell theory. If the charge were dependent on the particular point of spacetime we consider, we would not be able to construct gauge invariant actions for the matter fields.\\
Moreover the electromagnetic field itself may be used to define the charge. Looking at the first of Maxwell equations, we see that the electric charge in a given region of space may also be defined as the flux of the electric field through the boundary of that region. This is the content of Gauss' Law. It is important to observe that the electromagnetic field does not carry any charge itself.

We would like to generalize this link to the case of Yang-Mills theories. We start by calculating the first variation of the action under an infinitesimal gauge transformation, which of course must be equal to zero.
\be
\delta S_{YM}=\int_{M}(\delta A^{a}_{\mu}\left(D_{\nu}F^{\mu\nu}\right)^{a}+\partial_{\mu}\left(\delta A_{\nu}^{a}(F^{\mu\nu})^{a}\right)=0
\ee
If we require that we are doing our variations around an effective motion the first term is zero. Moreover for an infinitesimal gauge transformation we have
\be
\delta A_{\nu}^{a}=\partial_{\nu}w^{a}+A_{\nu}^{b}w^{c}f^{abc}.
\ee
Substituting this expression in the previous equation we obtain
\be
\int_{M} \partial_{\mu}\left([F^{\mu\nu},A_{\nu}]^{a}w^{a}\right)=0.
\ee
From the arbitrariness of $w^{a}$, which in particular may be taken constant and brought outside the integral sign, we have the following Noether current
\ba
(j^{\mu})^{a}=[F^{\mu\nu},A_{\nu}]^{a}\\
\partial_{\mu}j^{\mu}=0.
\end{align}
The Noether current may also be obtained from the equations of motion, but that derivation does not show the link with gauge invariance\cite{Ramond:QFT}. In fact from
(\ref{YANG E MILLS}) we have
\be
0=\partial_{\mu}\partial_{\nu}F^{\mu\nu}=-\partial_{\mu}[A_{\nu},F^{\mu\nu}].
\ee
The integral over a compact region of a space-like hypersurface of the time component of the Noether current is the charge
\be\label{GAUSS NON ABELIANO}
Q^{a}=\int_{\Sigma}[F^{0j},A_{j}]^{a}=\int_{\Sigma}\partial_{j}(F^{0j})^{a}=\oint_{\partial\Sigma}(F^{0j})^{a}n_{j}.
\ee
The transformation properties of the current under gauge transformations are cumbersome, but the charge transforms nicely if we impose some restrictions on the possible bahaviour of the gauge transformations at infinity \cite{Ramond:QFT}. Indeed, if $U$ is a gauge transformation which is constant at spatial infinity, we can bring it out of the surface integral, thus obtaining
\be\label{TRASF Q NON ABELIANO}
Q\to U^{\dagger}QU.
\ee
This transformation law characterizes elements belonging to the adjoint representation of the gauge group. So we have as many different charges as the number of generators of the Lie group.

Formula (\ref{GAUSS NON ABELIANO}) looks pretty much the same as Gauss' Law. Though in the case of Maxwell's theory, when no sources are present inside $\Sigma$, the flux of the electric field vanishes. In the case of Yang-Mills theories the field itself acts as a source for itself. This is evident from the non-linearity of the Yang-Mills equations (\ref{YANG E MILLS}). As a consequence of this the field has to carry a charge.

Despite the formal similarity with electrodynamics, the definition we provided for the non-abelian charge may not be entirely satisfactory from a physical point of view. In fact, according to (\ref{GAUSS NON ABELIANO}) and (\ref{TRASF Q NON ABELIANO}), the charge  is a gauge-dependent quantity, in contrast to the case of electrodynamics where it is a gauge-singlet. Here we follow \cite{AbbottDeser} to give an alternative definition of the non-abelian charge.
Let us consider a gauge connection $A_{\mu}$ and write it in terms of a background field $\bar{A}_{\mu}$, which we take to be a solution of the sourceless field equations, plus corrections
\be
A_{\mu}=\bar{A}_{\mu}+a_{\mu}.
\ee
If the sources are in a spatially bounded region, we may take $\bar{A}_{\mu}$  to have the same asymtpotic behaviour of $A_{\mu}$ at spacelike infinity. The term $a_{\mu}$ is a correction to the sourceless field due to the presence of sources and need not be small.
From the Yang-Mills equations with a source
\be
D_{\mu}F^{\mu\nu}=J^{\nu}
\ee
it follows that the current
\be
j^{\nu}=J^{\nu}-(D_{\mu}F^{\mu\nu})_{N}
\ee
is covariantly conserved with respect to the background gauge connection
\be\label{COVARIANT CONSERVATION LAW}
\bar{D}_{\nu}j^{\nu}\equiv D_{\nu}(\bar{A})j^{\nu}=0.
\ee
The expression $(D_{\mu}F^{\mu\nu})_{N}$ contains only terms in $D_{\mu}F^{\mu\nu}$ which are second or higher order in $a_{\mu}$.
The covariant conservation law (\ref{COVARIANT CONSERVATION LAW}) is not useful as it is if we want to construct a charge which is a gauge singlet. Thus we need to make use of symmetries of the background field $A_{\mu}$. We say that the background has symmetries if there exists a collection of covariantly conserved scalars (Killing vectors)
\be\label{KILLING NON ABELIANO}
\bar{D}_{\mu}\bar{\xi}^{s}=0.
\ee
Killing vector fields are the generators of gauge transformations which leave $A_{\mu}$ invariant. They are labeled by a discrete index $s$.
The quantity $\mbox{Tr}(\bar{\xi}^{s}j^{\nu})$ is a gauge singlet and is a conserved current, as it follows from the covariant conservation laws (\ref{COVARIANT CONSERVATION LAW}), (\ref{KILLING NON ABELIANO}).
The charges are defined in terms of the time component of the gauge singlet current
\be
Q_{E}^{s}=\frac{1}{4\pi}\int d^{3}x \;\mbox{Tr}(\bar{\xi}^{s} j^{0})=\frac{1}{4\pi}\int d^{3}x\; \partial_{i}\mbox{Tr}(\bar{\xi}^{s} f^{i0}),
\ee
where
\be
f_{\mu\nu}=\bar{D}_{\mu}a_{\nu}-\bar{D}_{\nu}a_{\mu}.
\ee
This formula is quite similar to the definition of the electric charge in electrodynamics.

It is also possible to define magnetic charges, in much the same way as we did before for electric charges, starting from the Bianchi identity
\be
D_{\mu}*F^{\mu\nu}=0.
\ee
We have the following conserved charge
\be
Q_{M}^{s}=\frac{1}{4\pi}\int d^{3}x \partial_{i}\mbox{Tr}(\bar{\xi}^{s} *f^{i0}).
\ee
In order to have a non-vanishing magnetic charge the field $a_{i}$ should have string-like singularities (as it happens for instance in the case of monopoles). In contrast to Noether charges, magnetic charges do not arise from  a symmetry of the action, but they have a purely topological meaning. According to this definition of electric and magnetic charges in the non-Abelian case, symmetry properties of the background are crucial.

There are specific cases where the first definition seems to be inappropriate and one should use the second one, which can also be generalized to the case of gravity.
\end{subsection}

\begin{subsection}{Asymptotic conditions}\label{ASYMPTOTIC CONDITIONS}
Not all the gauge connections are suitable for constructing a configuration space for Yang-Mills theories \cite{DeWitt}. We require that physical connections are such that the non-Abelian charge and the spatial integral of the Lagrangian are both finite. In the case in which the spacetime $M$ is homeomorphic to $\mathbb{R}^{4}$, we can  choose $\Sigma\cong\mathbb{R}^{3}$ and introduce polar coordinates on it. Thus we have
\be
F^{0j}\approx\frac{1}{r^{2+\varepsilon}}
\ee
from the finiteness of the charge, and
\be
F^{\mu\nu}\approx\frac{1}{r^{3/2+\varepsilon}}
\ee
from that of the space integral of the Lagrangian. The second condition is weaker than the first one, so the space-space components will decay more slowly than the time-space components.
From these we obtain the asymptotic behaviour for the components of the connection
\begin{align}
A_0\approx&\frac{1}{r^{1+\varepsilon}}\\
A_{j}\approx&\frac{1}{r^{1/2+\varepsilon}}.
\end{align}
When imposing these conditions on the physical connections, we also get as a bonus that the spatial integral of the time-time component of the stress-energy tensor, \emph{i.e.} the energy, is finite.
\end{subsection}

\begin{subsection}{A non-Yang-Mills gauge theory}
In this final section\footnote{According to De Witt's classification of gauge theories, Chern-Simons theory is a type I gauge theory. As in the Yang-Mills case, the infinite-dimensional gauge group is constructed starting from a finite-dimensional Lie group.} we present an example of a gauge theory with an action different from the Yang-Mills case \cite{Reshetikin:QFT}. In the case of a three-dimensional spacetime a natural choice for the action is given by the integral of the Chern-Simons form
\be
S_{CS}=\int_{M} Tr\left[A\wedge dA+\frac{2}{3}A\wedge\left(A\wedge A\right)\right].
\ee
To build this theory we do not need to introduce a metric on $M$. The theory is purely topological both on the classical and on the quantum level.
The first variation of the action under an infinitesimal gauge transformation is given by th formula
\be\label{GAUGE VAR CS} 
\delta S_{CS}(w)=\int_{\partial M} Tr(w (dA+A\wedge A)),
\ee
where $w$ is the parameter specifying the transformation
\be
\delta A = D(A)w.
\ee
We will say in a minute under which conditions the action is gauge invariant. In order to find the equation of motion, let us calculate the first variation of the action for an arbitrary variation of the gauge connection.
\be\label{CHERN SIMONS VAR}
\delta S_{CS}=\int_{M}Tr\left(\delta A \wedge (dA+A\wedge A)\right)+\int_{\partial M} Tr(A\wedge \delta A)
\ee
The equation of motion is the flatness condition 
\be
F(A)\equiv dA+A\wedge A=0.
\ee
The boundary term in (\ref{CHERN SIMONS VAR}) defines a canonical one-form $\Theta$ on the boundary $\partial M$. Its action on a vector field $\delta A$ tangent to the boundary is given by the formula
\be
\Theta(\delta A)=\int_{\partial M} Tr(A\wedge \delta A).
\ee
The kernel of this one-form defines a Lagrangian submanifold $L(\partial M)$. The most general boundary conditions one may choose to guarantee the stationarity of the action for a solution of the equation of motion is that the variation field at the boundary belongs to $L(\partial M)$. This also guarantees that the action is gauge invariant, \emph{i.e.} the RHS of equation (\ref{GAUGE VAR CS}) vanishes.

The Chern-Simons form may be used to make gauge fields acquire a mass without breaking gauge invariance\cite{Nakahara:GTP}\cite{Jackiw}\cite{Deser}. We can restrict our attention to the simplest case of the Maxwell field and consider the action
\be
S=-\frac{1}{4}\int_{M}Tr(F_{\mu\nu}F^{\mu\nu}))+m^{2}\int_{M}\varepsilon^{\mu\rho\sigma}(A_{\mu} \partial_{\rho} A_{\sigma}+\frac{2}{3}A_{\mu}A_{\rho}A_{\sigma}).
\ee
The Chern-Simons term is purely topological, while the Yang-Mills term needs the introduction of a metric. Then the theory will display dependence on local data. Though the presence of the topological term gives important modifications to the theory.

The equations of motion written in components are
\be
\partial_{\mu}F^{\mu\nu}+m^2 (*F)^{\mu}=0.
\ee
With a little bit of manipulations on this equation we get
\be
(\Box +m^2)(*F)^{\sigma}=0,
\ee
so that a massive gauge field is present in the theory.
\end{subsection}

\begin{subsection}{Principal Bundles}
Let us consider a differentiable manifold $M$, which we will call the base space, and a Lie group $G$. Let $P$ be another differentiable manifold and $\pi:P\rightarrow M$ an application called the projection. The list of objects $(P,M,G,\pi)$ is defined to be a \emph{principal bundle} with structure group $G$ if the following properties holds:
\\
(\emph{a}) $P$ is locally diffeomorphic\footnote{Some authors prefer to say that $P$ is locally diffeomorphic to $M\times \tilde{G}$, where $\tilde{G}$ is a Lie group isomorphic to $G$} to $M\times G$. Given an open covering $\{U_{i}\}$ of $M$, for each $U_{i}$ there exists a map
\be
\phi:\pi^{-1}(U_{i})\rightarrow U_{i}\times G,
\ee
which is a diffeomorphism and satisfies the property
\be
\pi\phi^{-1}(p,f)=p.
\ee
Here $p$ and $f$ denote respectively a point in $M$ and in $G$. The map $\phi$ is called the \emph{local trivialization}. Clearly the \emph{typical fibre} at $p$, defined as $\pi^{-1}(p)$, is diffeomorphic to $G$.\\
\\
(\emph{b}) For each pair $U_{i}$, $U_{j}$ so that $U_{i}\cap U_{j}\neq\varnothing$ there is an application $u_{ij}(p)$ relating the two corresponding trivializations.
\be
f_{i}=u_{ij}(p)f_{j}
\ee
It is called the \emph{transition function}. For a fixed $p$ it is a map from $G$ to itself, and we require that it belongs to the group $G$ too. 
It has to verify the following properties $\forall p$ ($e$ denotes the identity in $G$):
\begin{align}\label{PROPRIETÀ TRANSITION FUNCTION}
u_{ii}(p)=&e\\
u_{ij}(p)u_{ji}(p)=&e\\
u_{ij}(p)u_{jk}(p)=&u_{ik}(p)
\end{align}
If a set of local trivializations $\{\phi_{i}\}$ is given, the transition functions may be expressed by the formula
\be
u_{ij}(p)=\phi_{i}^{-1}(p)\phi_{j}(p).
\ee
Written in this form it is clear that it satisfies properties (\ref{PROPRIETÀ TRANSITION FUNCTION}). Given a different set of local trivializations $\{\tilde{\phi}_{i}\}$ we have
\begin{align}
u_{ij}(p)=&\phi^{-1}_{i}(p)\phi_{j}(p)\\
\tilde{u}_{ij}=&\tilde{\phi}^{-1}_{i}(p)\tilde{\phi}_{j}(p)
\end{align}
We define the functions
\be
\tau_{i}(p)=\phi^{-1}_{i}(p)\tilde{\phi}_{j}(p)
\ee
and require that they belong to the group $G$. These can be used to relate the transition functions corresponding to different local trivializations
\be\label{DIFF LOCAL TRIV}
\tilde{u}_{ij}=\tilde{\phi}^{-1}_{i}\tilde{\phi}_{j}=\tilde{\phi}^{-1}_{i}(\phi_{i}\phi_{i}^{-1})(\phi_{j}\phi_{j}^{-1})\tilde{\phi}_{j}=\tau^{-1}_{i}u_{ij}\tau_{j}.
\ee
We can give a physical interpretation to this machinery, regarding the transition functions $u_{ij}$ as the gauge transformations required for pasting two local charts together and the $\tau_{i}$ functions as those corresponding to the gauge degrees of freedom within a given chart.

How can we recognize a trivial bundle given its transition functions? We defined a trivial bundle as a principal bundle globally diffeomorphic to the product $M\times G$. Then it is possible to choose a set of local trivializations such that $\phi_{i}=\phi_{j}$ in the overlap between two charts $U_{i}\cap U_{j}$. Then it follows that all the transition functions are trivial
\be
u_{ij}=e.
\ee
As a consequence of formula (\ref{DIFF LOCAL TRIV}), if we consider a different set $\{\tilde{\phi}_{i}\}$, we  have in this case
\be\label{TRANS FUNCTION TRIVIAL}
\tilde{u}_{ij}=\tau_{i}^{-1}\tau_{j}.
\ee
The converse can be shown by the same argument. Thus a bundle is trivial if and only the transition functions have the form (\ref{TRANS FUNCTION TRIVIAL})

\begin{center}
%\centering
\includegraphics[width=0.6\columnwidth]{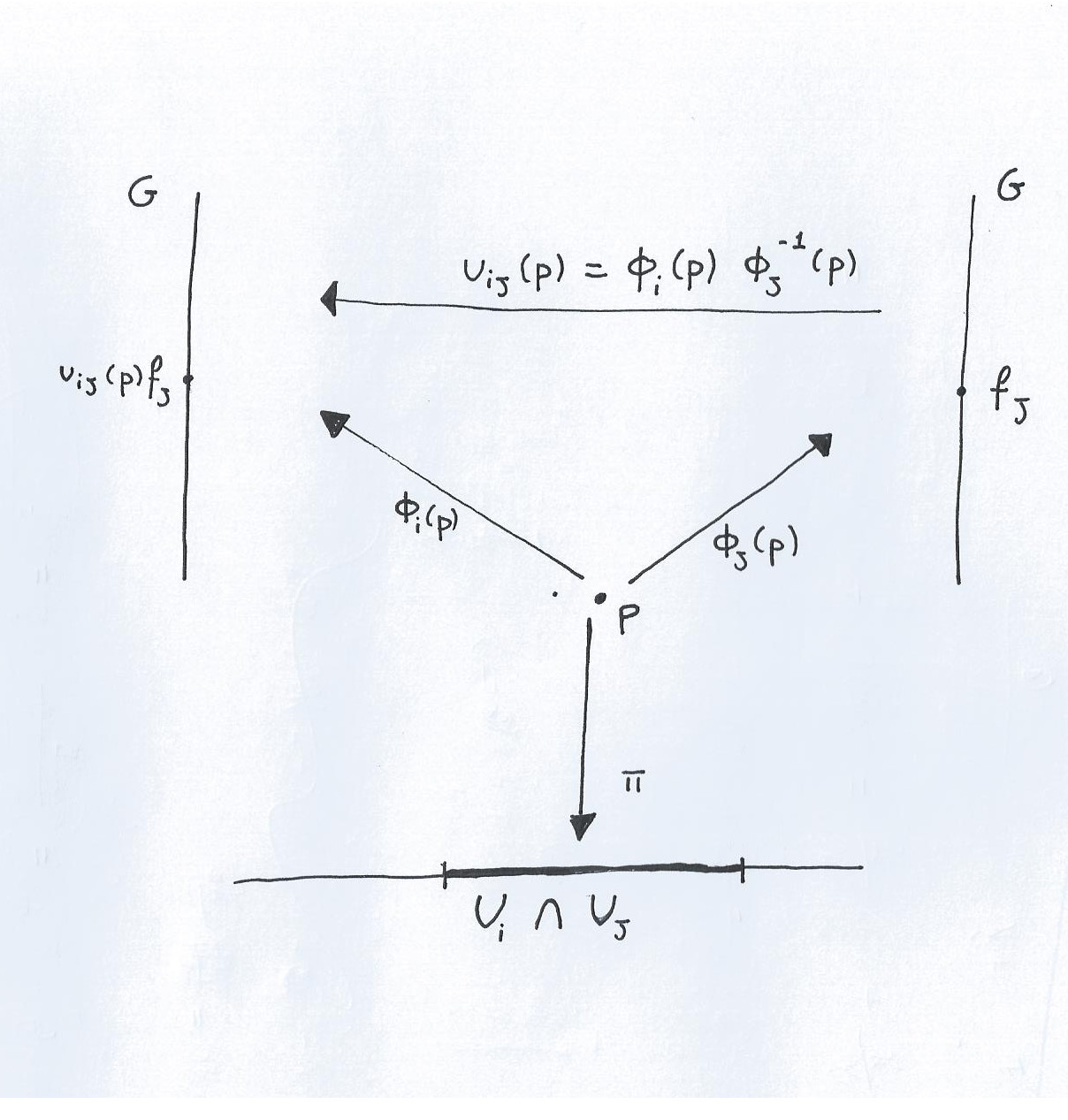}
%\captionof{figure}{From Gribov's paper \cite{Gribov:78}}
\end{center}
More often a principal bundle is denoted simply by $P \xrightarrow{\pi} M$.
The geometrical setting of principal fibre bundles allows us to express in precise terms what we mean by assigning smoothly a \emph{phase} on $M$. In our previous discussion we assigned a phase to \emph{each point} of euclidean space, but in this more general setting it is more prudent to talk about the assignment of a smoothly varying phase on an open set of $M$.\\
We define an application $s:U\subseteq M\rightarrow P$ to be a section if $\pi s(p)=p$. When $U=M$ it is called a \emph{global section}. This is a generalization of the concept of functions defined on a open set in $M$ with values in $G$.\\
On principal bundles it is also possible to define the right action of the Lie group $G$ on the fibres. Given $a\in G$ and a point $w\in P$, the right action is denoted by $R_{a}w=wa$. To make sense of this we must refer to a local trivialization, $\phi(w)=((\pi(w),f)$. The right action of a Lie group on itself is well defined so we can use it in the definition of the right action $R_{a}w=\phi^{-1}(\pi(w),fa)$. This definition is actually independent of the particular trivialization used. Indeed we have
\be
R_{a}w=\phi^{-1}_{i}(\pi(w),f_{i}a)=\phi^{-1}_{j}\phi_{j}\phi^{-1}_{i}(\pi(w),f_{i}a)=\phi^{-1}_{j}(\pi(w),u_{ij}^{-1}f_{i}a)
\ee
A fibre bundle is said to be trivial when the manifold $P$ is diffeomorphic to the direct product $M\times G$. An important theorem holds which states that a principal bundle is trivial if and only if it admits a global section. The proof of the ``only if'' statement is trivial, as a section in this case is simply a $G$-valued function on $M$. Then we just have to prove the converse. Let $s(p)$ be a section on $P$, where $p$ is a point on $M$. We can use the right action to reach $s(p)a$. Given a point $g$ along the fibre $\pi^{-1}(p)$ at $p$, there exists an $a$ such that for any $w$, $\pi(w)=p$, $R_{a}s(p)=w$ because the right action of a Lie group is \emph{transitive}. Moreover this $a$ is unique because the right action is also \emph{free}. We can then define a global trivialization $\phi:P\rightarrow M\times G$ such that $\phi(s(p)a)=(p,a)$.

%%definisci le sezioni
%%
%%cos'è un fibrato banale
%%
%%teoremino sui fibrati principali banali
%%
%%azione destra di un gruppo di Lie su un fibrato
%%
%%
%%
%%
%%
%%
%%
%%
%%
%%
%%
%%
%%
%%
%%

\end{subsection}
\begin{subsection}{Ehreshmann connection}
On a non-trivial principal bundle there's no way to globally separate the spatial and Lie group degrees of freedom. The concept of connection is necessary to provide us with a local separation in this sense.\\
%%It also allow us to define a notion of parallel transport.\\
At each point $u$ of $P$ the tangent space $T_{u}P$ has a naturally defined subspace called the \emph{vertical subspace}. Let us take an element $A\in\mathfrak{g}$ the Lie algebra of $G$, and consider the group element $\exp{tA}$. By the right action of this object on $u$ we have
\be
R_{\exp{tA}}=u \exp{tA}.
\ee
As $t$ varies we move along a curve which lies in the fibre at $p=\pi(u)$. Indeed $\pi(R_{\exp{tA}})=\pi(u)=p$ as we can immediately see in a local trivialization. The tangent vector $A^{\#}$ to that curve will be called the \emph{fundamental vector field} generated by $A$.
\be
A^{\#}f(u)=\frac{d}{dt}f(u \exp{tA})|_{t=0}
\ee
Thus the set of fundamental vector fields at a point is isomorphic to the Lie-algebra $\mathfrak{g}$.

We can think of a particle with an internal structure described by a $G$ index, and associate to each point in space a vertical direction 
corresponding to its internal degrees of freedom.
\begin{center}
%\centering
\includegraphics[width=0.3\columnwidth]{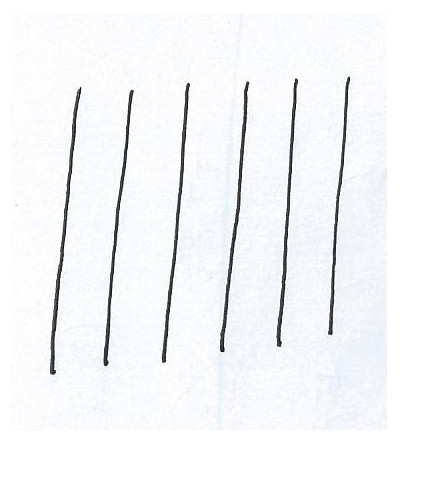}
\captionof{figure}{}
\end{center}
Though, this is not the end of the story, as we need to link vertical lines passing through different points in space.
\begin{center}\label{atiyah}
%\centering
\includegraphics[width=0.5\columnwidth]{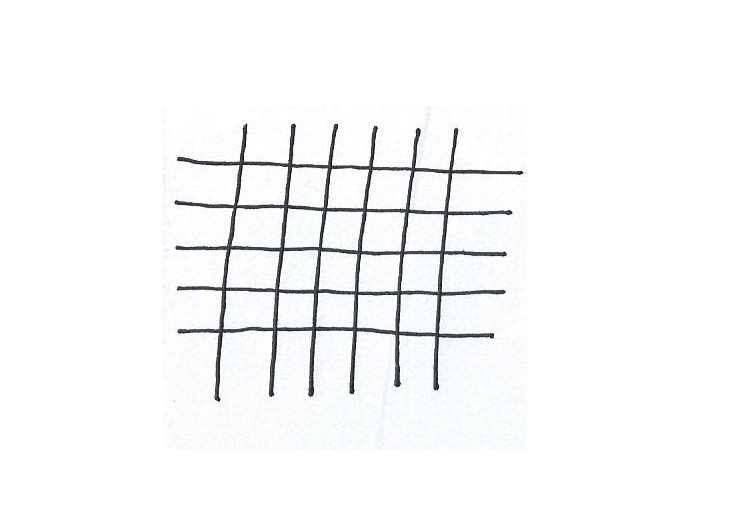}
\captionof{figure}{}
\end{center}
The picture represents a particular situation that may be given, where fibres are obtained from each other by a simple horizontal translation. Globally this is not the most general situation, but locally we can always reduce to this case. 
We define a connection on the bundle by specifying a smooth assignment of an horizontal subspace at each point.
\be
T_{u}P=H_{u}P\oplus V_{u}P
\ee
So any smooth vector field $X$ on $P$ is decomposed in a unique way into a sum of an horizontal and a vertical vector field.
\be
X=X_{H}+X_{V}
\ee
Evaluating $X$ at the point $u$, we have that $X_{H}\in H_{u}P$ and $X_{V}\in V_{u}P$. The assignment is smooth if and only if $X_{H}$ and $X_{V}$ are smooth for any $X$.\\
Horizontal subspaces are related by right translation.
\be
H_{ug}P=R_{g*}H_{u}P
\ee
The meaning of this property is that horizontal subspaces pertaining to points on the same fibre are linearly related. Right translation allows us to move from one subspace to the others. This is the way to express gauge transformations in this mathematical framework. It follows that locally the situation may be represented not only as in figure \ref{atiyah} but also equivalently by the following picture:
\begin{center}\label{atiyah3}
%\centering
\includegraphics[width=0.3\columnwidth]{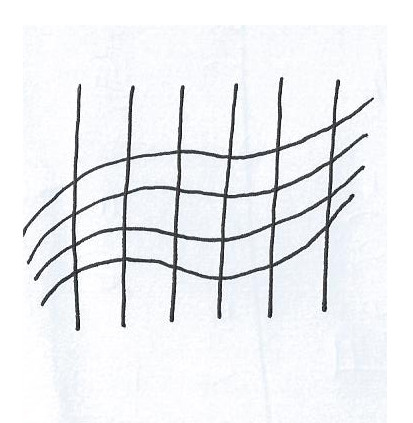}
\captionof{figure}{}
\end{center}
In this case fibres are identified coherently but in a different way. No privileged choice exists and each one can be obtained from a given one by means of a gauge transformation. We would like to stress that though pictures can be useful to help our intuition, they have only a local significance. On the other hand the mathematical definition of the connection is global.

We now want to introduce a geometric object which contains all the information about a connection and can render the concept operative. This is a Lie-algebra-valued one form, which we denote by $\omega$. In other words it is a section of $\mathfrak{g}\otimes T^{*}P$. $\omega$ is referred to as to as the \emph{connection one-form}. It satisfies the following properties:\\
\emph{(a)} when acting on the fundamental vector field generated by $A$ it returns $A$
\be
\omega(A^{\#})=A;
\ee
\emph{(b)} the right action of $G$ on the bundle explicates on $\omega$ as an adjoint action
\be
R_{g}^{*}\omega=Ad_{g^{-1}}\omega,
\ee
that is
\be
R_{g}^{*}\omega_{ug}(X)=\omega_{ug}(R_{g*}X)=g^{-1}\omega_{u}(X)g.
\ee
We can alternatively define the horizontal subspace at a point $u$ as the kernel of the one-form $\omega$
\be
H_{u}P\equiv\{X\in T_{u}P|\omega(X)=0\}.
\ee
It could be proved that the two definitions are equivalent.
%%We now prove that if and only if a vector field is in the kernel of $\omega$ it satisfies the previously seen property (b) for an horizonal vector field. We start by proving the \emph{if} clause.
%%Let $X$ be in the kernel of $\omega$ and $g$ a group element. Then by the transformation property of $\omega$ under transport along a fiber we have
%%\be
%%\omega(R_{g*})X)=R_{g}^{*}\omega(X)=g^{-1}\omega(X)g=0.
%%\ee
%%Then from $X_{u}\in H_{u}P$ it follows $R_{g*}X_{u}\in H_{ug}P$.
%%Let us now turn to the \emph{only if} sentence.
%%Using property (b) we have
%%\be
%%\omega_{ug}(X_{ug})=\omega_{ug}(R_{g*})X_{u})=R_{g}^{*}\omega_{u}(X_{u})=g^{-1}\omega_{u}(X_{u}),
%%\ee
%%and this should be valid for any $g$.

%%
%%definizione di connessione su un fibrato
%%
%%picture di Atyiah
%%
%%proprietà varie di una connessione e definizione di trasporto parallelo
%%
%%curvatura di una connessione
%%
%%potenziali e campi di gauge
%%
%%
%%
%%
%%
%%
%%
%%
%%
%%

\end{subsection}

%%
%%azione di yang e mills
%%
%%equazioni del moto per yang e mills
%%
%%self-dual e anti-self-dual solutions
%%
%%azioni gauge invarianti di altro tipo(chern simons)
%%
%%
%%
%%
%%
%%
%%
%%
%%
%%

\end{section}
\begin{section}{The Quantum Theory}
%%
%%definizione dello spazio delle storie e introduzione della notazione di de witt
%%
\begin{subsection}{Space of Histories}
We now turn to the study of the quantum dynamics of the Yang-Mills field. In order to do that it is important in first place to define the configuration space of our system.
The configuration space for a field theory is the space whose points are fields on spacetime. This is an infinite-dimensional space called the \emph{space of histories}, denoted by $\mathcal{U}$, whose topology and geometric structure may be quite complicated \cite{DeWitt}. We can introduce local charts on this space in order to make contact with the local expressions for the fields. In De Witt's notation we can express a point in the space of histories as $\phi^{i}$. Here $i$ is an index with a discrete as well as a continuous part, namely a spacetime point. In this way we can imagine to represent a field configuration by an infinite (actually continuous infinite) sequence, containing the values it assumes at each point of space-time.

Once we have built the configuration space we can define an action functional\footnote{In the case of fermions defining the space of histories is not enough. In order to construct a Lagrangian formulation one should also introduce an adjoint spinor field and its own space of histories.} on it $S[\phi]|\;\mathcal{U}\rightarrow\mathbb{R}$. The equations of motion are given by
\be
\frac{\delta S[\phi]}{\delta \phi(x)}=0,
\ee
which can be re-expressed in De Witt notation
\be
_{i,}S[\phi]=0,
\ee
where comma denotes functional differentiation. The derivative here is taken from the left. In the case of a bosonic field this actually makes no difference, but it matters when fermions are involved. These equations describe the classical dynamics of the system. We can use the action to define the generating functional
\be\label{PARTITION FUNCTION}
Z[J]=\int\mathcal{D}\phi e^{iS[\phi]+iJ_{i}\phi^{i}},
\ee
where $J$ denotes a fictitious field coupled to $\phi$. A generalized Einstein convention applies to repeated indices: we must sum over the discrete part of the index $i$ and integrate over the continuous part.
\be
J_{i}\phi^{i}=\int J(x)^{a}\phi(x)^{a}
\ee
Actually the integral in (\ref{PARTITION FUNCTION}) is not a properly defined integral, as it cannot be associated with any measure function. It is just defined on a lattice in spacetime, and the continuous limit is recovered by letting the lattice spacing go to zero \cite{Ryder:QFT} \cite{Feynman} \cite{Reshetikin:QFT}. 
The generating functional can be interpreted as the vacuum-to-vacuum transition amplitude in the presence of the source $J$.
\be
Z[J]=\langle \mbox{out}|\mbox{in} \rangle_{J}
\ee
The kets $|\mbox{in}\rangle$ and $|\mbox{out}\rangle$ are asymptotic states of zero energy obtained from imposing suitable boundary conditions on the field $\phi$ at time $t=-\infty$ and $t=+\infty$. Usually Feynman boundary conditions are imposed, which amount to the requirement that at $-\infty$ only positive frequency waves should be present in the Fourier representation of $\phi$, while negative frequencies only are allowed at $+\infty$. We can also choose not to fix the boundary conditions from the beginning. In this more general case we can consider a set of boundary conditions labeled by two parameters, say $\alpha$, $\beta$. The correct formula for the partition function would then be given by further integrating the RHS of (\ref{PARTITION FUNCTION}) over these two parameters \cite{DeWitt}
\be\label{funzionale Z + condizioni al contorno}
Z[J]=\int d\alpha\int d\beta\int\mathcal{D}\phi e^{iS[\phi;\alpha,\beta]+iJ_{i}\phi^{i}}.
\ee
We can calculate all the correlation functions in terms of functional derivatives of $Z[J]$ with respect to the value of the source $J$ at a given spacetime point
\be
\langle\phi^{i_{1}}\phi^{i_{2}}\dots\phi^{i_{n}}\rangle=i^{-n}Z[J]^{,i_{1}i_{2}\dots i_{n}}.
\ee
These are used to construct transition amplitudes, according to the LSZ theorem. In the case of an interacting theory the correlation functions obtained from perturbation theory are often divergent. The divergences need to be cured by an appropriate regularization procedure, \emph{e.g.} dimensional regularization or zeta function regularization. 
\end{subsection}
\begin{subsection}{Gauge theories (three different ones)}
The concept of gauge invariance is more general than the one we used in constructing Yang-Mills theories. In that case we constructed the gauge group starting from a finite-dimensional Lie group. This is too restrictive to be of any use for a general definition. For example we can consider the Einstein-Hilbert action for general relativity. It is invariant under the group of diffeomorphisms on spacetime, which is of course a group of \emph{local} transformations, but it does not arise from a finite-dimensional group. Such a definition would then rule out general relativity, besides the deep similarity with Yang-Mills' theory. We then have to be more careful and define a theory to be gauge invariant if the action is invariant under some suitable flows, following \cite{DeWitt}.  We denote the vector fields which generate these flows by $\bold{Q}_{\alpha}$, where $\alpha$ is an index with a discrete as well as a continuous part. These vector fields are such that they annihilate the action:
\be
Q_{\alpha}^{i}\, {_{i,}S}=0.
\ee
We define a (super)Lie-bracket on $\Phi$ by the formula
\be
[\bold{Q}_{\alpha},\bold{Q}_{\beta}]=\bold{Q}_{\alpha}\bold{Q}_{\beta}-(-1)^{\alpha\beta}\bold{Q}_{\beta}\bold{Q}_{\alpha}.
\ee
The value of an index at the exponent of $(-1)$ is $0$ or $1$ respectively in the case of a boson or a fermion. Thus the bracket is a commutator in the case of two bosons or a boson and a fermion, and it is an anticommutator in the case of two fermions. The concept of supersymmetry is necessary if we want to discuss gauge theories in full generality. We require the bracket of two flows generators to be also the generator of a flow
\be
[\bold{Q}_{\alpha},\bold{Q}_{\beta}]S=0.
\ee
For the sake of simplicity from now on we will call flow what is actually the generator of a flow. It is important to observe that for all field theories we can construct vector fields leaving the action invariant. For example
\be
V^{i}=S_{,j}\, {^{j} T^{i}}
\ee
where $T$ is any anti-supersymmetric tensor field
\be
^{i} T^{j}=-(-1)^{ij+T(i+j)}\, {^{j} T^{i}}.
\ee
These fields vanish on the dynamical shell, hence they are not true flows. They are called skew fields. We assume that any true flow can be expressed as a linear combination of the fields $\bold{Q}_{\alpha}$ and of the skew fields.

The fields $\bold{Q}_{\alpha}$ form a complete set of flows modulo skew fields.
\be
[\bold{Q}_{\alpha},\bold{Q}_{\beta}]=\bold{Q}_{\gamma}c^{\gamma}_{\alpha\beta}+S_{1}\bold{T}_{\alpha\beta}
\ee
The coefficients $c^{\gamma}_{\alpha\beta}$ on the RHS are anti-supersymmetric
\be
c^{\gamma}_{\alpha\beta}=-(-1)^{\alpha\beta}c^{\gamma}_{\beta\alpha}.
\ee
$T$ is a set of skew fields anti-supersymmetric with respect to the indices $\alpha$ and $\beta$
\be
^{i}T^{j}_{\alpha\beta}=-(-1)^{\alpha\beta}\, {^{j} T^{i}_{\alpha\beta}}.
\ee
The flows satisfy a supersymmetric version of the Jacobi identity
\be
[\bold{Q}_{\alpha},[\bold{Q}_{\beta},\bold{Q}_{\gamma}]]\varepsilon^{\gamma\beta\alpha}=0,
\ee
where
\be
\varepsilon^{\alpha\beta\gamma}=-(-1)^{\alpha\beta}\varepsilon^{\beta\alpha\gamma}=-(-1)^{\beta\gamma}\varepsilon^{\alpha\gamma\beta}.
\ee
The Lie bracket of two skew fields is skew, in the same way as the Lie bracket of a skew field and a $\bold{Q}_{\alpha}$. Thus what is really fundamental to know are the commutators of the flows $\bold{Q}_{\alpha}$. Vector fields of the form
\be
^{i}Q_{\alpha}\xi^{\alpha}+{^{i}T^{j}} _{j,}S
\ee
close an algebra. When true flows exist this is called the gauge algebra.
The classification of gauge theories is based on the possible form that the commutator of the fields $\bold{Q}_{\alpha}$ may assume. Three different possibilities are given.

\textbf{CASE I} It is possible to find a complete set of flows $\bold{Q}_{\alpha}$ such that they close the gauge algebra
\be\label{caso 1}
[\bold{Q}_{\alpha},\bold{Q}_{\beta}]=\bold{Q}_{\gamma}c^{\gamma}_{\alpha\beta}.
\ee
The coefficients $c^{\gamma}_{\alpha\beta}$ are constant and are called structure constants. Yang-Mills' theories and general relativity are particular cases of this class of theories.

The proper gauge group is obtained by taking the exponential of the generators. The full gauge group is obtained by adding to the gauge group all the transformations of $\Phi$ into itself which do not depend on $\phi$, leave the action invariant and do not arise from global symmetries. This is important in the case of general relativity, where the gauge group is not obtained from a finite-dimensional group, and there are transformations close to the identity which cannot be obtained by exponentiation of the the generators\footnote{This statement constitutes the content of a theorem due to Freifeld.}.

The closure property (\ref{caso 1}) implies that the gauge group decomposes $\Phi$ into subspaces to which the flows $\bold{Q}_{\alpha}$ are tangent. These are by definition the gauge orbits and are isomporphic to the gauge group.

\textbf{CASE II} It is still possible to find a complete set of flows $\bold{Q}_{\alpha}$ such that they close the gauge algebra, but the coefficients in (\ref{caso 1}) are no longer constant. The $c^{\gamma}_{\alpha\beta}$ are now called structure functions. As a consequence the orbits are no longer isomorphic to the gauge group.

\textbf{CASE III} It is not possible to close the gauge algebra with skew fields vanishing everywhere on $\Phi$. The flows $\bold{Q}_{\alpha}$ close an algebra by themselves but only on-shell \emph{i.e.}, the commutator of two such fields contains also terms proportional to the equations of motion. It is just the dynamical shell that happens to be decomposed into orbits. Although physics takes place in the space of orbits as usual, the dynamics cannot be obtained from a functional on this space: the whole of $\Phi$ is needed.
It is shown in \cite{DeWitt} that there is an infinite hierarchy of structure functions.
\end{subsection}
\begin{subsection}{FDFP Method}
The space of histories is no longer a good candidate as a configuration space when we consider gauge theories. That's because a gauge theory is basically a theory with redundancies. In order to properly deal with them we have to divide the space of histories by the group of local gauge transformations. The space obtained in this way is called the space of orbits $\mathcal{R}\equiv\mathcal{U}/\mathcal{G}$. This may consist of more disconnected components, depending on topological properties of the fields (gauge connections) such as the winding number. In doing perturbation theory though, only the component of the orbit space to which the zero field (the perturbative vacuum) belongs need to be considered, so that when we talk about the space of orbits we will usually mean this subset. We see that the space of histories is actually a fibre bundle with base space $\mathcal{R}$ and structure group $\mathcal{U}$. Indeed there exists a natural projection $\pi:\mathcal{U}\rightarrow\mathcal{R}$, which associates to 
each field the gauge-orbit to which it belongs. Moreover it is possible to define local trivializations and use gauge transformations as structure functions, so that the defining properties of a fibre bundle are satisfied.

It is a consequence of what we said that for quantum gauge theories the integral appearing in formula \ref{PARTITION FUNCTION} should actually be performed on $\mathcal{R}$. Though this is an intractable space, so we have to find a way to circumvent the problem. A possible solution would be to fix the gauge. A gauge fixing condition is expressed by an equation of the type
\be
\mathcal{G}(A)=0,
\ee
where $\mathcal{G}(A)$ is a functional on the space of histories. The gauge field is denoted by $A$ as usual. An example of a gauge fixing condition is provided by the Lorenz gauge
\be
\partial^{\mu}A_{\mu}=0.
\ee
The importance of this condition lies in its linearity and covariance. Other local gauge fixing conditions are commonly used, like the axial gauge, the temporal gauge and the Coulomb gauge. Each of these choiches has its merits but also its drawbacks: in the Coulomb gauge for the electromagnetic field only the physical degrees of freedom survive but it is not covariant; the axial gauge is useful in non-Abelian theories for some reason which will become clear in the following, but it is affected by some technical pathology. The Coulomb and Lorenz gauges are also called transverse gauges because they amount to the requirement that the longitudinal component of the field equal zero. One would like to use the gauge fixing condition to select one and only one representative along each gauge orbit. This is the idea underlying the method which was developed by Feynman, De Witt, Faddeev and Popov to deal with gauge theories. It is useful to define the quantity
\be
\Delta^{-1}_{\mathcal{G}}[A]=\int\mathcal{D}U\delta[\mathcal{G}(A^{U})].
\ee
$\mathcal{D}U$ is an invariant measure on the group $\mathcal{G}$, so that if we have
\be
U''=U'U
\ee
then
\be
\mathcal{D}U''=\mathcal{D}U.
\ee
We can then prove that also $\Delta^{-1}_{\mathcal{G}}[A]$ is gauge invariant. We have
\be
\Delta^{-1}_{\mathcal{G}}[A^{U'}]=\int\mathcal{D}U\delta[\mathcal{G}(A^{U'U})]=\int\mathcal{D}U''\delta[\mathcal{G}(A^{U''})]=\Delta^{-1}_{\mathcal{G}}[A].
\ee
We can then write
\be
\Delta_{\mathcal{G}}[A]\int\mathcal{D}U\delta[\mathcal{G}(A^{U})]=1,
\ee
where both factors on the LHS are gauge invariant. We can write down naively the generating functional formula as an integral over $\mathcal{U}$, then insert this expression for one in it
\be
\int\mathcal{D}A e^{iS[A]}=\int\mathcal{D}A \Delta_{\mathcal{G}}[A]\int\mathcal{D}U\delta[\mathcal{G}(A^{U})] e^{iS[A]}.
\ee
Performing a gauge transformation we go from $A^{U}$ to $A$ without affecting the action because it is gauge invariant
\begin{multline}\label{FDFP}
\int\mathcal{D}A e^{iS[A]}=\int\mathcal{D}A \Delta_{\mathcal{G}}[A]\int\mathcal{D}U\delta[\mathcal{G}(A)] e^{iS[A]}=\\
=\int\mathcal{D}U\int\mathcal{D}A \Delta_{\mathcal{G}}[A]\delta[\mathcal{G}(A)] e^{iS[A]}.
\end{multline}
The integrand does not depend on the group element $U$, so the integration over $U$ just gives a multiplicative constant equal to the volume of the gauge group $\mathcal{G}$. Actually this constant is infinite but this is not a real problem. If we want to be more rigorous we'd better do the same machinery on a lattice, where the gauge group is finite-dimensional, so we are allowed to throw away its volume and only then go to the continuum limit \cite{Reshetikin:QFT}. The result we obtain in either case is just the same. To complete our discussion we have to find a manageable expression for the term $\Delta_{\mathcal{G}}[A]$. We assume that the change of variables from $U$ to $\mathcal{G}$ is nowhere singular, so that we can write
\be
\Delta^{-1}_{\mathcal{G}}[A]=\int\mathcal{D}\mathcal{G}\mbox{det}\frac{\delta U}{\delta\mathcal{G}}\delta[\mathcal{G}(A^{U})]=
\frac{\delta U}{\delta\mathcal{G}}\lvert_{\mathcal{G}=0}.
\ee
The desired expression is
\be\label{DELTA FDFP}
\Delta_{\mathcal{G}}[A]=\mbox{det}\frac{\delta\mathcal{G}}{\delta U}\lvert_{\mathcal{G}=0}.
\ee
We want to stress that formula (\ref{FDFP}) was obtained starting from the assumption that each $U$ corresponds to one and only one $\mathcal{G}$. This is the key point which will lead us to the non-perturbative modifications presented in the next chapters. In fact gauge orbits may intersect the gauge fixing surface more than once or never. The picture represents possible situations that may be given in the case of the Lorenz gauge. Transverse and vertical components of the gauge connection are plotted respectively on the horizontal and the vertical axis.
\begin{center}
%\centering
\includegraphics[width=\columnwidth]{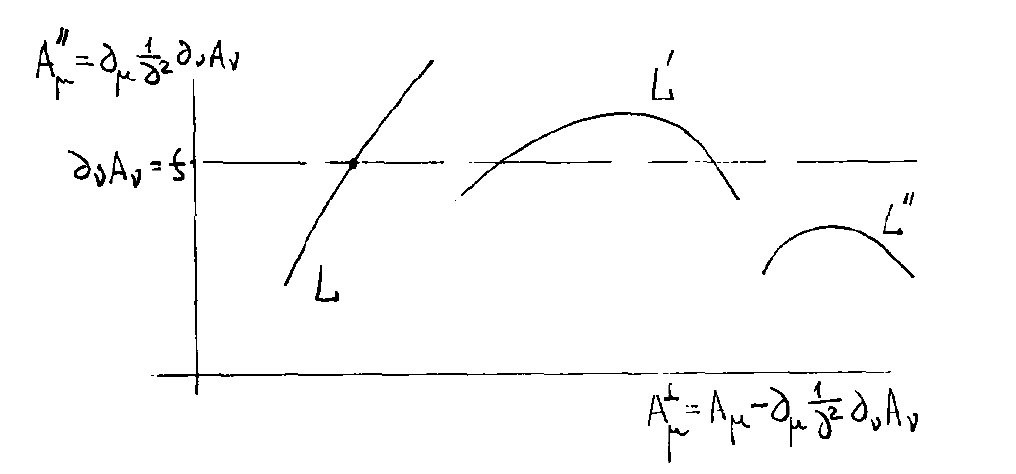}
\captionof{figure}{From Gribov's paper \cite{Gribov:78}}
\end{center}

When we limit our considerations to perturbation theory we can simply ignore the problem and assume that this one-to-one correspondence holds. 

In the use of the FDFP the gauge fixing term is usually smeared out. If we take the gauge condition to be linear, we can simply substitute in the above formulae $e^{i\frac{(\mathcal{G})^2}{2\alpha}}$ for the delta function. This is also accomplished by choosing gauge conditions $\mathcal{G}-f=0$ and performing a gaussian average over all the possible choices of $f$. Usually different choices of the parameter $\alpha$ are said to correspond to different ``gauges''. Though this term may be misleading as the gauge fixing functional is $\mathcal{G}$ and the freedom in the choice of this parameter must not be confused with the gauge freedom, that we have already lost. The parameter $\alpha$ corresponds instead to an ultra-local choice for the matrix used in the gaussian average \cite{DeWitt}\cite{DeWitt1981}. Common choices are the Feynman gauge $\alpha=1$ and the Landau gauge $\alpha=0$. The delta function is recovered taking the limit $\alpha\to 0$ at the end of the calculations. Some correlation functions, \
emph{e.g} the two-point function, 
depend on the choice of the gauge parameter, while the amplitudes for physical processes are obviously $\alpha$-independent.  

The expression $\frac{\delta\mathcal{G}}{\delta U}\lvert_{\mathcal{G}=0}$ defines the Faddeev-Popov operator, denoted by $FP(A)$. In the case of the Lorenz gauge the variation of the gauge condition under an infinitesimal gauge transformations reads
\be
\delta \partial^{\mu}A_{\mu}=\partial^{\mu}\delta A_{\mu}=\partial^{\mu}\left(\partial_{\mu}+[A_{\mu},\quad]\right)w.
\ee
So the Faddeev-Popov operator is given by the formula
\be\label{FP LORENTZ}
FP(A)_{Lorenz}=\partial^{\mu}\left(\partial_{\mu}+[A_{\mu},\quad]\right).
\ee
In the case of the axial gauge $\mathcal{G}(A)=A_3$, we have
\be\label{FP AXIAL}
FP(A)_{axial}=\partial_{z}.
\ee
It is evident that in general it is an operator acting on fields in the adjoint representation of the Lie Algebra of the finite dimensional Lie Group associated to $\mathcal{G}$.
\end{subsection}
\begin{subsection}{Ghost fields}
We can express the determinant in formula (\ref{DELTA FDFP}) as a functional integral over Grassmann variables. Disregarding unessential constant factors, which may be included in the normalization of the measure, we have
\be
\mbox{det}\frac{\delta\mathcal{G}}{\delta U}=\int\mathcal{D}\omega\mathcal{D}\bar{\omega}e^{i\bar{\omega}\frac{\delta\mathcal{G}}{\delta U}\omega}.
\ee
The fields $\omega$, $\bar{\omega}$ are called respectively ghost and antighost fields. The name should not be misleading as antighosts are not the complex conjugates of ghost fields, they are actually independent. A new quantum number may be introduced, being $+1$ for ghosts, $-1$ for antighosts, $0$ in all other cases. The action and the physical observables must have vanishing ghost number, as it will be clear from the discussion of BRST symmetry.

Ghost fields are fermionic and have Lie algebra indices. Though they do not obey the spin-statistics connection, as the operator $FP(A)$ which accounts for their dynamics is not a Dirac operator as can be seen in formulae (\ref{FP LORENTZ}) and (\ref{FP AXIAL}). Ghost and antighost fields seem to be unphysical for this reason, so it would be nice to have a strong argument to separate the ghosts from physical states. This will be provided by BRST as we will see in the following. Nevertheless internal ghost lines contribute to the scattering amplitude of physical processes. The axial gauge is in this sense a good choice because in that case the ghosts are not coupled to the gauge field, so they do not contibute in the calculation of amplitudes. Though useful for this reason, its use is at least questionable because of a technical observation made by De Witt \cite{DeWitt}. We start from a gauge connection $A$ and look for another gauge-equivalent connection which satisfies the axial gauge condition:
\be
U^{-1}A_{3}U+U^{-1}\partial_{3}U=A'_{3}=0.
\ee
This equation can be recast in the form
\be
\partial_{3}U=-A_{3}U.
\ee
Then it is evident that in general, even if the field $A$ satisfies the asymptotic conditions discussed in (\ref{ASYMPTOTIC CONDITIONS}), there can be no solution to this equation satisfying the boundary condition
\be
U\to\mathbb{I},
\ee
as $x\to\infty$ in space.

Ghost fields were first introduced by Feynman, though not in this terms, as a way to preserve the unitarity of the $S$ matrix in quantum gravitation                                                                            at one-loop level \cite{Feynman}. Summing over all one-loop diagrams which describe a given process is not enough. A new fictitious vector particle has to be introduced in the theory, and diagrams with one such closed loop have to be included (with a minus sign) in the computation of the amplitude. He observed that the same lack of unitarity manifests itself when one tries to quantize Yang-Mills theories. Adding a mass to the gauge field, gauge invarinace is lost but unitarity is recovered, so it is easier to guess the modifications needed in order to make sense of the massless limit. The correct rules for quantizing Yang-Mills theories were discovered while attempting to quantize gravitation, thanks to the striking similarity between Einstein's theory and Yang-Mills'.

\end{subsection}

\begin{subsection}{BRST symmetry}
In this section we follow closely the beautiful exposition of the subject made by Weinberg in his book \cite{weinberg1996quantum}.
Using the FDFP method gauge invariance of the action is lost. Actually this is fundamental in order for the method to be successful! Nevertheless gauge invariance leaves an important legacy to the quantum theory, which can only be understood by taking into account not only the gauge field but also ghosts and antighosts. The effective Lagrangian is the sum of the Yang-Mills Lagrangian, the gauge fixing term and the ghost term
\be
\mathcal{L}=\mathcal{L}_{YM}+\mathcal{L}_{gf}+\mathcal{L}_{ghost}
\ee
We use Fourier transform to rewrite the gauge fixing term in terms of an auxiliary field $h$.
\be
\int\mathcal{D}he^{ih\mathcal{G}}e^{i\frac{\alpha h^2}{2}}.
\ee
Thus the effective Lagrangian with the inclusion of the auxiliary field reads as
\be
\mathcal{L}=\mathcal{L}_{YM}+h_{a}\mathcal{G}_{a}+\frac{\alpha}{2}\mathcal{G}_{a}\mathcal{G}_{a}+\bar{\omega}_{a}(FP(A)\omega)_{a}.
\ee
It is invariant under the following BRST transformation\footnote{After Becchi, Rouet, Stora who discovered it and Tyutin who did it independently from the others.}
\begin{align}\label{BRST}
\delta_{\theta}A_{a\mu}=&\theta D_{\mu}\omega_{a}=\theta\left(\partial_{\mu}\omega_{a}+f_{abc}A_{b\mu}\omega_{c}\right),\\
\delta_{\theta}\omega_{a}^{*}=&-\theta h_{a},\\
\delta_{\theta}\omega_{a}=&-\frac{1}{2}\theta f_{abc}\omega_{b}\omega_{c},\\
\delta_{\theta} h_{a}=&0.
\end{align}
The parameter $\theta$ is taken to be Grassmann.
When matter fields are present, say a Dirac spinor $\psi$, they transform according to the rule
\be
\delta_{\theta}\psi=i\tau_{a}\theta\omega_{a}\psi.
\ee
It is evident that a BRST transformation acts on the gauge and matter fields as an infinitesimal gauge transformation with gauge parameters $\theta\omega_{a}$. Thus the action, being gauge invariant, is automatically BRST invariant.

Given any functional of the fields, which we can collectively denote by $\Phi$, call it $F[\Phi]$, we can represent the action of a BRST transformation on it as follows
\be\label{operatore BRST s}
\delta_{\theta}F[\Phi]=\theta sF[\Phi],
\ee
where $s$ is a nihilpotent operator
\be
\delta_{\theta}sF[\Phi]=\theta s(sF[\Phi])=0.
\ee
We can prove that this property holds separately for the various fields appearing in the Lagrangian. We start with the spinor
\begin{multline}
\delta_{\theta}\psi=it_{a}\delta_{\theta}(\omega_{a}\psi)=-\frac{i}{2}f_{abc}\tau_{a}\theta\omega_{b}\omega_{c}\psi-\tau_{a}\tau_{b}\omega_{a}\theta\omega_{b}\psi=\\
=-\frac{i}{2}f_{abc}\tau_{a}\theta\omega_{b}\omega_{c}\psi+\tau_{a}\tau_{b}\theta\omega_{a}\omega_{b}\psi=0.
\end{multline}
We made use of the anticommuting nature of $\theta$ and the ghost and of the definition of the structure constants.
For the gauge field
\begin{multline}
\delta_{\theta}sA_{a\mu}=\delta_{\theta}D_{\mu}\omega_{a}=\partial_{\mu}\delta\omega_{a}+f_{abc}\delta_{\theta}A_{b\mu}\omega_{c}+f_{abc}A_{b\mu}\delta_{\theta}\omega_{c}=\\
=\theta\left( -\frac{1}{2}f_{abc}\partial_{\mu}(\omega_{b}\omega_{c})+f_{abc}(\partial_{\mu}\omega_{b})\omega_{c}+f_{abc}f_{bde}A_{d\mu}\omega_{e}\omega_{c}-\frac{1}{2}f_{abc}f_{cde}A_{b\mu}\omega_{d}\omega_{e}\right) =\\
=\theta\left( \frac{1}{2}f_{abc}(\partial_{\mu}\omega_{b})\omega_{c}+\frac{1}{2}f_{abc}(\partial_{\mu}\omega_{c})\omega_{b}
 -f_{abc}f_{cde}A_{d\mu}\omega_{e}\omega_{b}-\frac{1}{2}f_{abc}f_{cde}A_{b\mu}\omega_{d}\omega_{e}\right)
\end{multline}
The first and second term cancel because $f_{abc}$ is antisymmetric in $a$ and $b$, while the other two terms cancel because of the Jacobi identity.
The verification of the nilpotency of $s$ on the antighost and on $h$ is immediate.

We now consider a product of two fields $\phi_{1}$ and $\phi_{2}$, whose transformation properties under BRST are known. The action of a BRST transformation on the product is
\be
\delta_{\theta}(\phi_1\phi_2)=\theta s\phi_1\phi_2+\phi_1\theta s\phi_2=\theta\left[s\phi_1\phi_2\pm\phi_1 s\phi_2\right].
\ee
In the case $\phi_1$ is a fermion a minus sign appears when moving $\theta$ to the left. We can now calculate the behaviour of $s\theta$ under BRST.
\be
\delta_{\theta}s(\phi_1\phi_2)=s\phi_1\theta s\phi_2 \pm \theta s\phi_1 s\phi_2.
\ee
The statistics obeyed by $s\phi$ is the opposite of that obeyed by $\phi$, as a consequence of the fermionic nature of the operator $s$. Then it follows that the RHS of the last equation is zero as we can see by moving $\theta$ to the left. The argument can be easily generalized to the case of any product of fields
\be
\delta_{\theta}s(\phi_1\phi_2\dotsm).
\ee
Hence it follows  that for any functional $F[\phi]$ of a given set of fields collectively denoted by $\phi$, we have
\be
\delta_{\theta}sF[\phi]=\theta ssF[\phi]=0.
\ee
Which is what we originally wanted to show.

The terms in the Lagrangian involving the ghosts, antighosts and auxiliary fields can be rewritten in a clever way. We start from evaluating the change of the gauge fixing term under BRST, which, as we have seen, acts on it as a gauge transformation with gauge parameter $\lambda_{a}=\theta\omega_{a}$
\be
\delta_{\theta}\mathcal{G}_{a}=\int\frac{\delta\mathcal{G}_{a}}{\delta\lambda_{b}}\lvert_{\lambda=0}\theta\omega_{b}=\theta\int FP(A)_{ab}\omega_{b}
\ee
Using this result together with the transformation properties of the antighost and the auxiliary field, we can thus write
\be
\omega_{a}^{*}FP_{ab}\omega_{b}+h_{a}\mathcal{G}_{a}+\frac{1}{2}\zeta\mathcal{G}_{a}\mathcal{G}_{a}=-s\left(\omega_{a}^{*}\mathcal{G}_{a}+\frac{1}{2}\zeta\omega_{a}^{*}\mathcal{G}_{a}\right)
\ee
Thus the effective action for Yang-Mills theories can be written as the sum of two terms, a gauge invariant one and another one given by the result of the action of $s$ on a certain functional.
\be
S_{eff}=S_{YM}+s\Psi
\ee
It follows from the nihilpotency of the BRST transformation that this action is indeed BRST-invariant.

Matrix elements between physical states must be independent from our choice of the gauge-fixing functional \emph{i.e.}, from the choice of the functional $\Psi$. Using Schwinger's principle, we can compute the change of a matrix element $\langle\alpha|\beta\rangle$ under an infinitesimal change $\tilde{\delta}\Psi$:
\be\label{Schwinger Psi}
\tilde{\delta}\langle\alpha|\beta\rangle=i\langle\alpha|s\tilde{\delta}\Psi|\beta\rangle.
\ee
We use a tilde here to make a clear distinction between changes induced by a change in the functional $\Phi$ and a BRST transformation. 
In order to describe the change of a field operator $\Phi$ under a BRST transformation we can introduce a fermionic charge $Q$ such that
\be
\delta_{\theta}\Phi=-i[\theta Q,\Phi]=-i\theta [Q,\Phi]_{\mp}.
\ee
The minus sign denotes the commutator for a bosonic $\Phi$, while the plus denotes the anticommutator for a fermionic $\Phi$. According to our definition of the operator $s$ (\ref{operatore BRST s}), the last equation can also be rewritten as
\be
[Q,\Phi]_{\mp}=is\Phi.
\ee
Starting from this expression we can see which kind of restrictions the nihilpotency of the BRST transformation puts on $Q$
\be\label{nihilpotent Q}
0=-ss\Phi=[Q,[Q,\Phi]_{\mp}]_{\pm}=[Q^2,\Phi]_{-}.
\ee
The last equality follows from the graded Lie algebra identities
\begin{align}
[A,[B,C]_{-}]_{+}=&[[A,B]_{+},C]_{-}-[B,[A,C]_{-}]_{+},\\
[A,[B,C]_{+}]_{-}=&[[A,B]_{+},C]_{-}-[B,[A,C]_{+}]_{-}.
\end{align}
Equation (\ref{nihilpotent Q}) must hold for all operators $\Phi$, so $Q^2$ must either vanish or be proportional to the identity operator. Though the latter possibility has to be excluded because $Q$ carries a non-zero ghost quantum number. So we have
\be
Q^2=0.
\ee
From equation (\ref{Schwinger Psi}) we have
\be
\tilde{\delta}\langle\alpha|\beta\rangle=\langle\alpha|[Q,\tilde{\delta}\Psi]_{\mp}|\beta\rangle.
\ee
This should vanish for any couple of arbitrarily chosen physical states. As a consequence, physical states satisfy the relations
\be
Q|\beta\rangle=\langle\alpha|Q=0.
\ee
We can thus state that in Yang-Mills theories \footnote{actually this remains true when considering more general gauge theories} the space of physical states is given by the cohomology of the BRST charge Q \emph{i.e.}, the kernel of $Q$ modulo the image of $Q$.

Besides allowing us to construct a Fock space of physical states, BRST also shed some light on the geometrical meaning of ghost fields. From fromula (\ref{BRST}) we see that the transformation law for the ghost field shows much resemblance to the Maurer-Cartan equation. We can thus identify the ghost field with the Maurer-Cartan one-form of the gauge group $\mathcal{G}$, provided that $s$ is identified with the external differential on the group manifold \cite{bonora}. 
\end{subsection}

\begin{subsection}{BRST and boundary conditions}
In (\ref{capitolo gauge}) we discussed the asymptotic conditions which must be imposed on the gauge field in the case of an infinite space such as $\mathbb{R}^4$ or $\mathbb{R}^3$. We saw that these conditions are such that the space is compactified into a sphere. No attention was paid to the behaviour of the ghost field at infinity. In the case of a compact space the presence of a boundary radically changes the situation. We have to be more careful and impose appropriate boundary conditions on \emph{all} the fields on which the action depends. These conditions have to reflect the BRST invariance of the theory\cite{MossSilva}\cite{esposito2005spectral}\cite{Moss}. The choice of the right boundary conditions is of great importance both from the classical and the quantum point of view. From a classical perspective we know that the gauge connection is not an observable, as we can perform a gauge transformation and yet obtain the same physical situation. When the boundary conditions are not gauge invariant, 
performing a gauge transformation 
actually we change the solution of the dynamical equations. On the quantum side we can see from equation (\ref{funzionale Z + condizioni al contorno}) (the parameters $\alpha$ and $\beta$ label sets of boundary conditions) that if we want the generating functional to be gauge invariant then the action as a whole has to be gauge invariant. Thus we have to restrict our attention to sets of boundary conditions possessing the same gauge symmetry of the Lagrangian.

A quite general class of boundary conditions is given by the so called \emph{mixed boundary conditions}. In order to formulate these conditions we need to introduce a bit of mathematical machinery.

Near the boundary $\partial M =\Sigma$ we can decompose the field in its tangential and normal components. Let $n$ be the unit normal to $\Sigma$ and $e^{i}_{a}$ a basis of tangential vectors on $\Sigma$. We have
\be
A_{a}=A_{i}e^{i}_{a}+An_{a}.
\ee
The metric structure on $M$ induces a metric structure on $\Sigma$. Moreover the Levi-Civita connection on $M$ induces a Levi-Civita connection on $\Sigma$. Let us see this explicitly. We can write the line element on $M$ near $\Sigma$ as
\be
ds^2=dn^2+f(n)d\tilde{s}^2,
\ee
where $d\tilde{s}^2$ is the line element on $\Sigma$. We can then compute the Christoffel symbols in the basis $\{e^{i}_{a},n\}$.
We are actually interested just in some of these, namely those with at least a $n$ index. Denoting by a semicolon the covariant derivative on $M$, by a stroke that on $\Sigma$ and by a dot the derivative along $n$, we have
\begin{align}\label{Semicolon vs Stroke}
A_{i;j}=&\partial_{j}A_{i}+\Gamma^{k}_{ji}A_{k}=A_{i|j}+\Gamma^{n}_{ji}A_{n},\\
A_{n;j}=&\partial_{j}A_{n}+\Gamma^{k}_{jn}A_{k}=A_{|j}+\Gamma^{k}_{jn}A_{k},\\
A_{n;n}=&\partial_{n}A_{n}+\Gamma^{k}_{nn}A_{k}=\dot{A}+\Gamma^{k}_{nn}A_{k}.\\
\end{align}
We do not need to consider $A_{i;n}$ because it is equal to $\dot{A_{i}}$ by definition. The covariant derivative of $n$ is called the extrinsic curvature of the boundary $\Sigma$.
\be
K_{ij}\equiv n_{i;j}
\ee
Its symmetry comes from that of the Christoffel symbols:
\be
n_{i;j}=\Gamma^{k}_{ji}=n_{j;i}.
\ee
From the definition of the Christoffel symbols
\be
\Gamma^{a}_{bc}=\frac{g^{ad}}{2}\left(\partial_{b}g_{dc}+\partial_{c}g_{db}-\partial_{d}g_{bc}\right)
\ee
we have
\begin{align}
\Gamma^{n}_{ji}=&-\frac{g^{nn}}{2}\partial_{n}g_{ji}=-\frac{1}{2}\partial_{n}g_{ji}=K_{ij},\\
\Gamma^{k}_{jn}=&\frac{g^{kl}}{2}\partial_{n}g_{lj}=-g^{kl}\Gamma^{n}_{lj}=-K_{j}^{\;k},\\
\Gamma^{k}_{nn}=&0.
\end{align}
We can finally express the relation between the two metric connections with the following equations
\begin{align}\label{Connessione sul bordo}
A_{i;j}=&A_{i|j}+K_{ij}A,\\
A_{n;j}=&A_{|j}-K_{j}^{\;k}A_{k},\\
A_{i;n}=&\dot{A}_{i},\\
A_{n;n}=&\dot{A}.
\end{align}
Let $P_{\pm}$ denote two projectors, one along the normal and the other tangential to the boundary. We purposely do not specify which is which for the sake of generality. The mixed boundary conditions are expressed by the following equations
\begin{align}
\left(\nabla_{n}+K\right)P_{+}A=&0\\
P_{-}A=&0.
\end{align}
This is a generalization both of Dirichlet and Neumann boundary conditions. Actually we should expect this to be a good candidate from our experience with the electromagnetic field in flat space. Studying the electromagnetic field in a box we know that if we impose Dirichlet boundary conditions on each of the components of the potential, then the only solution of Maxwell equations is the trivial one. So we have to relax this condition a bit. If we consider the electromagnetic field in a box with perfectly conducting walls, the tangential component of the electric field vanishes on the walls while that of the magnetic field will have extrema on it. With an appropriate choice of the gauge this is equivalent to say that the tangential components of the potential vanish on the walls, while the normal derivative of the time component of the potential is zero. 
%because in the latter it may appear that we are neglecting the four-dimensionality of the dynamics of the electromagnetic field. This is not the case as we are allowed to set the time component of the connection to zero by choosing \emph{e.g.}, the radiation gauge or the temporal gauge, and we can also separate between initial and boundary conditions, thus reducing the problem of choosing the boundary conditions to a purely three-dimensional one.} 

Let us see how the mixed boundary conditions present themselves in a natural way. We choose to impose Dirichlet boundary conditions on the tangential components of the gauge field $A_{i}$. Performing a BRST transformation (\ref{BRST}) and evaluating the result on the boundary, we have
\be
\delta_{\theta} A_{i}=\theta\omega_{;i}.
%%\theta(D_{i}\omega)^a=\theta\left(\omega_{;i}^{a}+[A_{i},omega]^a\right)\lvert_{\Sigma}=\theta\omega_{;i}^{a}
%%\delta_{\theta} A^a=\theta(D_{n}\omega)^{a}=\theta\left(\dot{\omega}^{a}+[A,omega]^a\right)\lvert_{\Sigma}=\theta\dot{\omega}^{a}.
\ee
Thus, in order for the boundary conditions to be compatible with BRST, we have to impose Dirichlet boundary conditions on the ghost. If we assume that the Faddeev-Popov operator is self-adjoint with this choice of the boundary conditions, then the antighost too has to satisfy the same boundary conditions. We are not finished yet, as we need also to know how to properly deal with the remaining components of the gauge field. We can eliminate the auxiliary field by imposing $b=\mathcal{G}[A]$. This does not spoil the BRST symmetry of the action, but BRST transformations are now nihilpotent only when acting on solutions of the equations of motion. Then we can set the gauge fixing $\mathcal{G}[A]$ to zero on $\Sigma$. A final check of consistency is now in order: we have to verify that the condition
\be
\mathcal{G}[A]=0
\ee
goes in itself under BRST.
\be
\delta\mathcal{G}[A]=\mathcal{G}[A+\delta A]-\mathcal{G}[A]=\theta FP(A)\omega=0
\ee
The consistency is guaranteed if the last equality holds. We assume that the ghost can be expanded in a basis of eigenvectors of the Faddeev-Popov operator
\begin{align}
\omega=&\sum_{k}c^{k}u^{k},\\
FP(A)u^{k}=&\lambda^{k}u^{k}.
\end{align}
If the eigenvectors are such that they vanish at the boundary $\Sigma$ the consistency is proved.

In the case of the Lorenz gauge fixing condition, using (\ref{Connessione sul bordo}) we have
\be
\partial^{\mu}A_{\mu}=\dot{A}+A_{i}^{\; ;i}=\dot{A}+A_{i}^{\;|i}+K_{i}^{\;i}A=\dot{A}+KA\quad\mbox{on $\Sigma$}
\ee
In the last equality we used the fact that $A_{i}=0$ on $\Sigma$. In conclusion the following set of boundary conditions is BRST invariant
\begin{align}
A_{i}=\omega=\bar{\omega}=0,\\
\dot{A}+KA=0.
\end{align}
These are called the magnetic boundary conditions, because they correspond to fixing the magnetic field on the boundary if the $n$ direction is timelike.

Another possibility would be to impose Dirichlet boundary conditions on the normal component of the gauge field.
\be
\delta_{\theta} A=\theta\dot{\omega}=0
%%\theta(D_{n}\omega)^{a}=\theta\left(\dot{\omega}^{a}+[A,omega]^a\right)\lvert_{\Sigma}=\theta\dot{\omega}^{a}
\ee
As a consequence, the ghost has to satisfy Neumann boundary conditions together with the antighost. We can fix the electric field on the boundary
\be
F_{in}=A_{n;i}-A_{i;n}=A_{|i}-K_{i}^{\;k}A_{k}-\dot{A}_{i}=0.
\ee
The electric field is obviously gauge-invariant, so this condition is automatically BRST invariant. The electric boundary conditions are thus
\begin{align}
A=\dot{\omega}=\dot{\bar{\omega}}=0\\
\dot{A}_{i}+K_{i}^{\;k}A_{k}=0.
\end{align}

Now we just want to give an outline of the work by Moss and Silva \cite{MossSilva}, where they give a method to generate whole new sets of BRST invariant boundary conditions. As we have seen in the previous section, physical configurations are in the cohomology of the BRST charge. This implies that they are annihilated by the charge
\be
Q|\Psi\rangle=0.
\ee 
Instead of working with equivalence classes of state vectors we can impose another independent condition on the states, for example:
\be
G|\Psi\rangle=0,
\ee
where $G$ is the ghost density operator. If physical states are annihilated by these operators, then the same must be true for the amplitudes. In order to accomplish this we just have to impose these conditions as classical equations on the fields at the boundary. The BRST charge is obtained from the Noether current associated to BRST invariance of the theory. The ghost density operator is
\be
G=cp-\bar{c}\bar{p},
\ee
where $p$ and $\bar{p}$ are the conjugate momenta respectively to the ghost and the antighost fields. Canonical methods are then used to obtain the boundary conditions. We do not go into details, because it would require an exposition of the theory of constrained Hamiltonian systems. It is important to stress that the division of phase space into ghosts and their conjugate momenta is not preserved under canonical transformations. Starting from this observation it is possible to develop a method to generate whole new sets of BRST invariant boundary conditions.
\end{subsection}

%%definizione dell'integrale funzionale nel formalismo di de witt
%%
%%formulazione delle teorie di gauge
%%
%%metodo di Faddeev Popov
%%
%%discussione sui campi di ghost, sul loro significato e sulla neccessità di introdurli
%%
%%BRST
%%
%%invarianza di gauge e condizioni al contorno
%%
%%gauge fixing globali e condizioni asintotiche
%%
%%
%%
%%
%%
%%
%%
\end{section}

\begin{section}{Gribov ambiguity}
Despite the extraordinary success of perturbative quantum Yang-Mills theories in describing the ultra-violet behaviour of the strong force, there are still some issues which have to be investigated in order to gain a full understanding of this fundamental interaction at all scales. From the mathematical point of view even the existence of a quantum Yang-Mills theory has yet to be proved \cite{JaffeWitten:Millennium}! But there are problems which are of interest for both the mathematician and the physicist, which a full theory should account for, like the existence of a \emph{mass gap} and \emph{colour confinement}.
%%spiegazione del problema della mass gap e del confinamento del colore
There is a peculiar phenomenon in non-Abelian Yang-Mills theories, which could hopefully shed some light on the infrared-behaviour of the corresponding quantum theories. This phenomenon is referred to as the \emph{Gribov ambiguity}, after Gribov who first pointed out the problem \cite{Gribov:78}.
The origin of the ambiguity is quite subtle and its easier to understand it from a Lagrangian point of view\footnote{The same problem was approached by Gribov also from an Hamiltonian point of view \cite{Gribov:Hamiltonian}}. When quantizing Yang-Mills theories with the Faddeev-Popov method we need to introduce a \emph{gauge-fixing}. Tipically the gauge-fixing condition is linear and so defines an hyperplane in the space of histories for the theory. This is the case for transverse gauges like the Landau gauge and the Coulomb gauge. It is vital to guarantee the success of the method that this hyperplane intersects each gauge-orbit exactly once. Thus fixing the gauge is equivalent to consider sections of the space of histories. Actually it was shown by Singer \cite{Singer:GA} that the space of histories has indeed a much complicated topological structure and does not admit a global section, so we are not allowed to use the Faddeev-Popov method.
\begin{subsection}{Gribov pendulum}\label{Gribov pendulum}
It was discovered by Gribov in the case of the Coulomb gauge for the $SU(2)$ gauge theory, that there exist gauge-equivalent potentials satisfying the same gauge-fixing condition.
These potentials are for this reason called Gribov copies. To see how this could happen, let us consider on $\mathbb{R}^{4}$ the four-potential $A_{\mu}$, where $A_{0}=0$ and $\partial^{i}A_{i}=0$. If there exists any other potential satisying the same gauge-fixing, it must be a solution of the equation
\be
\partial^{i}(U^{-1}A_{i}U+U^{-1}\partial_{i}U)=0.
\ee

%Putting $A_{i}=0$ in this equation and considering a spherically symmetric $U$ we have the Gribov pendulum equation.
We make an ansatz for the gauge transformation $U=e^{\frac{\alpha(r)}{2}\hat{n}}$, with $\hat{n}=in^{a}\sigma_{a}$.
In order for $U$ to be regular at the origin we must require that $\alpha(r)\to 2l\pi$ as $r\to 0$ with $l\in\mathbb{Z}$. Moreover, if the non Abelian charge has to remain unchanged, we must require that  the limit of $U$ at $\infty$ belongs to the center of $SU(2)$, namely
\be
U\to \pm\mathbb{I}.
\ee
In terms of the function $\alpha$ this is expressed as
\be
\alpha(r)\to m\pi \quad \mbox{for}\; r\to\infty.
\ee
$m$ denotes an integer number. The case $m=2k$ is referred to as the strong boundary condition. In contrast the case $m=2k+1$ corresponds to the weak boundary conditions. We will see the reason for this terminology in what follows.
%The most general Lie Algebra valued one-form  for a sperical symmetry reads in the case of $SU(2)$ as
\be
A_{i}=f_{1}(r)\frac{\partial\hat{n}}{\partial x_{i}}+f_{2}(r)\hat{n}\frac{\partial\hat{n}}{\partial x_{i}}+f_{3}(r)\hat{n}n_{i}
\ee
Denoting the gauge transformed field as $\tilde{A}_{i}$ and requiring the divergence of the field to be invariant under a gauge transformation
\be
\partial_{i}\tilde{A}_{i}=\partial_{i}A_{i}
\ee
we get the equation
\be
\alpha''+\frac{2}\alpha'-\frac{4}{r^2}\left(\left(f_{2}+\frac{1}{2}\right)\sin\alpha+f_1(\cos\alpha-1)\right)=0.
\ee
Under the change of variables $\tau=\log r$ this can be recast in the form
\be
\ddot{\alpha}+\dot{\alpha}-4\left(\left(f_2+\frac{1}{2}\right)\sin\alpha+f_1(\cos\alpha-1)\right)=0
\ee
which is the same equation that describes the motion of a forced damped pendulum. The appearance of a minus sign in front of the sine means that the angles are measured starting from the point of unstable equilibrium of the free pendulum.
We see that by choosing suitable expressions for $f_1$ and $f_2$ we can satisfy the boundary condition $\alpha(r)\to2m\pi$ at $\infty$, corresponding to an integer number of round trips or to oscillations around the position of unstable equilibrium. Though not all the possible choices of the parameters correspond to transverse potentials. A particularly interesting case is obtained in the case in which $f_1=f_3=0$, so that the potential is transverse and satisfies De Witt's asymptotic conditions for every choice of $f_2=f$.
In this case the pendulum equation reduces to
\be
\ddot{\alpha}+\dot{\alpha}-4\left(f+\frac{1}{2}\sin\alpha\right)=0.
\ee
The expression for the transformed field is
\be
\tilde{A}_{i}=\left(f+\frac{1}{2}\right)\sin\alpha\frac{\partial\hat{n}}{\partial x_{i}}+\left(f+\frac{1}{2}\cos\alpha-\frac{1}{2}\right)\hat{n}\frac{\partial\hat{n}}{\partial x_{i}}+\frac{1}{2}\alpha'(r)\hat{n}n_{i}.
\ee
Choosing an $f$ such that $f\to 0$ at infinity, we have in the case of strong boundary conditions that
\be
\lim_{r\to\infty}rA_{i}=0.
\ee
In the case of weak boundary conditions, corresponding to the pendulum falling to the position of stable equilibrium, we have that the gauge transformed potential is slowly decaying at infinity
\be
\tilde{A}_{i}\approx\frac{1}{r}.
\ee
Thus in the case of weak boundary conditions, De Witt's asymptotic conditions cannot be satisfied. The case of the Gribov copies of the vacuum falls in this class. Indeed the vacuum corresponds to the choice $f=0$. In the absence of a forcing term the pendulum necessarily reaches the bottom and stays there at infinity.

We can calculate the Pontrjagyn index for the copies.
\be
\nu=\frac{1}{64\pi^2}\int Tr\left((g^{-1}\partial_{i}g)(g^{-1}\partial_{j}g)(g^{-1}\partial_{k}g)\right)\varepsilon_{ijk}.
\ee
What we obtain is that strong copies have integer index, while for weak copies it is half-odd.
\end{subsection}
\begin{subsection}{The Faddeev-Popov operator}\label{FP}
Going back to the  equation for the copies and taking a gauge transformation $U$ differing by an infinitesimal from the identity, say $U=\mathbb{I}+\alpha$, we see that it reduces to
\be\label{zero-modes}
FP(A)\alpha=-\partial^{\mu}(\partial_{\mu}+[A_{\mu},\; ])\alpha=0.
\ee
This is indeed the equation for the zero modes of the Faddeev-Popov operator, which is Laplace-type.

We can make an analogy with the time-independent Schroedinger equation, interpreting the Laplacian $-\triangle=-\partial^{\mu}\partial_{\mu}$ as the usual kinetic term and the term containing the gauge connection as a sort of potential. For $A_{\mu}=0$ equation (\ref{zero-modes}) reduces to Laplace equation. If we impose Dirichlet boundary conditions there is no non-trivial solution to the problem on $\mathbb{R}^{n}$. For a non-zero but small $A_{\mu}$ the eigenvalues of the operator constitute a sequence of positive numbers bounded from below. As we increase the amplitude of the potential we can expect the lowest eigenvalue to reach zero, and then turn negative. The region in the space of histories corresponding to transverse potentials such that equation (\ref{zero-modes}) admits non-trivial solutions is called the \emph{Gribov horizon}. As we further increase the amplitude of the potential there will be more and more eigenvalues turning negative.

In order to pursue the analogy with the Schroedinger equation we must require the Faddeev-Popov operator to be self-adjoint. Then we can treat it as a true Hamiltonian operator. We start by observing that the aforementioned operator is symmetric for any connection $A_{\mu}$. We assume our spacetime to be a region in $\mathbb{R}^{n}$ and do the calculations separately for the Laplace operator and for the term containing the connection.
\begin{multline}
\langle \phi,-\triangle\psi\rangle=-\int_{\partial V}\phi^{a*}n^{\mu}\partial_{\mu}\psi^{a}+\int_{V}\partial^{\mu}\phi^{a*}\partial_{\mu}\psi^{a}=\\
=\int_{\partial V}(n^{\mu}\partial_{\mu}\phi^{a*}\psi^{a}-\phi^{a*}n^{\mu}\partial_{\mu}\psi^{a})-\int_{V}(\triangle\phi^{a})^{*}\psi^{a}
\end{multline}
\begin{multline}\label{COMM SIMM}
\langle\phi,[A_{\mu},\partial_{\mu}\psi]\rangle=\int_{V}(\phi^{a})^{*}(f^{bca}A_{\mu}^{b}\partial_{\mu}\psi^{c})+\int_{\partial V}n^{\mu}\phi^{a*}\left([A_{\mu},\psi]\right)^{a}=\\
=\int_{V} [A_{\mu},\partial_{\mu}\phi]^{*}\psi + boundary\; term 
\end{multline}
In order to find the  domain of self-adjointness we have to pay attention to the boundary terms we get upon integrating by parts, and find for which class of boundary conditions the operator is self-adjoint. We see that Neumann boundary conditions are not appropriate in this case as they cannot make the boundary term in (\ref{COMM SIMM}) vanish. On the other hand if we choose Dirichlet boundary conditions the operator is self adjoint. We can see this by studying the boundary term in the one-dimensional case.
\be\label{BOUNDARY TERM}
(\dot{\phi}^{*}\psi-\phi^{*}\dot{\psi})\lvert_{0}^{2\pi}=0
\ee
Imposing Dirichlet boundary conditions we have $\psi(0)=\psi(2\pi)=0$ we have
\be
(\phi^{*}\dot{\psi})\lvert_{0}^{2\pi}=0.
\ee
The value of the derivative of $\psi$ at the boundary is not fixed and so $\psi(0)$ and $\psi(2\pi)$ are completely arbitrary numbers. Thus the only way to satisfy equation (\ref{BOUNDARY TERM}) is to take $\phi(0)=\phi(2\pi)=0$.

We now take $V$ to be a rectangle in $\mathbb{R}^{n}$ and identify opposite faces, so that we get a flat torus $T^{n}$. One may think the most natural choice for the domain of the Laplace operator on the torus being the space $C^{\infty}_0(V)$ of smooth functions with compact support on $V$. Though the Laplace operator is not essentially self-adjoint on this domain as we can see from equation (\ref{COMM SIMM}), in the simplest case in which $n=1$. For a function $\psi\in C^{\infty}_0(V)$ we have that both $\psi$ and its derivative vanish at the boundary of $V$, so that  the value of $\phi$ and that of $\phi'$ there are not constrained at all. Then the domain of the adjoint operator is larger than $C^{\infty}_0(V)$. Nevertheless one can construct self-adjoint extensions of the Laplace operator on $C^{\infty}_0(V)$; actually they are infinite in number and can be put in one-to-one correspondence with the group $U(1,1)$. Indeed we can write the boundary conditions in the form
\be\label{TRAS BOUNDARY}
\Psi_{2\pi}=A\Psi_0,
\ee
where $A$ is an operator acting on the column vector $\Psi$ whose elements are the value of the function and its derivative at the boundary.
\be
\Psi=\begin{pmatrix}\psi\\ \dot{\psi}\end{pmatrix}.
\ee
Equation (\ref{BOUNDARY TERM}) can be recast in matrix form
\be
\Phi^{\dagger}_0\Omega\Psi_0=\Phi^{\dagger}_{2\pi}\Omega\Psi_{2\pi},
\ee
where
\be\label{INVA}
\Omega=\begin{pmatrix} 0 & -1\\ 1 & 0\end{pmatrix}.
\ee
Using (\ref{TRAS BOUNDARY}) and the arbitrariness in $\Phi_0$ and $\Psi_0$ we obtain the condition of invariance of the bilinear form $\Omega$ under the linear transformation $A$
\be\label{definizione matrice A}
A^{\dagger}\Omega A=\Omega.
\ee
The matrix $\Omega$ can be diagonalized. We denote by $S$ the matrix that diagonalizes $\Omega$
\be
S^{\dagger}\Omega S=i\sigma_3.
\ee
If we define the matrix
\be
Q=S^{\dagger}A S,
\ee
equation (\ref{definizione matrice A}) transforms into
\be
\sigma_3=Q^{\dagger}\sigma_3 Q.
\ee
Thus boundary conditions of the form (\ref{TRAS BOUNDARY}) with $A\in U(1,1)$ are in a one-to-one correspondence to self-adjoint extenxions of the Laplace operator on $C^{\infty}_0(V)$. One can also prove that other extensions do not exist. Notice that Dirichlet or Neumann boundary conditions are not natural choices on a torus since they would require the choice of a privileged point on it ($n$ privileged cuts in the $n$ dimensional case). The generalization to the multidimensional case is immediate. We need an operator $A^{(j)}$ for each of the $n$ cuts we have to make on the torus. Using the notation $(x;x_{j})\equiv(x_1,\dots x_{j}\dots x_{n})$ to single out the only coordinate which is going to vary, we have
\be
\Psi_{(x;x_{j}=2\pi)}=A^{(j)}\Psi_{(x,x_{j}=0)},
\ee
Now that we have found self-adjoint extensions of the Laplacian, the problem remains of finding those of the Faddeev-Popov operator. We would like to get rid of the boundary term in (\ref{COMM SIMM}), so we choose among the $A^{j}$ operators at our disposal those belonging to $U(1)\times U(1)$. In this case indeed the phase changes in $\phi$ and $\psi$ then cancel each other and, as the $j$-th component of the normal versor to the boundary has opposite signs on opposite faces, the integral is zero.
When the phases are chosen to be the same we have periodic-up-to-a-phase boundary conditions
\be
\Psi_{(x;x_{j}=2\pi)}=e^{i\alpha}\Psi_{(x,x_{j}=0)}.
\ee
The physical interpretation of these is that solenoids are present inside each of the $S^{1}$ factors of the torus, and $\alpha$ is the circulation of the corresponding vector potential.\\
In the following when discussing the properties of some operator we will implicitly assume that suitable boundary conditions are chosen in order to guarantee the self-adjointness.

The spectrum of the Laplace operator on a compact manifold\footnote{We also assume that the manifold has just one connected component.} is an unbounded sequence of non-negative numbers $0\leq\lambda_1\leq\lambda_2\leq\dots$. It is a consequence of Hodge theorem that functions annihilated by the Laplacian are constant. Indeed the space of harmonic functions is isomorphic to the $0$-th cohomology class
\be
\mbox{Harm}^0(M)\simeq H^0(M).
\ee
Then we can simply ignore these trivial zero-modes and consider just the positive part of the spectrum. The Faddeev-Popov operator corresponding to the vacuum $A_{\mu}=0$ is just the Laplacian. For a small amplitude potential $A_{\mu}$ we expect that the positivity property of the spectrum still holds. We can then define the Gribov region as a subset in the Space of Histories whose points are transverse connections such that the corresponding Faddeev-Popov operator is positive definite, \emph{i.e.} all its eigenvalues are positive. The boundary of this region consists of those connections such that the least eigenvalue vanishes. As the amplitude of the connection increases further, this eigenvalue turns negative  and keeps decreasing until the next eigenvalue vanishes. This region is called the second Gribov region, whose boundary is the union of the (first) Gribov horizon and a second horizon. Increasing the amplitude more and more we encounter a sequence of regions and horizons. If in a given region the 
determinant of the Faddeev-Popov operator is positive, then in the next one it will be negative and viceversa.
\begin{center}
\includegraphics[width=0.6\columnwidth]{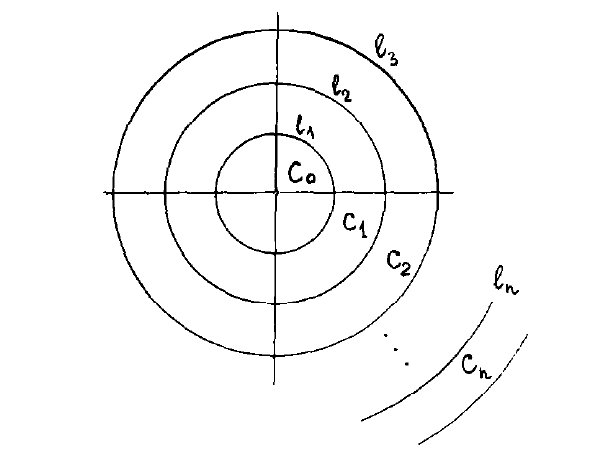}
\end{center}
%%autoaggiuntezza di FP, condizioni al contorno e topologia dello spazio
%%Definizione di regioni e orizzonti. parallelo con l'equazione di schrodinger
\end{subsection}
\begin{subsection}{Copies outside the horizon}
In this section we will demonstrate that potentials lying close to the horizon have a copy lying outside of it.
Let $C_{\mu}$ be a potential on the Gribov horizon. By definition the corresponding Faddeev-Popov operator has a non-trivial zero eigenvector, say $\phi_0$.
\be
\partial^{\mu}(D_{\mu}(C)\phi_0)=0
\ee
This generates an infinitesimal gauge transformation which preserves the gauge condition.
\be
C_{\mu}[e^{\phi_0}]\simeq C_{\mu}+D_{\mu}(C)\phi_0
\ee
 Let us consider a transverse potential lying close to the horizon $A_{\mu}=C_{\mu}+a_{\mu}$, $\partial^{\mu}A_{\mu}=0$ and a gauge transformation $S$ which we can express for a small Lie-algebra element $\alpha$  as
 \be
 S=\mathbb{I}+\alpha+\frac{\alpha^2}{2}+o(\alpha^2).
 \ee
 The gauge transformed potential has the expression
 \be
 A'_{\mu}=S^{\dagger}\partial_{\mu}S+ S^{\dagger}A_{\mu}S=A_{\mu}+D_{\mu}(A)\alpha+\frac{1}{2}[D_{\mu}(A)\alpha,\alpha]+o(\alpha^3).
 \ee
 If we require that this satisfies the Landau gauge we have the equation
 \be
 \partial^{\mu}D_{\mu}(A)\alpha+\frac{1}{2}\partial^{\mu}[D_{\mu}(A)\alpha,\alpha]=0,
 \ee
 which is valid up to higher order terms.
 We can express for convenience the solution $\alpha$ as $\phi_0$ plus a small perturbation. We now introduce an expansion parameter $\lambda$ which is useful in order to collect terms of the same order. $\lambda$ need not be small, and we can set it to $1$ at the end of the calculations. 
 \begin{align}
 A_{\mu}=C_{\mu}+\lambda a_{\mu}\\
 \alpha=\lambda\phi_0+\lambda^2\tilde{\alpha}
 \end{align}
 Inserting these expressions in the equation written above and collecting terms which are second order in $\alpha$ we get
 \be
 \partial^{\mu}D_{\mu}(C)\tilde\alpha=-\partial^{\mu}[a_{\mu},\phi_0]-\frac{1}{2}\partial^{\mu}[D_{\mu}(C)\phi_0,\phi_0].
 \ee
 From the orthogonality of the LHS to $\phi_0$ we get a normalization condition for $\phi_0$.
 \be\label{NORM PER PHI0}
 \int Tr\left(\phi_0\partial^{\mu}D_{\mu}(C)\tilde{\alpha}\right)=\int Tr\left(\partial^{\mu}D_{\mu}(C)\phi_0\tilde{\alpha}\right)=0
 \ee
 Whence it follows
 \be
 \int Tr\left(\phi_0(\partial^{\mu}[a_{\mu},\phi_0]+\frac{1}{2}\partial^{\mu}[D_{\mu}(C)\phi_0,\phi_0])\right)=0.
 \ee
 The eigenvalue equation for the Faddeev-Popov operator corresponding to the field $A_{\mu}$ is written as
 \be
 -\partial^{\mu}(\partial_{\mu}\psi+[C_{\mu},\psi]+[a_{\mu},\psi])=\varepsilon(a)\psi.
 \ee
 If $a$ is small we can determine the eigenvalues in perturbation theory. Obviously the unperturbed state of zero energy is $\phi_0$, so the correction of the energy to first order is simply given by the formula
 \be
 \varepsilon(a)=-\frac{\int Tr(\phi_0[a_{\mu},\partial^{\mu}\phi_0])}{\int Tr(\phi_0 \phi_0)}.
 \ee
 If we write $A'_{\mu}=C_{\mu}+a'_{\mu}$ with $a'_{\mu}=a_{\mu}+D_{\mu}(C)\phi_0$, we have
 \be
 \varepsilon(a')=-\frac{\int Tr(\phi_0([a_{\mu},\partial^{\mu}\phi_0]+\partial^{\mu}[D_{\mu}(C)\phi_0,\phi_0])}{\int Tr(\phi_0 \phi_0)}=-\varepsilon(a).
 \ee
 Here we made use of equation (\ref{NORM PER PHI0}). If $A_{\mu}$ is inside the first Gribov region then $A'_{\mu}$ lies necessarily outside of it.
 %%Discorso di Gribov sulle copie che si trovano un po' fuori AMPLIARE
 We want to discuss a particular situation which may be given. In the figure is depicted a potential $A'_{\mu}$ which is a copy of two distinct potentials $A_{\mu}$ and $A''_{\mu}$ inside the Gribov region. Then we can have two gauge equivalent potentials inside the Gribov region. Obviously to find the gauge transformation relating them we must take into account higher order terms in $\alpha$.
 \begin{center}
\includegraphics[width=0.6\columnwidth]{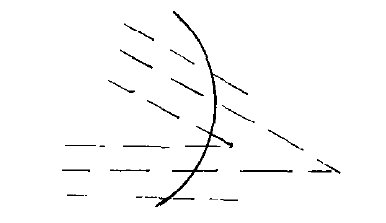}
\end{center}
 %%%%%%%%%          INSERIRE FIGURA!!!!!!!!!
 %%%%%%		CAPIRE QUEL DISCORSO SUI MINIMI DELLA LAGRANGIANA EFFETTIVA
\end{subsection}

\begin{subsection}{Henyey's example}
An example due to Henyey exhibit a copy for a non-zero $A_{j}$. Choose a connection in the form
\be\label{HENYEY}
A_{j}=ia_{j}\sigma_3
\ee
The transverality condition requires that $\partial_{j}a_{j}=0$.
We look for a gauge transformation in the form
\be
S=e^{i\alpha f\cdot\sigma},
\ee
where $f=e_{x}\cos\beta+e_{y}\sin\beta$ and $e_{x},e_{y}$ are two orthonormal vectors respectively directed along the $x$ and the $y$ axis.
We have for the gauge transformed field
\begin{multline}
\tilde{A_{j}}=S^{-1}A_{j}S+S^{-1}\partial_{j}S=i\sigma_3(a_{j}\cos(2\alpha)+\partial_{j}\beta\sin^2\alpha)+\\
ig\cdot\sigma\left(\partial_{j}\beta\frac{\sin(2\alpha)}{2}-a_{j}\sin(2\alpha)\right)+if\cdot\sigma\partial_{j}\alpha,
\end{multline}
where we defined $g=-e_{x}\sin\beta+e_{y}\cos\beta$.
Imposing the Coulomb gauge condition we get the equation
\begin{multline}
\partial_{j}\tilde{A_{j}}=i\sigma_3(\triangle\beta\sin^2\alpha+\sin(2\alpha)\partial_{j}\beta\partial_{j}\alpha-2a_{j}\sin(2\alpha)\partial_{j}\alpha)+\\
ig\cdot\sigma\left(\triangle\beta\frac{\sin(2\alpha)}{2}+(\partial_{j}\beta-2a_{j})\cos(2\alpha)\partial_{j}\alpha\right)-\\
if\cdot\sigma\partial_{j}\left(\partial_{j}\beta\frac{\sin(2\alpha)}{2}-a_{j}\sin(2\alpha)+ig\cdot\sigma\partial_{j}\beta\partial_{j}\alpha\right).
\end{multline}
In order to find a solution to this equation we have to put it into a simpler form. We then make the following ansatz
\be
\triangle\beta=0, \quad \alpha_{j}\partial_{j}\alpha=0, \quad \partial_{j}\beta\partial_{j}\alpha=0 
\ee
The equation reduces then to
\be
\triangle\alpha+a_{j}\partial_{j}\beta\sin(2\alpha)-\frac{1}{2}\partial_{j}\beta\partial_{j}\beta\sin(2\alpha)=0.
\ee
If we change to polar coordinates and make further assumptions
\be\label{ANSATZ POLARI}
a_{j}=a(r,\theta)e_{\phi}, \quad \beta=\beta(\phi),\quad \alpha=\alpha(r,\theta),
\ee
the first two ansatz are automatically satisfied. To satisfy also the third one
\be
\triangle\beta=\frac{1}{r^2\sin^2\theta}\partial^{2}_{\phi}\beta=0,
\ee
we may choose $\beta=\phi$. Finally taking $\alpha(r,\theta)=rb(r)\sin\theta$ we obtain an expression which can immediately be solved for $a$ in terms of $b$
\be\label{SOL HENYEY}
a=\frac{1}{2r\sin\theta}-\frac{b+(4rb'+r^2b'')\sin^2\theta}{\sin(2rb\sin\theta)}.
\ee
Thus we may assume a certain functional form for $b$ and obtain by means of this formula the corresponding expression for $a$. In other words we assign a gauge transformation and find for which connection it generates a Gribov copy. 
Now we have to impose some conditions on $b$ to guarantee the regularity of $a$ and a sufficiently fast decay at infinity. Looking at equations (\ref{ANSATZ POLARI}) and (\ref{HENYEY}) we see that the connection $A_{j}$ may be singular on the $z$ axis because its expression contains the vector field $e_{\phi}$. So we have to require that $a$ is zero at $\theta=0$ and $r=0$. The regularity at $\theta=0$ is automatically satisfied by expression (\ref{SOL HENYEY}) for any $r\neq0$. To ensure the regularity at $r=0$ we must require that $b(0)\neq 0$ and $b'(0)=0$. We also demand that $2rb<\pi$ in order to eliminate the possibility of a singularity coming from the vanishing of the denominator in the second term.
Dealing with the behaviour of the solution at infinity, we assume that $b$ decays as an inverse power of the distance from the origin
\be
b(r)\approx_{r\to\infty}\frac{1}{r^{n}}.
\ee
Then we get
\be
a\approx_{r\to\infty}o(r^{-2n+3})-n(n-3)\frac{\sin\theta}{2r}.
\ee
We want $a$ to decay faster than $\frac{1}{r^2}$
\be
r^2a\to0\quad \mbox{as}\,r\to\infty.
\ee
Taking for example $n=3$ in the formulae above we have
\begin{align}
b\approx\frac{1}{r^3}\\
r^2a\approx o\left(\frac{1}{r}\right).
\end{align}

Van Baal \cite{VanBaal:MGC} proved that following this construction we find copies on different sides of the Gribov horizon. Multiplying the function $b$ by any constant $\gamma$ such that $\gamma\leq 1$, $\gamma b$ satisfies the same conditions of $b$. Moreover $a(\gamma)$ does not depend on the sign of $\gamma$, so that we have two different copies for each connection of the family. As $\gamma$ varies, $A_{i}(\gamma)$ moves from the outside towards the Gribov horizon and reaches it for $\gamma=0$, where it coalesces with its copies. As a direct consequence $A_{i}(0)$ lies on the Gribov horizon. A zero mode for $FP(A(0))$ is given by $\frac{\partial g}{\partial\gamma}(0)$.
Moreover it is possible to show that the topological index of the copies vanishes, which makes sense because $g(\pm\gamma)$ are two continuous one-parameter families of gauge transformations and $g(0)=\mathbb{I}$. By definition they are of the same homotopy class of the identity, hence the topological index should be the same, \emph{i.e.} zero.
%%Mostrare l'esempio di Henyey e la discussione di Gribov sulle forze agenti su un pendolo
\end{subsection}
\begin{subsection}{Geometry of the Gribov region}
In this section we prove some theorems, following the paper by Zwanziger \cite{Zwanziger:BAN}.\\
The first important result in this sense is that the Gribov region is convex. For the proof take two flat connections $A_{1,2}$, such that the corresponding operators $FP(A_{1,2})$ are positive definite. Consider now the Faddeev-Popov operator corresponding to a connection lying on the segment from $A_1$ to $A_2$, $\bar{A}=\alpha A_2+(1-\alpha)A_1$ with $0\leq\alpha\leq 1$. Using the linearity of the Faddeev-Popov with respect to the connection we have
\be
FP(\bar{A})=\alpha FP(A_2)+(1-\alpha) FP(A_2).
\ee
The positive-definiteness of the LHS follows from that of the single terms on the RHS.

We now want to make rigorous the statement made before, when we made the analogy with the Schroedinger equation and reached the conclusion that the Gribov region is bounded in every direction. The argument goes as follows: we fix a non-zero connection $A$ and consider the Faddeev-Popov operator associated to $\mu A$, then find a vector $w$ such that the expectation value of the Faddeev-Popov operator is negative for a sufficiently large $\mu$. Let  $M$ be a compact manifold and $A$ a gauge connection not identically zero on it. There exists an open neighborood $\mathcal{O}$ on $M$ such that $A_{\mu}^{b}\neq 0$ in $\mathcal{O}$ and keeps the same sign. Let $\phi$ be a smooth function with compact support on $\mathcal{O}$ and let it be normalized
\be
\int_{M}|\phi|^2=1.
\ee
Then we can define
\be
B^{a}_{\mu}=\int_{M}\phi^{*}A_{\mu}^{b}\phi\neq 0.
\ee
Let $\hat{k}^{\nu}$ be a unit vector such that $n^{a}=\hat{k}^{\nu}B^{a}_{\mu}\neq 0$. We choose $w^{a}=e^{ikx}\phi(x)u^{a}$, $u$ being a normalized colour vector $u^{a*}u^{a}=1$ and $k=c\hat{k}$. $c$ is a constant at our disposal.
The expectation value of the Faddeev-Popov is made up of two terms
\begin{align}
(w,FP(\mu A)w)=&(w,-\triangle w)+\mu M(A),\\
M(A)=&-(w,[A_{\mu},\partial^{\mu}w]).
\end{align}
The first one is the expectation value of the Laplacian, which is obviously a non-negative number. We now turn to the evaluation of the second one
\be
M(A)=-\int_{M}w^{a}[A_{\mu},\partial^{\mu}w]^{a}=cu^{*a}(if^{abc}n^{b})u^{c}-(\phi u,[A_{\mu},\partial^{\mu}(\phi u)]).
\ee
The matrix $if^{abc}$ is hermitian and has at least a non-vanishing eigenvalue, say $\lambda$, in the case of a unitary gauge group. Then choosing $c$ opposite in sign to $\lambda$ we can obtain $M(A)<0$. Then the proof is finished because we just have to increase $\mu$ until $\mu M(A)$ is larger in magnitude than the expectation value of the Laplace operator.

This theorems remain valid also in the curved case and can be proved exactly in the same way. In fact in order to prove the boundedness of the Gribov region we just need to limit our considerations to $M(A)$. In a curved space partial derivatives are replaced by covariant derivatives, but the form of the expectation value $M(A)$ does not change because $w$ is a scalar.

The boundedness of the Gribov region can be proved also in a more straightforward way\footnote{We owe this proof to Daniel Zwanziger.}. It is sufficient to observe that the term $M(A)$ has a vanishing trace on the color indices, so it has vanishing trace overall. From the vanishing of the trace it follows that the operator in bracket has at least one negative eigenvalue. Call the corresponding eigenvector $w$, so $M(A)=-(w,[A_{\mu},\partial^{\mu}w])< 0$.
\end{subsection}
\begin{subsection}{Norm functional and bifurcation picture}
Given a gauge connection $A$ on $M$, we can define the following functional on the group of local gauge transformations: 
\be
I_{A}[g]=-\int_{M}Tr\left(g^{-1}A g +g^{-1}\partial g\right)^2.
\ee
It follows from the anti-hermiticity of the gauge connection that it is positive definite. The integrand is just the trace of the Euclidean norm of a connection on the same orbit of $A$.
 
Some connections may have a non-trivial stabilizer, \emph{i.e.} a subgroup of the group of local gauge transformations which leaves the connection invariant. A basic example is given by the vacuum $A=0$ on a compact space, whose stabilizer is the group of constant gauge transformations. The functional $I_0[g]$ is degenerate under this group of transformations.

In order to study the critical points of this functional we write the gauge transformation in the form
\be
g=e^{X},
\ee
where $X$ is a non-constant Lie-Algebra element, and expand near the identity \cite{DellAntonioZwanziger:EGO} \cite{VanBaal:MGC}.
\begin{multline}
I_{A}[g]=-\int_{M}Tr(A^2)+2\int_{M}Tr\left(X^{\dagger}\partial A\right)+\int_{M}Tr\left(X^{\dagger}FP(A)X\right)\\
-\frac{1}{3}\int_{M}Tr\left(X^{\dagger}\partial[[A,X],X]\right)+\frac{1}{12}\int_{M}Tr\left([D(A),X][\partial X,X]\right)+o(X^5)
\end{multline}
From this expression we see that the identity is a critical point for the functional if and only if the first order term vanishes, \emph{i.e.} the connection $A$ is transverse. Moreover the identity is a point of relative minimum provided $FP(A)$ is positive-definite. Thus we can give an alternative definition of the first Gribov region as the set of transverse connections whose Faddeev-Popov operator is positive-definite. Though we cannot exclude the existence of more than one minimum along a given gauge orbit. This is equivalent to the statement that there may exist copies even inside the first Gribov region. Moreover the number of minima may vary as we change the orbit, so that the number of copies is not necessarily a constant.
We define the fundamental modular region $\Lambda$  as the set of transverse connections with no copies. In order to pick one and only one representative on each gauge orbit we have to restrict to a subset of the Gribov region made up of connections such that the functional $I_{A}[g]$ attains its \emph{absolute} minimum at the identity. We will suppose with Dell'Antonio and Zwanziger that this minimum is unique along a given gauge orbit \cite{DellAntonioZwanziger:EGO}. In the same paper the authors proved that each gauge orbit passes inside the Gribov region and that the orbits are closed in a suitable topology. The topology of $\Lambda$ is quite complicated as identifications between boundary points may occur.

The orbit space is obtained dividing $\Lambda$ by the group of constant gauge transformations
\be
\Lambda/G=\mathcal{R}/\mathcal{G}.
\ee
We want to stress here an important point: in order to construct the Gribov region and the fundamental modular region we have to get rid of constant gauge transformations and constant Lie-Algebra vectors \cite{VanBaal:MGC}. This eliminates the degeneracy and makes the Faddeev-Popov operator positive-definite in the case of the vacuum $A=0$. If we don't do this we might be tempted to say that the vacuum lies on the Gribov horizon and so it is a configuration banished by the quantum dynamics as in \cite{Zwanziger:BAN}.

On a compact manifold there are solutions of the equation for the copies of the vacuum
\be
\partial^{\mu}(g^{-1}\partial_{\mu}g)=0
\ee
as we can take $\tilde{A}_{\mu}=g^{-1}\partial_{\mu}g$ to be a constant with $g$ homotopically non-trivial. Though this copies are not a problem at all as they lies on the Gribov horizon. The equation
\be
-\partial^{\mu}(\partial_{\mu}\psi+[A_{\mu},\psi])=0
\ee
is solved by any $\psi=g^{-1}Xg$, where $X$ is a \emph{constant} Lie-Algebra vector. Our prescription of discarding constant zero-modes of $FP(A)$ can certainly be satisfied, as $\psi$ cannot be constant for all constant $X$. For example we can take the flat three-torus with coordinates\footnote{Of course this is not a global chart as the torus is not omeomorphic to the rectangle.} $(x,y,z)$ and a gauge transformation $g=e^{\omega z\tau_3}$, with $\tau_{k}=i\frac{\sigma_{k}}{2}$. We can make it periodic by requiring $\omega$ to be an even integer number $\omega=2k$. Then we have $\tilde{A}_{x}=\tilde{A}_{y}=0$ and $\tilde{A}_{z}=\omega\tau_3$. We can choose  for example $X=\tau_{1}$ so that $\psi=\sigma_2\cos(\omega z)-\sigma_1\sin(\omega z)$, which is obviously non-constant. So on a compact space copies of the vacuum are allowed to exist but they lie on the horizons.
\end{subsection}

\begin{subsection}{Gribov-Zwanziger functional integral formula}
It is time to revert to the beginning of our discussion. We started pointing out that the usual way to quantize gauge theories relies on the gauge-fixing procedure, which for non-Abelian theories leads to the Gribov ambiguity. Thus if we want to successfully quantize the theory we have to find a way to get rid of Gribov copies. The hypersurface $\partial A=0$ to which the FDFP technique restrict the domain of functional integration is still full of redundancies. Gribov proposed to restrict to the subset of this hypersurface where the Faddeev-Popov operator is positive definite. Though as we have seen in the previous sections there are copies even inside it. It is possible to avoid overcounting due to these copies by using stochastic quantization as a way to fix the Landau gauge non-perturbatively. We present here the original approach to the problem proposed in the papers by Gribov and Zwanziger. 
%This approach has been recently improved (see (bla bla bla)) to eliminate the effect of the presence of copies inside the Gribov horizon.Though we will not be concerned with this and simply limit our discussion to the restriction of the domain of 
%functional integration to the first Gribov region $\Omega$.
What we have to do is to multiply the integrand appearing in FDFP formula by the characteristic function of $\Omega$. The Heaviside function of the least eigenvalue of $FP(A)$ has exactly the property we are looking for, being equal to $1$ inside $\Omega$ while it vanishes identically outside of it. We can thus express the average of an observable $\phi(A)$ using Gribov-Zwanziger formula \cite{Zwanziger:BAN,Gribov:78,Sorella:rev}
\be\label{Gribov-Zwanziger}
\langle\phi\rangle=\int \mathcal{D}A\; N(A)^{-1} \mbox{det}_{+}(FP(A)) \delta(\partial A)e^{-S[A]}\phi(A).
\ee
Here $N(A)$ is the number of copies of $A$ inside $\Omega$ and $\mbox{det}_{+}(FP(A))=\theta(\lambda_{0})\mbox{det}(FP(A))$ is the term implementing the restriction.
There is no reason to expect $N(A)$ to be a constant, so in order to get sensible results from this formula we should find a better way to express it.

Following references \cite{Gribov:78,Sorella:rev} we can nevertheless treat $N(A)$ as if it were a constant and get a flavour of what kind of effects arise from the presence of the horizon term, \emph{i.e.} the Heaviside function. The argument is heuristic and goes as follows. We can express the FP determinant as an integral over ghost fields
\be
\mbox{det}(FP(A))=\int\mathcal{D}c\mathcal{D}\bar{c}e^{-\int_{M}\bar{c}FP(A) c}.
\ee
We can study the dynamics of the ghost treating $A$ as a background field. The ghost propagator is the inverse of $FP(A)$, so it blows up when $A$ reaches the horizon. The point now is to understand where this infinite comes from. We know from perturbation theory that the ghost propagator, after integration over the background field is performed and renormalization is done, has the expression:
\be
\mathcal{G}(k)=\frac{1}{k^2\left(1-\frac{11g^{2}C_2}{48\pi^2}\ln\left(\frac{\Lambda^2}{k^2}\right)\right)^{\frac{3}{22}\left(\frac{3}{2}-\frac{\alpha}{2}\right)}}.
\ee
For $\alpha<3$ this function has two poles: one at $k^2=0$ and the other one at a value of $k_0^2$ such that the term in round brackets vanishes. Though $k_0^2$ cannot be non-zero, otherwise for smaller values $k^2<k_0^2$ the propagator $\mathcal{G}(k)$ is complex, implying that $FP(A)$ is not positive definite\footnote{This is the observation made by Gribov. Though we can also observe that $\mathcal{G}(k)$ cannot be complex because of the self-adjointness property of $FP(A)$}, meaning that we have left $\Omega$. Thus the only pole that the ghost propagator can have is at zero momentum, where the ghost \emph{feels} the fields on the horizon. We can make this argument a concise statement by expressing the ghost propagator in the presence of a background field in the form
\be
\mathcal{G}(k,A)=\frac{1}{k^2\left(1-\sigma(k,A)\right)},
\ee
 then  requiring that the \emph{no-pole condition} be satisfied
\be
\sigma(k,A)<1.
\ee
To calculate the function $\sigma(k,A)$ we turn to perturbation theory. We calculate the colour singlet ghost propagator up to third order terms in the coupling constant $g$. To have a better control of the quantities involved in the various formulae we do our calculations in a box of volume $V$. Eventually $V$ is allowed to become infinite. 
\begin{center}
\includegraphics[width=0.6\columnwidth]{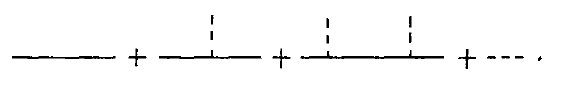}
\end{center}
The first order term is zero in the Landau gauge so to second order we have
\begin{multline}
\mathcal{G}(k,A)=\frac{\delta^{ab}}{N^2-1}\langle \bar{c}^{a}c^{b}\rangle=\\
=\frac{1}{k^2}\left(1+\frac{N}{N^2-1}\frac{1}{k^2}\sum_{p}\frac{(k_{\mu}-p_{\mu})p_{\nu}}{(k-p)^2}\frac{A^{a}_{\mu}(-p)A^{a}_{\nu}(p)}{(p)}\right).
\end{multline}
$N^2-1$ is the number of isotopical states in the adjoint representation of $SU(N)$ and it appears in the formula just because we are averaging over  all possible colour states. 
Assuming the second order correction to be small, which is correct because we are doing perturbation theory and calculating only tree-diagrams, we have
\be
\sigma(k,A)=\frac{1}{k^2}\frac{N}{N^2-1}\frac{1}{k^2}\sum_{p}\frac{(k_{\mu}-p_{\mu})p_{\nu}}{(k-p)^2}A^{a}_{\mu}(-p)A^{a}_{\nu}(p).
\ee
We are considering the Landau gauge so the potential is transverse
\be
q^{\mu}A^{a}_{\mu}=0.
\ee
From this equation follows that the product $A^{a}_{\mu}(-q)A^{a}_{\nu}(q)$ is indeed a projection operator onto the space orthogonal to $q^{\mu}$.
\be
q^{\mu}A^{a}_{\mu}(-q)A^{a}_{\nu}(q)=A^{a}_{\mu}(-q)A^{a}_{\nu}(q)q^{\nu}=0
\ee
The most general form for such a tensor is
\be
A^{a}_{\mu}(-q)A^{a}_{\nu}(q)=C\left(\delta_{\mu\nu}-\frac{q_{\mu}q_{\nu}}{q^2}\right).
\ee
Contracting both sides of this expression with the Kronecker delta $\delta_{\mu\nu}$ we have $C=\frac{1}{3}A^{a}_{\mu}(-q)A^{a}_{\mu}(q)$. Inserting this expression into that for $\sigma(k,A)$ we have
\begin{align}\label{ESPRESSIONE SIGMA(K)}
\sigma(k,A)=&\frac{1}{3}\frac{N}{N^2-1}\frac{k_{\mu}k_{\nu}}{k^2}I^{\mu\nu}(k),\\
I^{\mu\nu}(k)=&\frac{1}{V}\sum_{q}\frac{1}{(k-q)^2}\left(A^{a}_{\lambda}(q)A^{a}_{\lambda}(-q)\right)\left(\delta_{\mu\nu}-\frac{q_{\mu}q_{\nu}}{q^2}\right).
\end{align}
We will show later on that the quantity $A^{a}_{\mu}(q)A^{a}_{\mu}(-q)$ decreases as the inverse square of $q$ as $q$ increases. Then $\sigma(k,A)$ is decreasing as a function of $k$ and we can express the no-pole condition as
\be
\sigma(0,A)<1.
\ee
Putting $k=0$ in equation (\ref{ESPRESSIONE SIGMA(K)}) we obtain
\be
\sigma(0,A)=\frac{1}{4V}\frac{N}{N^2-1}\sum_{q}\frac{1}{q^2}\left(A^{a}_{\mu}(q)A^{a}_{\mu}(-q)\right).
\ee
We use the no-pole condition as the argument of the Heaviside function in the functional integral formula (\ref{Gribov-Zwanziger}), so that the partition function is given by the formula
\be
Z=\int \mathcal{D}A\; \mbox{det}(FP(A)) \delta(\partial A)e^{-S[A]} \theta(1-\sigma(0,A)).
\ee
Using the integral representation of the Heaviside function
\be
\theta(x)=\int \frac{d\beta}{2\pi i\beta} e^{\beta x},
\ee
we may rewrite the partition function as follows
\be
Z=\int \mathcal{D}A\int \frac{d\beta}{2\pi i\beta}\; \mbox{det}(FP(A)) \delta(\partial A)e^{-S[A]} e^{\beta(1-\sigma(0,A))}.
\ee
%Qui ho aggiunto un po' di chiacchiere per giustificare lo "scarto" del determinante
For the sake of simplicity we will omit the determinant in this formula as in \cite{Gribov:78}. We want to stress that in doing so we are not neglecting the coupling of the gauge field to the ghost, as this information is already in the no-pole condition. Passing to the Fourier representation we have
\be
Z=\int \frac{d\beta}{2\pi i\beta}e^{\beta}\int \mathcal{D}A\; \exp{\left(-\frac{1}{2g^2}\sum_{q}q^2|A^{a}_{\mu}(q)|^2-\frac{N\beta}{4(N^2-1)V}\sum_{q}\frac{|A^{a}_{\mu}(q)|^2}{q^2}\right)}.
\ee
Performing the functional integration on $A$
\be
Z=\int \frac{d\beta}{2\pi i\beta} e^{\beta}\prod_{q}\frac{1}{\left(q^2+\frac{\beta g^2}{q^2 V}\right)^{\frac{3}{2}(N^2-1)}}.
\ee
The integral over $\beta$ can be evaluated at the saddle-point $\beta_0$, given by the equation
\be
1-\frac{1}{\beta_0}-\frac{3Ng^2}{4V}\sum_{q}\left(\frac{1}{q^4}+\frac{Ng^2\beta_0}{(N^2-1)2V}\right)=0.
\ee
Neglecting the term $\frac{1}{\beta_0}$ in the thermodynamic limit $V\to\infty$ and defining
\be
\gamma^4=\frac{Ng^2\beta}{(N^2-1)2V},
\ee
we get the \emph{gap equation}
\be
\frac{3Ng^2}{4}\int \frac{d^4q}{(2\pi)^4}\frac{1}{q^4+\gamma^4}=1.
\ee
This equation, which of course must be understood in a regularized sense, gives the value $\beta_0$ we should use in the functional integral in order to incorporate corrections due do the presence of the horizon term.
When we do this we obtain for the gluon propagator
\be\label{GLUON PROPAGATOR}
D_{\mu\nu}^{ab}(q)=\langle A^{a}_{\mu}(-q)A^{b}_{\nu}(q)\rangle=\delta^{ab}g^2\frac{q^2}{q^4+\gamma^4}\left(\delta_{\mu\nu}-\frac{q_{\mu}q_{\nu}}{q^2}\right).
\ee
We now have to calculate the ghost propagator after integrating over the gauge field.
\be
\sigma(k)=\frac{1}{3V}\frac{N}{N^2-1}\frac{k_{\mu}k_{\nu}}{k^2}\sum_{q}\frac{1}{(k-q)^2}\langle A^{a}_{\lambda}(q)A^{a}_{\lambda}(-q)\rangle\left(\delta_{\mu\nu}-\frac{q_{\mu}q_{\nu}}{q^2}\right)
\ee
We thus have, making use of the gap equation,
\be
(1-\sigma(k))=Ng^{2}\frac{k^{\mu}k^{\nu}}{k^2}\mathcal{P}_{\mu\nu}(k),
\ee
where
\be
\mathcal{P}_{\mu\nu}(k)=\int\frac{d^4q}{(2\pi^4)}\frac{k^2-2kq}{(k-q)^2}\frac{q^2}{q^4+\gamma^4}\left(\delta_{\mu\nu}-\frac{q_{\mu}q_{\nu}}{q^2}\right).
\ee
We evaluate it to the lowest non-vanishing order near $k=0$
\be
\mathcal{P}_{\mu\nu}(k)=\delta_{\mu\nu}k^2\frac{3}{128\gamma^4}+o(k^4).
\ee
So the ghost propagator behaviour at low momenta is given by the asymptotic formula
\be\label{GHOST PROPAGATOR}
\mathcal{G}(k)\approx\frac{128\pi^2\gamma^2}{3Ng^2}\frac{1}{k^4}.
\ee
We see that the gluon propagator (\ref{GLUON PROPAGATOR}) is suppressed in the infrared where it vanishes instead of having a pole. The reason for this is the appearance of a mass term due to the presence of the Gribov horizon. This may be an evidence that actually the gluons are not asymptotic states, \emph{i.e.} there is a mass gap in the theory. On the other hand the ghost propagator (\ref{GHOST PROPAGATOR}) is enhanced in the infrared. The corrections contain a non-perturbative parameter $\gamma$ which appears as a result of the no-pole condition. The method used is apparently not rigorous and is not fully non-perturbative. In fact perturbation theory is used to some extent and the non-perturbative ingredient results from the Heaviside function.

%%calcoli di Sorella
%%soppressione infrarossa del propagatore del gluone e modifica di quello del ghost.

%%teorema di singer e discorsi sulla topologia non banale dello spazio delle orbite
%%
%%copie di gribov e modi zero dell'operatore di FP. Orizzonte di gribov
%%
%%restrizione suggerita da gribov
%%
%%teoremi di zwanziger e dell'antonio (ogni orbita di gauge passa all'interno dell'orizzonte, convessità e limitatezza in ogni direzione della regione di gribov)
%%
%%picture di biforcazione di van baal (il suo argomento sulla teoria di morse non regge?)
%%
%%
%%
%%
%%
%%
%%
\end{subsection}
\end{section}
\begin{section}{Some aspects of Yang-Mills theory in curved space}
A unified theory of all the interactions is not yet available. We have three forces (\emph{i.e.} the strong, the electromagnetic and the weak force) described by a quantum theory and the gravitational force, far weaker than the others, which still endures any attempt to quantize it. If our aim is that of understanding Physics at the energies available in collider experiments, we can simply ignore any gravitational effect. Nevertheless at a more fundamental level we cannot ignore the presence of gravitation, however small its effects may be. The study of quantum field theories on curved backgrounds is a first step in this direction. The adopted scheme is spurious: the gravitational field is treated as a classical background on which quantum dynamics takes place. Then Einstein equations are used to calculate the back-reaction of quantum fields on the background.
\be
G_{\mu\nu}=8\pi T_{\mu\nu}
\ee
Thus the principal object in this scheme is the regularized and renormalized stress-energy tensor $T_{\mu\nu}$. We are not going into details of all the problems involved in a formulation of quantum field theory on curved backgrounds. However it is important to observe that all theories which are renormalizable in flat space are still renormalizable when formulated in curved spaces \cite{Wald:QFTCS}. If we regard also the gravitational field as dynamical we can see that the curvature of the background breaks the celebrated perturbative renormalizability of Yang-Mills theory, which remains valid only in Minkowskian spacetime \cite{DeWitt}. 

We present some recent results \cite{Canfora} obtained in the study of the problem of Gribov copies in curved spaces. A brief exposition is given of an algorithmic technique which allows one to calculate the heat-kernel expansion in the non-minimal case \cite{Gusynin}. The possibility that gravity may modify the gauge field propagator in the infrared is discussed.
\begin{subsection}{Spherically symmetric spacetimes}
We have seen in the previous chapter that the Gribov ambiguity must be studied in Euclidean space. Euclidean space is obtained by a Wick rotation from Minkowski space. However a metric with Lorentzian signature cannot be turned into a Euclidean one by a Wick rotation in general. This can be done only in particular cases. Locally we can choose a system of coordinates $\{x^{\mu}\}$ such that the coordinate lines $x^0$ are timelike, while the others are spacelike. Then the squared line element can be written as follows
\be
ds^2=g_{00}(dx^0)^2+g_{0j}dx^0dx^{j}+g_{jk}dx^{j}dx^{k},
\ee
where $g_{00}<0$. All the components of the metric $g_{\mu\nu}$ may depend on coordinates in general. Under a Wick rotation $x^{0}\to ix^{0}$ we have
\be
d\tilde{s}^2=-g_{00}(dx^0)^2+ig_{0j}dx^0dx^{j}+g_{jk}dx^{j}dx^{k}.
\ee
In order for this to be a real Euclidean squared line element, we have to impose some requirements on the components of the metric $g_{\mu\nu}$. They have to be analytic in $x^0$ in order to perform the extension to the imaginary axis, where $g_{0j}$ is a pure imaginary, $g_{00}$ is negative and the other diagonal elements $g_{jj}$ are positive. However the positive definiteness of the new metric always has to be checked. By choosing a synchronous reference frame we can always make $g_{0j}$ vanish and the other components independent of $x^0$. Though this is a local property and cannot be true globally for any spacetime.

When there is a timelike Killing vector field, the spacetime is said to be \emph{stationary}. When there is a three-dimensional spacelike surface to which the Killing vector field is orthogonal the spacetime is said to be \emph{static}. The integral curves of the Killing field can be used as $x^0$ coordinate lines. In the case of a stationary spacetime the metric components do not depend on $x_0$ and $g_{0j}$ vanish, so that any static spacetime can be euclideanized. For this reason we will restrict our attention to this class of spacetimes.

In order for the Euclidean and the Lorentzian formulations of field theory to be equivalent, we must also require that the Euclidean correlation functions are obtained by analytic prolongation of the Lorentzian ones. The prolongation should be made without encountering any singularity \cite{Strocchi}.

Following \cite{Canfora} we will consider static, spherically symmetric spacetimes in four dimensions. After the Wick rotation is performed the squared line element is given by
\be
ds^2=g^2(r)dt^2+f^2(r)dr^2+r^2d\Omega^2.
\ee
Here $t$ is the Euclidean time, $r$ a radial coordinate and $\Omega$ the solid angle. On such spacetimes there exists a natural separation between space and time so that we have a curved space analogue of the Coulomb gauge
\begin{align}
A_{0}^a=&0\\
\nabla^{i}A_{i}^{a}=&0\label{Coulomb gauge}.
\end{align}
$\nabla$ denotes the metric-compatible connection and $a$ denotes a colour index. We see that all that matters in the Coulomb gauge is the metric on the spacelike hypersurface $\Sigma$ orthogonal to the Killing vector field $\frac{\partial}{\partial t}$.
\be
ds^2_{\Sigma}=(g_{\Sigma})_{ij}dx^{i}dx^{j}=f^2(r)dr^2+r^2d\Omega^2
\ee
Equation (\ref{Coulomb gauge}) can then be rewritten as
\be
\nabla^{i}A_{i}^{a}=\frac{1}{\sqrt{\mbox{det}g_{\Sigma}}}\partial_{i}\left(\sqrt{\mbox{det}g_{\Sigma}}(g_{\Sigma})^{ij}A_{j}^{a}\right)=0.
\ee
As in the flat case we consider the gauge group $SU(2)$. Let us consider a gauge transformation of a form similar to that we considered in (\ref{Gribov pendulum})
\be\label{gauge transf}
U=e^{i\frac{\alpha(r)}{2}x^{i}\sigma_{i}}.
\ee
Here $\sigma_{i}$ are flat Pauli matrices and $x^{i}$ is a unit radial vector. The combination $x^{i}\sigma_{i}$ is not a true scalar product, as $\sigma_{i}$ does not transform as a vector. We need to define it with respect to a particular reference frame. At a given point $p\in\Sigma$ with radial coordinates $(\theta,\phi)$ we take a Cartesian frame in $T_{p}\Sigma$ such that $x^{i}$ has components $(\sin\theta\cos\phi,\sin\theta\sin\phi,\cos\theta)$. To each of the Cartesian axes we associate one of the three Pauli matrices. Thus we have
\be
x^{i}\sigma_{i}=\sigma_{1}\sin\theta\cos\phi+\sigma_{2}\sin\theta\sin\phi+\sigma_{3}\cos\theta.
\ee
Using the algebra of Pauli matrices we have
\be
(x^{i}\sigma_{i})(x^{j}\sigma_{j})=(x^{i}x_{i})\mathbb{I}=\mathbb{I}.
\ee
We choose a gauge field whose expression in this Cartesian reference system is
\be
A_{h}=i\varepsilon_{hjk}\frac{x^{j}\sigma^{k}}{r^2} \varphi(r).
\ee
The Cartesian components have the explicit form
\begin{align}
A_1=&\frac{i}{r^2}\phi(r)(\sin\theta\sin\phi\sigma_3-\cos\theta\sigma_2)\\
A_2=&\frac{i}{r^2}\phi(r)(\cos\theta\sigma_1-\sin\theta\cos\theta\sigma_3)\\
A_3=&\frac{i}{r^2}\phi(r)(\sin\theta\cos\phi-\sin\theta\sin\phi\sigma_1)
\end{align}
Changing to the radial coordinate system we obtain
\begin{align}\label{gauge conn}
A_{r}=&0\\
A_{\theta}=&\frac{i}{r}\phi(r)(\sin\phi\sigma_1-\cos\phi\sigma_2)\\
A_{\phi}=&\frac{i}{r}\phi(r)\sin\theta(\cos\theta\cos\phi\sigma_1+\cos\theta\sin\phi\sigma_2-\sin\theta\sigma_3)
\end{align}
It is immediate to verify that this field is divergenceless for any choice of the function $\varphi(r)$.

The gauge group element $U$ maps the transverse gauge field $A$ into another field $A'$, given as usual by the formula
\be
A'=U^{\dagger}AU+U^{\dagger}\partial U.
\ee
If $A'$ satisfies the Coulomb gauge condition we have
\be\label{grib eq}
\nabla^{j}\left(U^{\dagger}A_{j}U+U^{\dagger}\partial_{j} U\right)=0.
\ee
This is an equation in the unknown function $\alpha(r)$. Boundary conditions have to be imposed on it as in (\ref{Gribov pendulum}). 
Inserting equations (\ref{gauge transf}) and (\ref{gauge conn}) into (\ref{grib eq}) we obtain
\be\label{Curved pendulum}
\begin{pmatrix}
\cos\theta & \sin\theta e^{-i\phi}\\
\sin\theta e^{i\phi} & -\cos\theta
\end{pmatrix}
\left(\left(\frac{r^2}{2f}\alpha'\right)' -\left(1+\frac{2\varphi}{r}\right)f\sin\alpha\right)=0.
\ee
We observe that (\ref{grib eq}) is \emph{a priori} a system of four coupled differential equations. The ansatz (\ref{gauge transf}) simplifies a lot the problem and it turns out that (\ref{grib eq}) leads just to an equation in $\alpha(r)$. This turns out to be a more general feature of the equation for Gribov copies and it can be exploited to study the problem also in cases with no spherical symmetry \cite{Canfora:Hedgehog}.

For $f=1$ equation (\ref{Curved pendulum}) obviously reduces to the flat case Gribov pendulum equation. For a generic $f$ and with respect to the variable $t=\log r$ this is the equation of a pendulum with time dependent damping and gravitational field.

In \cite{Canfora} it is shown that the naive vacuum $A=0$ has no strong copies in spherically symmetric spaces with a $C^1$ metric. Copies of the vacuum are allowed to exist when derivatives of the metric components are discontinuous and the spacetime is asymptotically anti-de Sitter. When copies of the vacuum are allowed to exist, even the definition of perturbation theory becomes troublesome.

On a general spacetime equation (\ref{grib eq}) can be recast in the form
\be
U^{\dagger}\Box U+\nabla^{\mu}U^{\dagger}\nabla_{\mu}U+U^{\dagger}A_{\mu}\nabla^{\mu}U+\nabla^{\mu}U^{\dagger}A_{\mu}U=0.
\ee
This is a quasi-linear second order partial differential equation in the unknown matrix $U$. Only natural operators are involved such as the Laplacian and the gradient. Another useful form is the following \cite{Gribov:Hamiltonian}:
\be
D_{\mu}(A)\left((\nabla^{\mu}U)U^{\dagger}\right)=0.
\ee
In the same reference \cite{Gribov:Hamiltonian} it is shown that the equation for Gribov copies in the flat case may be derived from an action functional. The generalization to the curved case is straightforward, the functional being
\be
W=\int_{M}\mbox{Tr}\left(\nabla^{\mu}U^{\dagger}\nabla_{\mu}U-2(\nabla^{\mu}U)U^{\dagger}A_{\mu}\right).
\ee
Equation (\ref{grib eq}) is the Euler-Lagrange equation for the functional $W$ subject to the constraint $U^{\dagger}U=\mathbb{I}$.
\end{subsection}
\begin{subsection}{Effective Schroedinger equation method}
Upon changing variables in equation (\ref{Curved pendulum})
\be\label{change tau}
\frac{\partial\tau}{\partial r}=\frac{f}{r^2},
\ee
we have that it assumes the following form:
\be\label{equazione copie tipo schrodinger}
\frac{\partial^2\alpha}{\partial\tau^2}=2r^2\left(1+2\frac{\varphi}{r}\right)\sin\alpha.
\ee
Here $r$ must  be understood as a function of $\tau$, according to (\ref{change tau}). We can define
\be
\tilde\varphi(\tau)=\frac{\varphi}{r}.
\ee
When $f=1$ we recover the case of Minkowski spacetime. In this case the equation for the copies is the following
\be
\frac{\partial^2\alpha}{\partial\tau^2}=2\left(1+2\tilde\varphi(\tau)\right)\frac{1}{\tau^2}\sin\alpha.
\ee
Another interesting case is represented by anti-de Sitter spacetime. The expression for the function $f$ is given in this case by
\be
f(r)=\frac{1}{\sqrt{1+\frac{r^2}{l^2}}}.
\ee
In this case equation (\ref{equazione copie tipo schrodinger}) assumes the following form
\be
\frac{\partial^2\alpha}{\partial\tau^2}=2\left(1+2\tilde\varphi(\tau)\right)\frac{l^2}{l^2\tau^2-1}\sin\alpha.
\ee

The linearized version of (\ref{equazione copie tipo schrodinger}) may be of help in studying a particular class of Gribov copies for a generic $\varphi$ \cite{Canfora}. We have seen in (\ref{Gribov pendulum}) that $\alpha$ must limit to a $2n\pi$ at the origin and to $m\pi$ at infinity. In order for the linearization to make sense me must require that these multiples are the same. We notice that, setting $\bar\alpha=\alpha-2n\pi$, we have
\be
\sin\alpha=\bar\alpha+o(\bar\alpha^3), \mbox{for $\alpha\simeq 2n\pi$}.
\ee
Then (\ref{equazione copie tipo schrodinger}) becomes
\be\label{pendolo tipo Schroedinger}
\frac{\partial^2\bar\alpha}{\partial\tau^2}=\left(1+2\tilde\varphi(\tau)\right) V(\tau)\bar\alpha.
\ee
In Minkowski space we have
\be
V_{flat}(\tau)=2\frac{1}{\tau^2},
\ee
while in anti-de Sitter
\be
V_{adS}(\tau)=2\frac{l^2}{l^2\tau^2-1}.
\ee
In \cite{Canfora} it is proposed to treat equation (\ref{pendolo tipo Schroedinger}) as a stationary Schroedinger equation for zero energy. $V(\tau)$ is then interpreted as a potential whose shape is modified by the presence of $\tilde\varphi$.

It is important to stress that we are allowed to interpret (\ref{pendolo tipo Schroedinger}) as a Schroedinger equation if we can prove that the corresponding Hamiltonian operator is self-adjoint on a suitable domain. %We want to point out that the hamiltonian corresponding to the anti-de Sitter case in the case of a zero background connection
%\be
%H=-\frac{\partial^2}{\partial\tau^2}+V_{adS}(\tau)
%\ee
%is not essentially self-adjoint on the domain of $C^{\infty}_0$ defined on the whole range of the variable $\tau$. In order to prove this, we first have to change variables so that the new variable is defined on the positive half-line.
There is a criterion due to Weyl which allows us to check if a given operator of the Schroedinger type (\emph{i.e.} Laplacian plus potential) is essentially self-adjoint on $C^{\infty}_0(\mathbb{R}^{+})$ \cite{ReedSimon}. First a few definitions are in order. The potential $V(x)$ is in the limit circle\footnote{The reader interested in the origin of this terminology may consult \cite{CoddingtonLevinson}} case at infinity (respectively at zero) if for some $\lambda$ all solutions of the equation
\be
-u''+V(x)u=\lambda u
\ee
are square integrable at infinity (respectively at zero). It is possible to show that if this property holds for some $\lambda$, it is true for all $\lambda$. If $V(x)$ is not in the limit circle case it is said to be in the limit point case. Weyl's limit point-limit circle criterion states that a Schroedinger type Hamiltonian is essentially self-adjoint on the half-line if and only if the potential is in the limit point case both at zero and at infinity.

There are two theorems which make the verification of these properties straightforward.
Let V(x) be differentiable on $(0,\infty)$ and bounded above by $K$ on $[1,\infty)$. Suppose that
\begin{itemize}
\item $$\int_{1}^{\infty}\frac{dx}{\sqrt{K-V(x)}}=\infty$$
\item $$\frac{V'(x)}{|V(x)|^{\frac{3}{2}}}\mbox{is bounded near infinity.}$$
\end{itemize}
Then $V(x)$ is in the limit point case at infinity.

Let $V$ be continuous and positive near zero. If $V(x)\geq \frac{3}{4x^2}$ near zero then $V(x)$ is in the limit point case at zero. If for some $\varepsilon>0$, $V(x)\leq (\frac{3}{4}-\varepsilon)x^{-2}$ near zero, then $V$ is in the limit circle case.

In the flat case the hamiltonian can be mapped in another one defined over the positive half-line by the transformation $\tau\to x$. It is immediate to verify that the new potential is in the limit point case both at infinity and at zero, so the Hamiltonian is self-adjoint by Weyl's criterion.

In the anti-de Sitter case the variable $\tau$ ranges in the interval $(-\infty,-\frac{1}{l})$. We can map this interval on $(0,\infty)$ by the transformation $x=-\tau-\frac{1}{l}$. The second derivative operator does not change form under this transformation, but the potential assumes the following expression
\be
V_{adS}(x)=\frac{2}{x^2+2x}.
\ee
It is in the limit point case at infinity but not at zero. Then it follows again by Weyl's criterion that the Hamiltonian is not essentially self-adjoint on the domain $C^{\infty}_0(\mathbb{R}^{+})$. Thus the self-adjoint extension is not unique and one has to be careful in specifying the right extension corresponding to the physical problem. We have already encountered a problem of this kind when we studied the Faddeev-Popov operator on a torus (\ref{FP}). The strong boundary conditions on $\alpha$ lead to Dirichlet boundary conditions on $\bar\alpha$. With these boundary conditions the Hamiltonian operator is self-adjoint. At present we do not know if other self-adjoint extensions can be of physical interest in the anti-de Sitter case.
\end{subsection}
\begin{subsection}{Heat-kernel techniques}\label{heat}
Let $H$ be a positive elliptic operator of degree $2r$ on a $n$-dimensional Riemannian manifold $M$ endowed with a Riemannian structure and a bundle structure.
The heat-kernel of $H$ is defined as the solution for $t>0$ of the equation
\be
(\partial_{t}+H)K=0,
\ee
subject to the initial condition
\be
\lim_{t\to 0}K(x,x',t)=\delta(x,x').
\ee
It can be formally expressed through the operator  $e^{-tH}$. One has indeed
\be
K(x,x',t)\equiv\langle x|e^{-tH}|x'\rangle=\sum_{n}e^{-t\lambda_{n}}f_{n}(x)\bar{f}_{n}(x').
\ee
The heat-kernel diagonal has an asymptotic expansion for $t\to 0^{+}$
\be
\langle x|e^{-tH}|x\rangle=\sum_{m=0}^{\infty}E_{m}(x|H)t^{\frac{m-n}{2r}}.
\ee
Hadamardm, Minakshisundaram and Pleijel worked out this expansion studying the Laplacian on Riemannian manifold. The applications in quantum field theory were studied by Schwinger, De Witt and Seeley.
%In the Lorentzian case the operator $H$ is hyperbolic and the corresponding asymptotic series is called the expansion.
The asymptotic expansion of the heat-kernel has a plethora of applications, both in Mathematics and in Physics. It is used \emph{e.g.} in the proof of the index theorem, in spectral geometry and in anomaly calculations. It is a general result indeed that the trace (taken over internal and spacetime indices) of heat-kernel coefficients is expressed in terms of geometric invariants.

The spectral zeta function of the operator $H$ is defined according to the formula
\be\label{zeta def}
\zeta_{H}(s)=\sum_{j}\frac{1}{\lambda_{j}^{s}}.
\ee
In general this defines an analytic function in a subset of the complex plane. If $H$ is the Laplacian this is an analytic function for $\Re s>\frac{n}{2}$. For a generic elliptic operator the lower bound for the real part of $s$ will depend on the order of the operator and the manifold dimension. There is a remarkable relation between the heat-kernel and the zeta function of the operator $H$. The latter is indeed the Mellin transform of the former.
\be\label{Mellin}
\zeta_{H}(s)=\frac{1}{\Gamma(s)}\int_{0}^{\infty}dt\; t^{s-1}\int d^{n}x\; K(x,x,t)
\ee
Formula (\ref{Mellin}) actually provides an analytic extension of $\zeta(s)$. The regularized functional determinant of the operator $H$ can be defined in terms of the derivative of the zeta function at the point $s=0$
\be
\mbox{det}H=e^{-\zeta'(0)}.
\ee
%This formula is of fundamental importance for the study of effective theories at one-loop level,
The regularized determinant encodes all the information about the one-loop dynamics, as it can be seen by a saddle-point expansion of the functional integral.

It is possible to give another useful expression for the heat-kernel in terms of the resolvent\footnote{The expression for the resolvent $(H-\lambda)^{-1}$ should be properly written as $(H-\lambda\mathbb{I})^{-1}$, where $\mathbb{I}$ is the identity operator.} $(H-\lambda)^{-1}$ by means of the Cauchy formula
\be
e^{-tH}=\int_{C}\frac{id\lambda}{2\pi}e^{-t\lambda}(H-\lambda)^{-1}.
\ee
The path $C$ encloses the spectrum of the operator $H$. We notice that, taking $\lambda=0$, the resolvent reduces to the Green function. It would be useful to work out an expression for the resolvent in momentum space. Though we notice that in the curved case the ordinary definition of the Fourier transform gives rise to an object that depends on the choice of coordinates. However it is possible to avoid this problem, giving an alternative definition of the Fourier transform that is invariant under general coordinate transformations. This is the basis of Widom's pseudodifferential calculus and constitutes the starting point in Gusynin's approach to the calculation of the heat-kernel coefficients.

Following \cite{Gusynin} we can express the matrix elements of the resolvent of a positive elliptic operator $H$ by means of the formula
\be\label{Widom}
G(x,x',\lambda)\equiv\langle x|\frac{1}{H-\lambda}|x'\rangle=\int\frac{d^{n}k}{(2\pi)^{n}\sqrt{g(x')}}e^{il(x,x',k)}\sigma(x,x',k;\lambda).
\ee
Here $l(x,x',\lambda)$ is a biscalar under general coordinates transformations and constitutes a generalization of the phase $k_{\mu}(x-x')^{\mu}$ used in the flat case. This expression for the resolvent is manifestly covariant.
The generalization of the linearity property of $l(x,x',\lambda)$ valid in the flat case is obtained by requiring that symmetric combinations of covariant derivatives vanish in the coincidence limit
\be\label{cov der symm}
\nabla_{(\mu_1}\nabla_{\mu_2}\dots\nabla_{\mu_{m})}l|_{x=x'}\equiv\left[\nabla_{(\mu_1}\nabla_{\mu_2}\dots\nabla_{\mu_{m})}l\right]=0, \hspace{1em}\mbox{$m\neq 1$}.
\ee
along with
\be\label{prima cov der}
[\nabla_{\mu}l]=k_{\mu}.
\ee
The square bracket denotes the coincidence limit and symmetrization must be understood over indices enclosed by the round brackets. These conditions are sufficient to determine $l(x,x',\lambda)$ in a neighborhood of the point $x'$. Indeed the commutator of covariant derivatives acts on tensors as follows:
\begin{align}
[\nabla_{\mu},\nabla_{\nu}]f^{\nu_{1}\dots\nu_{k}}_{\mu_{1}\dots\mu_{n}}=&R_{\mu\nu\lambda}^{\quad\;\nu_{i}}f^{\nu_{1}\dots\nu_{i-1}\lambda\nu_{i+1}\dots\nu_{k}}_{\mu_{1}\dots\mu_{n}}-R_{\mu\nu\mu_{i}}^{\quad\;\lambda}f^{\nu_{1}\dots\nu_{k}}_{\mu_{1}\dots\mu_{i-1}\lambda\mu_{i+1}\dots\mu_{n}}+\\
+&T^{\lambda}_{\mu\nu}\nabla_{\lambda}f^{\nu_{1}\dots\nu_{k}}_{\mu_{1}\dots\mu_{n}}+W_{\mu\nu}f^{\nu_{1}\dots\nu_{k}}_{\mu_{1}\dots\mu_{n}}.
\end{align}
Using this formula and (\ref{cov der symm}), (\ref{prima cov der}) one can find the covariant derivatives of $l$ in the coincidence limit. 
The resolvent is a solution of the equation
\be
(H(x,\nabla_{\mu})-\lambda)G(x,x',\lambda)=\frac{1}{\sqrt{g}}\delta(x-x'),
\ee
subject to the boundary conditions which define the domain of the operator $H$. Putting expression (\ref{Widom}) into it one gets the equation
\be\label{equazione per sigma}
(H(x,\nabla_{\mu}+i\nabla_{\mu}l)-\lambda)\sigma(x,x',k;\lambda)=I(x,x').
\ee
The function $I(x,x')$ is a biscalar and is defined by conditions analogous to those satisfied by $l(x,x',k)$
\begin{align}
[I]=&E,\\
\left[\nabla_{(\mu_1}\nabla_{\mu_2}\dots\nabla_{\mu_{m})}I\right]=&0,
\end{align}
where $E$ is the unit matrix.

One then introduces an auxiliary parameter $\varepsilon$, which will be set to one at the end of the calculations, and expand $\sigma(x,x',k;\lambda)$ and $H(x,\nabla_{\mu}+i\nabla_{\mu}l)$ following the rules $l\to l/\varepsilon$, $\lambda\to\lambda/\varepsilon^{2r}$.
\begin{align}
\sigma_{\varepsilon}(x,x',k;\lambda)=&\sum_{m=0}^{\infty}\varepsilon^{2r+m}\sigma_{m}(x,x',k;\lambda)\\
H(x,\nabla_{\mu}+i\nabla_{\mu}l/\varepsilon)=&\sum_{m=0}^{2r}\varepsilon^{-2r+m}A_{m}(x,\nabla_{\mu},\nabla_{\mu}l)
\end{align}
Substituting these expansions into equation (\ref{equazione per sigma}) and collecting terms of the same order in $\varepsilon$ one gets a system of equations for the coefficients $\sigma_{m}$ which can be solved recursively.

The diagonal matrix elements of the heat-kernel are then given by the relation
\be
K(x,x,t)=\sum_{m=0}^{\infty}\int \frac{d^{n}k}{(2\pi)^{n}\sqrt{g}}\int_{C}\frac{id\lambda}{2\pi}e^{-t\lambda}[\sigma_{m}](x,k,\lambda).
\ee

One finds from the recursion relations satisfied by the coefficients $\sigma_{m}$, that these coefficients are generalized homogeneous functions in the variables $(k,\lambda)$
\be
[\sigma_{m}](x,tk,t^{2r}\lambda)]=t^{-(m+2r)}[\sigma_{m}](x,k,\lambda)].
\ee
Hence it follows that the heat-kernel expansion coefficients are obtained from those of the Fourier transform of the resolvent
\be
E_{m}(x|H)=\int \frac{d^{n}k}{(2\pi)^{n}\sqrt{g}}\int_{C}\frac{id\lambda}{2\pi}e^{-t\lambda}[\sigma_{m}](x,k,\lambda).
\ee

The advantage of this approach is that it gives an algorithm to calculate the coefficients $E_{m}(x|H)$ and it can be generalized to the case of non-minimal operators. Non-minimal second order operators consitute indeed a very interesting class of operators, whose general form is
\be
H^{\mu\nu}=-g^{\mu\nu}\Box+a\nabla^{\mu}\nabla^{\nu}+X^{\mu\nu}.
\ee
Here $\nabla^{\mu}$ is the covariant derivative, including both the Levi-Civita connection and the gauge connection. The tensor $X^{\mu\nu}$ is a matrix in the internal indices. The parameter $a$ may assume all real values, in particular for $a=0$ the operator reduces to a minimal one. The gauge field operator for Yang-Mills' theory falls in this class
\be\label{H operator YM}
H^{\mu\nu}_{YM}=-g^{\mu\nu}\Box+\left(1-\frac{1}{\alpha}\right)\nabla^{\mu}\nabla^{\nu}+R^{\mu\nu}.
\ee
This can also be expressed as an operator acting on one-forms
\be
H(\alpha)=\delta d+\frac{1}{\alpha} d\delta.
\ee
In the case $\alpha=1$ it reduces to the Laplace-Beltrami operator, whose action on one forms $\varphi_{\nu}dx^{\nu}$ is given by a Bochner-Lichnerowicz formula
\be
\left(\left(\delta d+\frac{1}{\alpha} d\delta\right)\varphi\right)_{\mu}=(-\delta_{\mu}^{\;\nu}\Box+R_{\mu}^{\;\nu})\varphi_{\nu}.
\ee

Endo found a formula which allows one to express the heat-kernel for a generic value of the parameter $\alpha$ in terms of that relative to  the minimal case $\alpha=1$ \cite{Bimonte}.
\be
K^{(\alpha)}_{\mu\nu'}(\tau)=K^{(1)}_{\mu\nu'}(\tau)+i\int_{\tau}^{\tau/\alpha}dy\;\nabla_{\mu}\nabla^{\lambda}K^{(1)}_{\lambda\nu'}(y)
\ee

It is important to stress that the heat-kernel expansion corresponding to (\ref{H operator YM}) does not make sense in the singular case $\alpha\to0$. In fact in this limit only some coefficients corresponding to a negative power of $t$ are finite, while the others diverge in this limit. Therefore the study of the gauge field propagator in the Landau gauge cannot be done in this way. 
%We give the formulas for the lowest three
%\begin{align}
%[\nabla_{\mu}l]=&k_{\mu}\\
%[\nabla_{\mu}\nabla_{\nu}l]=&\frac{1}{2}k_{\alpha}T^{\alpha}_{\mu\nu}\\
%[\nabla_{\mu}\nabla_{\nu}\nabla_{\lambda}l]=k_{\alpha}\left(-\frac{1}{3}R_{\mu\nu\lambda}^{\quad\;\alpha}-\frac{1}{3}R_{\mu\lambda\nu}^{\quad\;\alpha}+\frac{1}{2}\nabla_{\mu}T^{\alpha}_{\nu\lambda}+\frac{1}{6}\nabla_{\nu}T^{\alpha}_{\mu\lambda}+\frac{1}{6}

\end{subsection}

\begin{subsection}{Effects of the curvature on the size of the Gribov region}
Consider a point $p$ on a given spacetime $M$. We choose Riemannian normal coordinates in a neighborood of $p$. These coordinate system is constructed in the following way. For each $X\in U\subset T_{p}M$ consider the  affinely parametrized geodesic $\gamma_{X}$ starting from $p$ with initial velocity $X$. The exponential map is defined as $\exp(X) | X\in U\to q=\gamma_{X}(1)$. The set of points $q$ constitutes a neighborood $I$ of $p$. If $U$ is sufficiently small the exponential is invertible and one can use coordinates of the vector $X$ in $T_{p}M$ to identify the point $q$. In such a coordinate system, if one takes the coordinate lines to be orthogonal in $p$, the metric and the Christoffel symbols  are given by the following approximate formulae \cite{Bunch:Parker}
\begin{align}\label{Christoffel geodesic all}
g_{\mu\nu}=&\delta_{\mu\nu}-\frac{1}{3}R_{\mu\alpha\nu\beta}X^{\alpha}X^{\beta}+o(||X||^3)\\
\Gamma^{\lambda}_{\mu\nu}=&-\frac{1}{3}\left(R^{\lambda}_{\;\mu\nu\beta}+R^{\lambda}_{\;\nu\mu\beta}\right) X^{\beta}+o(||X||^2)\label{Christoffel geodesic}.
\end{align}
Thus in Riemannian geodesic coordinates, the spacetime is nearly flat. If the gravitational field is weak, we can use perturbation theory to study the small modifications of the theory introduced by the non-vanishing curvature tensor.

The ghost operator in curved spaces is the following
\be
FP(A)=-\nabla_{\mu}\nabla^{\mu}-[A_{\mu},\nabla^{\mu}].
\ee
It acts on scalar fields, so we have\footnote{For the covariant derivative we use the same convention as in \cite{Wald}, according to which $\nabla_{a}\omega_{b}=\partial_{a}\omega_{b}-\Gamma_{ab}^{c}\omega_{c}$.}
\be
FP(A)\omega=-g^{\mu\nu}\left(\partial_{\mu}\partial_{\nu}\omega-\Gamma_{\mu\nu}^{\lambda}\partial_{\lambda}\omega\right)-[A_{\mu},\partial^{\mu}\omega].
\ee
Using formula (\ref{Christoffel geodesic}) we have
\be
\delta^{\mu\nu}\Gamma_{\mu\nu}^{\lambda}=\frac{2}{3}Ric_{\alpha}^{\;\lambda}X^{\alpha},
\ee
so that
\be\label{FP corretto}
FP(A)\omega=\left(-\Box+\frac{2}{3}Ric_{\alpha}^{\;\lambda}X^{\alpha}\partial_{\lambda}-\frac{1}{3}R^{\mu\;\nu}_{\;\alpha\;\beta}X^{\alpha}X^{\beta}\partial_{\mu}\partial_{\nu}\right)\omega-[A_{\mu},\partial^{\mu}\omega].
\ee
The second and third term are corrections which account for the presence of the gravitational field. Now suppose that $\omega$  is a real zero-mode of the flat ghost operator. We can use perturbation theory to evaluate the shift to the zero energy level (we have indeed neglected the contribution of the third term in the previous equation, which seems to be legitimate for a small volume $V$ if $\omega$ and its second derivatives are of the same order)
\begin{multline}
\varepsilon=\frac{2}{3}\frac{\int Ric_{\alpha}^{\;\lambda}X^{\alpha} Tr\left(\omega\partial_{\lambda}\omega\right)}{\int Tr(\omega^2)}=\frac{2}{3}Ric_{\alpha}^{\;\lambda}\frac{\int X^{\alpha}\frac{1}{2}\partial_{\lambda}Tr(\omega^2)}{Tr(\omega^2)}=\\
=-\frac{1}{3}Ric^{\;\lambda}_{\alpha}\frac{(\int\partial_{\lambda}X^{\alpha})Tr(\omega^2)}{\int Tr(\omega^2)}=-\frac{1}{3}Ric^{\;\lambda}_{\alpha}\delta_{\lambda}^{\;\alpha}=-\frac{R}{3}.
\end{multline}
Choosing Dirichlet boundary conditions for the ghost field, we have been able to set to zero the boundary term after integrating by parts. %If we consider a small volume the third term in (\ref{FP corretto}) can be simply ignored.

%Einstein's equations in vacuum in the presence of a cosmological costant read as follows
%\be
%G_{ab}+\Lambda g_{ab}=0,
%\ee
%from which we have
%\be
%\Lambda=\frac{R}{4}.
%\ee
%Hence in vacuum the shift in the zero eigenvalue is simply given by
%\be
%\varepsilon=-\Lambda.
%\ee
The formula we have just derived can be interpreted in the following way: the curvature of space moves the Gribov horizon. This follows from the contribution of the Ricci term in (\ref{FP corretto}). The Riemann term in (\ref{FP corretto}) is now being studied in order to understand if it introduces modifications to the qualitative conclusions due to the Ricci term.  We recall that according to asymptotic freedom, as the energy decreases to zero, the amplitudes of field fluctuations  keep increasing until they reach the Gribov horizon. The presence of the Gribov horizon makes the gauge field propagator decrease to zero at zero momentum. If $R$ is positive the horizon moves inward, hence it is reached at a higher energy and the gauge field propagator should be more suppressed. If $R$ is negative the horizon moves outward, and the energy such that field fluctuations reach the horizon should be lower. Thus, the gauge field propagator should be less suppressed in the infrared in this case.% The argument is 
heuristic, nevertheless we shall prove the conclusion to be true by generalizing Gribov's calculation to the curved case.

In order to give a proof of such a statement we would have to generalize Gribov's construction to the curved case. This turns out to be a difficult task because the non-minimal operator corresponding to the Landau gauge is singular. In the flat case, working in momentum space is of great help. In the curved case, in order to work out the expression for the Green function, one can follow either \cite{Bunch:Parker} or \cite{Gusynin}. However, the success of the method developed in \cite{Bunch:Parker} relies on some remarkable cancellations that occur only in the scalar case. Hence the method can be generalized to the vector case only for minimal operators. On the other hand, following \cite{Gusynin}, some of the coefficients $[\sigma_{m\mu\nu}]$ which determine the asymptotic expansion of the Green function in momentum space are divergent in the limit $\alpha\to0$. Then it seems that, in order to study the consequences of the Gribov ambiguity in curved spaces and make comparisons with results obtained in the 
flat case, one has to work in position space. However, as we already pointed out, also the heat-kernel expansion is not defined in the limit $\alpha\to0$, therefore one cannot obtain the Green function from it. These difficulties pose serious limits to our ability to prove such a conjecture.
\end{subsection}
\end{section}

\begin{section}{Conclusions}
The Gribov ambiguity plays an important role in the quantum dynamics of non-Abelian Yang-Mills theories at low momenta, where the perturbative expansion is not reliable. It leads to the introduction of strong costraints on the configuration space of the theory. One finds out that the solution to the ambiguity in the Landau gauge automatically entails a solution to the mass gap problem (see again \cite{JaffeWitten:Millennium} for a discussion).

In more recent times the Gribov problem has been considered also in curved spaces. In \cite{Canfora},\cite{Canfora:Hedgehog} the equation for the Gribov copies  of the perturbative vacuum in transverse gauges has been studied in spherically symmetric backgrounds and topologically non-trivial backgrounds, respectively. In the Coulomb gauge copies of the vacuum are not admitted on a spherically symmetric spacetime with a smooth metric; though such copies do exist on asymptotically AdS spacetimes with a bulk. A generalization of the hedgehog ansatz, used in the study of the non-linear sigma model, leads to a simpler form for the Gribov copies equation, which is of great help in addressing the problem of Gribov copies of the vacuum on a generic spacetime manifold. As it has been pointed out in \cite{Canforabelian}, even in the Abelian case spacetime curvature may be responsible for the appearance of copies of the vacuum. Indeed there are some backgrounds such that these copies are infinite in number. Therefore, 
restricting to the orthogonal compliment of the subspace they span, could lead to a mutilation of a significant part of the vector space associated to the ghost field.

Research on the subject is receiving new input also from analysis performed in the flat case.  In particular, it has been shown in \cite{Lechtenfeld} that it is possible to include a Gribov horizon effective term in the action even in the case of a generic $R_{\xi}$ gauge. The generalized horizon term, which reduces to the usual one in the Landau gauge, is constructed by requiring the quantum theory to be invariant under field-dependent BRST transformations.

In the present paper we presented an attempt to derive the infrared modified gluon two-point function, following Gribov's original construction \cite{Gribov:78}. In this framework we need, as a starting point, to calculate the gluon two-point function for the original Yang-Mills theory. This is needed in order to take into account corrections, due to the Gribov horizon effective action, perturbatively. A powerful tool to calculate such objects is provided by the heat-kernel asymptotic expansion. However, this turns out to be not defined in the limit $\alpha\to 0$, in which the gauge field operator becomes singular, which is the one we have to stick to if we want to work in the Landau gauge. Moreover, the applicability of this method is at least questionable, since the heat-kernel expansion is used to calculate Green's functions according to a local construction, which is suitable for calculating UV divergences, while the presence of the Gribov horizon is primarily an IR problem. Therefore it seems more 
appropriate to use global methods, which are capable of capturing the long-distance behaviour of the theory. Further investigations are being carried on along this direction.
%rassegna breve dei lavori di Canfora in spazi curvi
%rssegna di alcuni lavori recenti in spazio piatto, per esempio Lechtenfeld
\end{section}

\appendix
\begin{section}{Elliptic operators} 
This appendix is devoted to the definition of the concept of elliptic operator, used in (\ref{heat}). We will follow \cite{Nakahara:GTP}.

Let $E$ and $F$ be two fibre bundles and denote the spaces of sections respectively by $\Gamma(M,E)$ and $\Gamma(M,F)$. Consider an operator $H$ acting on $\Gamma(M,E)$ with values in $\Gamma(M,F)$. In local coordinates its action on a section $s(x)\in\Gamma(M,E)$ assumes the form
\be
(Hs(x))^{a'}=\sum_{M\leq N}A^{a'}_{Ma}D_{M}s(x)^{a},
\ee
where $a$, $a'$ are fibre indices in $E$ and $F$, respectively. $D_{M}$ is a derivative of order $M$ and $N$ is the order of the operator. Replacing $D_{M}$ with a monomial of degree $M$ in the components of a covector $\xi$ one gets the \emph{symbol} of the operator $H$. This object is actually coordinate dependent, but the symbol of the highest order part does not depend on the choice of a coordinate system. The latter is called the \emph{leading symbol} of $H$, denoted by $\sigma(H,\xi)$. It can also be obtained from the following formula
\be
\sigma(H,\xi)\omega=\frac{1}{N!}H(f^{N}s)|_{p},
\ee
where $f^{N}$ is a function such that $f^{N}(p)=0$ and $df^{N}(p)=i\xi$, furthermore the section $s$ is chosen in a way such that $s(p)=\omega$. This definition of the leading symbol of $H$ at the point $p$ is manifestly coordinate independent. It is straightforward to verify that this definition agrees with that given above.

The symbol is a map from the fibre of $E$ at $p$ to the corresponding fibre of $F$
\be
\sigma(H,\xi): \pi_{E}^{-1}(p)\rightarrow\pi_{F}^{-1}(p).
\ee
The operator $H$ is said to be elliptic if and only if the symbol is invertible for each $p$ and $\xi\neq0$. Clearly a necessary condition is that the dimensions of the fibres are the same.

For the Laplace operator $E$ and $F$ are real line bundles. Its leading symbol is
\be
\sigma(-\triangle,\xi)=||\xi||^2.
\ee
The Dirac operator maps the space of sections of a Clifford bundle into itself. It is written as
\be
i\gamma^{\mu}\nabla_{\mu}=i\gamma^{\mu}(\partial_{\mu}+\omega_{\mu}),
\ee
where $\omega_{\mu}=\frac{1}{2}i\omega_{\mu}^{\;\alpha\beta}\Sigma_{\alpha\beta}$ is the spin connection and $\gamma^{\mu}$ is expressed in terms of the flat Dirac matrices\footnote{The Dirac matrices are taken to be hermitian.} and the vierbein  as $\gamma^{\mu}=\gamma^{\alpha}e_{\alpha}^{\;\mu}$.
The leading symbol is the following:
\be
\sigma(i\gamma^{\mu}\nabla_{\mu},\xi)=-\gamma^{\mu}\xi_{\mu}.
\ee
Operators whose leading symbol coincides with that of the Laplace or Dirac operator are called respectively Laplace-type or Dirac-type operators. Non-minimal second order operators are of great interest, because of their use in gauge theories, see (\ref{H operator YM}). The leading symbol of the operator $H_{YM}$ is
\be
\sigma(H_{YM},\xi)=g^{\mu\nu}||\xi||^2-\left(1-\frac{1}{\alpha}\right)\xi^{\mu}\xi^{\nu}.
\ee

The symbol of a differential operator is a polynomial in $\xi$. Considering non-polynomial functions of $\xi$ it is possible to construct a more general class of operators, called \emph{pseudo-differential} operators.  
\end{section}
\section*{Acknowledgments}
I wish to thank Dr Esposito for his kind guidance and for sharing his vast knowledge of the scientific literature. I acknowledge my indebtedness to Professor Fedele Lizzi and Dr Patrizia Vitale for their support, both scientific and human. I am grateful to Fabrizio Canfora who first introduced me to the Gribov ambiguity and for his many useful suggestions. I wish to thank Professors Ivan Avramidi, Valeriy Gusynin and Daniel Zwanziger for correspondence. Finally, and above all, I wish to thank Federica for her love and encouragement.

\printbibliography
\end{document}